\documentclass[a4paper,traditabstract]{aa}  
\usepackage{txfonts}

\usepackage{graphicx}	
\usepackage{amssymb}	
\usepackage{morefloats}
\usepackage{color}
\usepackage{footnote}
\usepackage{tablefootnote}
\usepackage{subfig}
\usepackage{appendix}
\usepackage{natbib,twoopt}
\usepackage{pdflscape}
\usepackage{pgfplots}
\usepackage{ftnxtra}
\usepackage{fnpos}
\usepackage{ulem}
\usepackage{rwu}
\usepackage[utf8]{inputenc}
\DeclareUnicodeCharacter{2212}{-}
\graphicspath{{./Figures/}}
\DeclareGraphicsExtensions{.png,.pdf,.jpg, .pgf}

\begin{document}
\title{Constraining physical conditions for the PDR of Trumpler 14 in the Carina Nebula\thanks{{\it Herschel} is an ESA space observatory with science instruments provided by European-led Principal Investigator consortia and with important participation from NASA.}}
\titlerunning{Carina Nebula}
\author{
Ronin Wu\inst{1, 2}
\and
Emeric Bron\inst{1, 3}
\and
Takashi Onaka\inst{2}
\and
Franck Le Petit\inst{1}
\and
Fr\'ed\'eric Galliano\inst{4}
\and
David Languignon\inst{1}
\and
Tomohiko Nakamura\inst{2, 5}
\and
Yoko Okada\inst{6}
}
\institute{
LERMA, Observatoire de Paris, PSL Research University, CNRS, Sorbonne Universit\'es, UPMC Univ. Paris 06, F-92190, Meudon, France;
\email{ronin.wu@obspm.fr}
\and
Department of Astronomy, the University of Tokyo, Bunkyo-ku, Tokyo 113-0033, Japan
\and
Grupo de Astrof\'isica Molecular. Instituto de Ciencia de Materiales de Madrid (ICMM, CSIC), Sor Juana Ines de la Cruz 3, 28049 Cantoblanco, Madrid, Spain
\and
IRFU, CEA, Universit\'e Paris-Saclay, F-91191 Gif-sur-Yvette, France \\    
Universit\'e Paris-Diderot, AIM, Sorbonne Paris Cit\'e, CEA, CNRS, F-91191 Gif-sur-Yvette, France
\and
Recruit Communications Co. Ltd, Tokyo, Japan
\and
I. Physikalisches Institut der Universit\"at zu K\"oln, Z\"ulpicher Strasse 77, 50937 K\"oln, Germany
}
\authorrunning{Ronin Wu et al.}

\abstract
{We investigate the physical conditions of the CO gas, based on the submillimeter imaging spectroscopy from a $2'\,\times\,7'$~($1.5\,\times\,5\,\mathrm{pc^{2}}$) area near the young star cluster, Trumpler\,14 of the Carina Nebula. The observations presented in this work are taken with the Fourier Transform Spectrometer~(FTS) of the Spectral and Photometric Imaging REceiver (SPIRE) onboard the \textsl{Herschel} Space Observatory. The newly observed spectral lines include $\cits$, $\cison$, and CO transitions from $\mathrm{J}=4-3$ to $\mathrm{J}=13-12$. Our field of view covers the edge of a cavity carved by Trumpler\,14 about $1\,\mathrm{Myr}$ ago and marks the transition from H\textsc{ii} regions to photo-dissociation regions. The observed CO intensities are the most prominent at the northwest region, Car I-E. With the state-of-the-art Meudon PDR code, we successfully derive the physical conditions, which include the thermal pressure~($P$) and the scaling factor of radiation fields~($G_{\mathrm{UV}}$), from the observed CO spectral line energy distributions~(SLEDs) in the observed region. The derived $G_{\mathrm{UV}}$ values generally show an excellent agreement with the UV radiation fields created by nearby OB-stars and thus confirm that the main excitation source of the observed CO emission are the UV-photons provided by the massive stars. The derived thermal pressure is between $0.5-3\,\times\,10^{8}\,\mathrm{K\,cm^{-3}}$ with the highest values found along the ionization front in Car I-E region facing Trumpler\,14, hinting that the cloud structure is similar to the recent observations of the Orion Bar. We also note a discrepancy at a local position~($<\,0.17\,\times\,0.17\,\mathrm{pc^{2}}$) between the PDR modeling result and the UV radiation fields estimated from nearby massive stars, which requires further investigation on nearby objects that could contribute to local heating, including outflow. Comparing the derived thermal pressure with the radiation fields, we report the first observationally-derived and spatially-resolved $P \sim 2\times10^4\,G_{\mathrm{UV}}$ relationship. As direct comparisons of the modeling results to the observed $^{13}\mathrm{CO}$, [O\textsc{i}]\,$63\,\mum$, and [C\textsc{ii}]\,$158\,\mum$ intensities are not straightforward, we urge the readers to be cautious when constraining the physical conditions of PDRs with combinations of $^{12}\mathrm{CO}$, $^{13}\mathrm{CO}$, [C\textsc{i}], [O\textsc{i}]\,$63\,\mum$, and [C\textsc{ii}]\,$158\,\mum$ observations.
}

\keywords{
ISM: individual objects: Carina Nebula -- Interstellar medium (ISM), nebulae -- ISM: clouds -- ISM: molecules -- ISM: structure -- (Galaxy:) open clusters and associations: individual: Trumpler 14
}
\maketitle

\section{Introduction}\label{tex:intro}
The radiative feedback of stars on their parent cloud is a key topic both in the context of evolution and composition of interstellar matter and to constrain star formation mechanisms. Emission lines from photo-dissociation regions (PDRs) have been studied for decades to understand the physical and chemical processes induced by this feedback \citep{Tielens1985, Sternberg1989}. In these regions, far ultraviolet (FUV) photons dissociate molecules and heat the gas at several hundred Kelvins via photo-electric effect on grains and, in dense gas, by H$_2$ pumping followed by collisional de-excitations~\citep{Roellig2006}. Several tracers are commonly used to study PDRs. Observations of atomic lines of O, C$^+$ and C probe the neutral hydrogen layer and give access to most of the gas cooling rate at the  edge of the cloud. The atomic-molecular transition can be observed thanks to the emission of H$_2$ infrared lines. Less abundant molecules, diatomic and complex ones as carbon chains, formaldehyde or methanol, have been detected at various depths in PDRs showing that complex chemistry takes place in this hostile environment \citep{YoungOwl2000, Lis2003, Guzman2013, Nagy2013, Guzman2014, Cuadrado2015}. Because H$_2$ is difficult to observe, except at the warm surface layer of cloud, due to its lack of electric dipole moment, many studies rely on the observation of CO and its isotopologs to constrain the mass of gas and the morphology of clouds.

Since its launch in 2009, \herschel~\textsl{Space Observatory}~\citep{Pilbratt2010} has allowed Galactic and extragalactic observations of excited CO in high-J states (J$_{\mathrm{up}}>10$) in PDRs of various objects~\citep{Kohler2014, Stock2015, Wu2015, Parikka2017}. In parallel, observations of CO emission in protostars and extragalactic environments showed highly excited CO. For example, emission from rotational levels J$_{\mathrm{up}}>40$ has been observed in the protostellar region Orion BL/KL, suggesting an excitation by outflows and shocks~\citep{Goicoechea2015a}. In extragalactic regions, the origin of CO excitation is unclear \citep{Yildiz2010, Hailey-Dunsheath2012, Greve2014, Kamenetzky2016, Rosenberg2015}. Candidate excitation sources include mechanical-heating, shocks, X-rays, and UV photons, depending on the objects and the adopted tools in the analyses. It has been a topic of interest whether it is possible to infer the UV radiation field intensity and star formation rate from high-J CO line intensities. The study of CO excitation in spatially resolved Galactic PDRs can help to answer this question. Presently, most analyses rely on the usage of non-local thermal equilibrium~(non-LTE) radiative transfer codes, such as large velocity gradient~(LVG) approximations, to infer gas density and temperature from CO line intensities~\citep{Kohler2014}. Very few detailed analyses based on PDR modeling, which constrains the thermal and chemical processes hidden behind the observed CO lines, can be found in the literature. Based on the non spatially-resolved CO observations from $\mathrm{J=4-3}$ up to $\mathrm{J=19-18}$ in NGC 7023 and up to $\mathrm{J=23-22}$ in the Orion Bar, as well as other atomic and molecular lines, and their comparisons with the Meudon PDR code \citep{LePetit2006}, it is suggested that the FUV radiation field produced by the Trapezium stars cluster and HD~200775 are responsible for a high pressure layer ($P\sim$10$^8$ K cm$^{-3}$) at the very edge of the two PDRs and is sufficient to support the high-J CO excitation observed in the two objects~\citep{Joblin2018}. These results are in resonance with the recent \textsl{ALMA} observations which confirm the presence of a thin layer with high pressure at the edge of the PDR in the Orion Bar~\citep{Goicoechea2016, Goicoechea2017}. More studies as such with detailed PDR modeling are necessary for understanding the excitation of CO in the interstellar environment.

\begin{figure*}[thbp!]
\centering
\includegraphics[width=\textwidth]{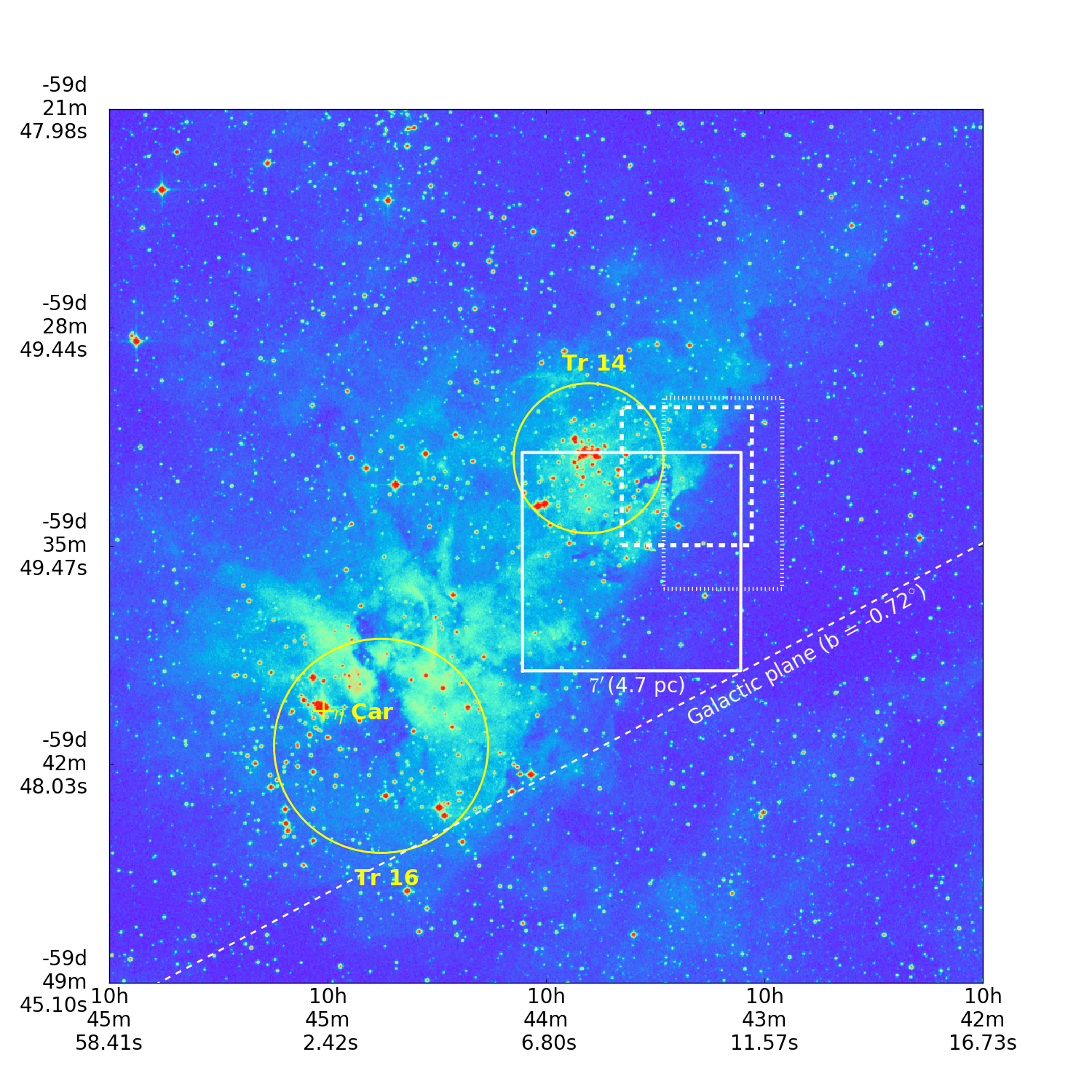}
\caption{
Overview of the region north of the dust-lane in the Carina Nebula. The background optical image is taken with the UK Schmidt Telescope in the DSS2-red band.   The locations and sizes of the two most massive star-clusters, Trumpler 14 and Trumpler 16, are marked with yellow circles~\citep{Tapia2003}. The areas enclosed by the white dotted and dashed lines indicate the regions previously observed by \citet{Brooks2003} and \citet{Kramer2008}. The long dashed line across the image marks the Galactic plane at $b=-0.72^{\circ}$. The square enclosed by the solid white line indicates the FOV in this work. It spans an area of $7'\times7'~(4.7\times4.7\,\mathrm{pc})$. A zoom-in view of the area included in the white square can be found in Figure~\ref{fig:carina_zoom}.
  \label{fig:carina_fov}
  }
\end{figure*}
\begin{figure*}[thbp!]
\centering
\includegraphics[width=\textwidth]{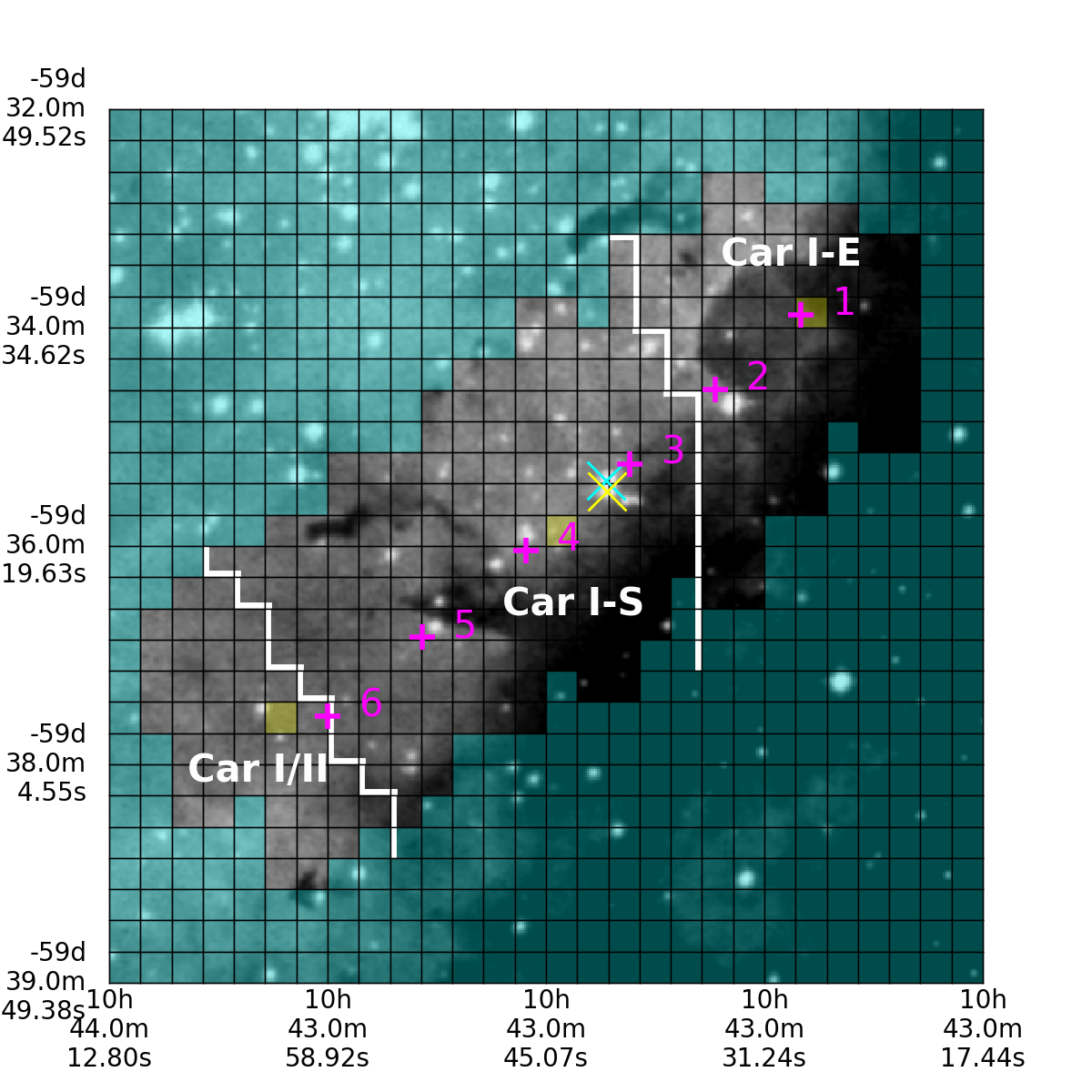}
\caption{
A zoom-in of our FOV (the white square in Figure~\ref{fig:carina_fov}). The effective FOV is indicated by the pixels in gray-colors. The background image is the optical image taken in the DSS2-red band. The positions of the central bolometer, SLWC4, at the six requested pointings are indicated with the magenta crosses and labeled accordingly as in Table~\ref{table:obs}. The boundaries of Car I-E, Car I-S, and Car I/II areas are indicated with the thick white lines. The locations of two massive stars, \textsl{ALS 15 204} and \textsl{ALS 15 203}, discussed in Section~\ref{sec:mod_results} are marked with the cyan and yellow ``$\times$'' in the figure. The CO SLED observed from the three yellow-masked pixels are displayed in Figure~\ref{fig:sled_3pix} and discussed in Section~\ref{subsec:co_structure}.
\label{fig:carina_zoom}
}
\end{figure*}

Another well known object that is suitable for studying the impact of radiative feedback of massive stars on their parent cloud is the Carina Nebula. Located at a close distance, $2.3\,\mathrm{kpc}$~\citep{Smith2006}, from us in the Milky Way, the Carina Nebula is the largest and brightest nebula in the southern sky. Some of its central regions are even brighter than the famous Orion Nebula. Aside from the famous luminous blue variable, $\eta$ Carinae~($\eta$\,Car), Carina Nebula hosts 70 known O stars, 3 Wolf-Rayet stars, and 127 B0 to B3 stars~\citep{Gagne2011}.

It was first revealed in the observations of radio continuum at $6\,\mathrm{cm}$ that the H\textsc{ii} regions of the Carina Nebula include two major components, Car I and Car II, separated by $\sim\,10'$~($\sim\,7\,\mathrm{pc}$) along the Galactic plane~\citep{McGee1968, Gardner1968}. Based on the distribution of local maxima of radio continuum brightness at $0.843\,\mathrm{GHz}$, \citet{Whiteoak1994} detailed the structure of Car I into Car I-W, Car I-E, and Car I-S, which are associated to the ionization front shaped by the UV-photons from Trumpler\,14 in the west, east and south of Car I. With the NASA Kuiper Airborne Observatory~(KAO) far-infrared~(FIR) and Fleurs Synthesis Telescope (FST) $1415\,\mathrm{MHz}$~observations, it is confirmed that the main heating sources for Car I and Car II are Trumpler\,14 and Trumpler\,16, respectively~\citep{Harvey1979, Retallack1983}. Figure~\ref{fig:carina_fov} gives an overview of the north region of the Carina Nebula observed by the UK Schmidt Telescope in the R-band of the Second Digital Sky Survey~(DSS2 red)\footnote{Based on photographic data obtained using The UK Schmidt Telescope. The UK Schmidt Telescope was operated by the Royal Observatory Edinburgh, with funding from the UK Science and Engineering Research Council, until 1988 June, and thereafter by the Anglo-Australian Observatory. Original plate material is copyright (c) of the Royal Observatory Edinburgh and the Anglo-Australian Observatory. The plates were processed into the present compressed digital form with their permission. The Digitized Sky Survey was produced at the Space Telescope Science Institute under US Government grant NAG W-2166.}.

In this paper, we study the CO and C emission observed by \herschel~SPIRE/FTS across an H\textsc{ii}--PDR interface located at the northwest in the Carina Nebula to better understand the origin of CO excitation in PDRs as well as the impact of radiative feedback of massive stars. The observed region~(enclosed by the white square in Figure~\ref{fig:carina_fov}) covers Car I-E~(upper right), Car I-S~(center), and a small region at the intersection of Car I and Car II~(Car I/II, lower left, see also Figure~\ref{fig:carina_zoom} for the region definition). Our field-of-view~(FOV) covers the edge of a cavity likely shaped by the young~($1-2\,\mathrm{Myr}$) OB-star cluster Trumpler\,14 since $\sim\,1\,\mathrm{Myr}$ ago with its ionization front sitting at a projected distance of $\sim\,2\,\mathrm{pc}$ from Trumpler 14. 

Trumpler\,14 hosts a dozen of O-stars and more than a hundred of B-stars. Its stellar components have been well studied~\citep{Vazquez1996, Carraro2004}. The estimated UV input by the stellar members of Trumpler\,14 to the ionization front of PDR is~$\simeq 10^{4}\,G_{0}$~\citep{Brooks2003} where $G_0$ = $1.6\times10^{-3}$ erg s$^{-1}$ cm$^{-2}$ according to~\citet{Habing1968}~(hereafter, the Habing unit). For comparison, this FUV flux is slightly lower than the one induced by the Trapezium star cluster on the famous Orion Bar ($\sim\,3\,\times\,10^{4}\,G_{0}$,~\citealt{Goicoechea2015b}) and comparable to the one found in extragalactic starburst regions~\citep{Malhotra2001, Chevance2016}. Even though the FUV flux is similar between the Carina Nebula and the Orion Bar, the distance of the Trapezium star cluster to the Bar is $\sim$ 10 times closer than in the Carina Nebula. As Trumpler\,14 is a more evolved star cluster than the Trapezium star cluster, the interaction zones at its PDR interface can be better approximated by pressure equilibrium in a stationary PDR model~\citep{Brooks2003}. Within the dense molecular gas at the northwest region, young stellar objects~(YSOs) are carving out ionized regions anew, proving the on-going and active star formation in this region~\citep{Tapia2015}.  Such a region is likely a representation of prototypical PDRs near star--forming sites that dominate the extragalactic observations of starburst systems where individual PDRs are mostly unresolved in the infrared and submm.

The PDRs in the Car I-E region have thus far been studied by several groups. The FUV flux in this region, as estimated from stellar composition and FIR observations, is $\sim\,10^{4}$ in the Habing unit~\citep{Brooks2003, Mizutani2004a}. However, estimations with PDR models have given somehow lower values of FUV flux~($1390$, \citealt{Oberst2011} and $3200$,~\citealt{Kramer2008} in the Habing unit). In a big picture, these results indicate that the UV-photons provided by nearby massive stars are more than sufficient in supporting the global heating of PDRs observed in this region. However, the inconsistency persists between the model-estimated values of FUV flux and the UV radiation fields estimated from stellar composition and FIR observations.

In this work, we adopt a distance of $2.35\pm0.5\,\mathrm{kpc}$ to the Carina Nebula~\citep{Smith2006}, although it is worth noting that the photometric analysis from pre-main sequence~(PMS) stars in the nebula gives a distance of $\sim2.9\,\mathrm{kpc}$ to the star clusters, Trumpler\,14 and 16~\citep{Hur2012}. All the sky coordinates quoted in this work are in the J2000 epoch. In the model analysis, we adopt the scaling factor according to the radiation field given by~\citet{Mathis1983}~(hereafter, the Mathis unit) unless otherwise specified. The difference between the Habing and Mathis units is about a factor of $1.3$, in the spectral range of $91.2$ to $111\,\mathrm{nm}$. Our observations and data reduction procedure are presented in section \ref{sec:obs}. We interpret these data with the Meudon PDR code in  section \ref{sec:interpretation}. Results are then presented and discussed in sections \ref{sec:results} and \ref{sec:discussion}, respectively.

\section{Observations}
\label{sec:obs}

The \herschel~data presented in this work have been observed on 2012-12-03 and 2013-02-04 as one of the Open Time~(OT1) programs~(P.I.: Takashi Onaka). Our results are mainly derived from the spectroscopic data observed by the \textsl{Fourier Transform Spectrometer}~(FTS) of the \textsl{Spectral and Photometric Imaging REceiver}~(SPIRE,~\citealt{Griffin2010}) on board the \herschel\ \textsl{Space Observatory}~\citep{Pilbratt2010}. We give a brief instrumentation overview of the \herschel\ SPIRE/FTS, and then layout the steps we take to map the observations.

\subsection{An overview of the \herschel\ SPIRE/FTS}
\label{subsec:fts}

The \herschel\ SPIRE/FTS is made of two bolometer arrays, the SPIRE Short Wavelength~(SSW,~$194-324\,\mum$) and SPIRE Long Wavelength~(SLW,~$316-672\,\mum$) arrays. Each of the SSW and SLW arrays, respectively, packs 19 and 7 unvignetted bolometers hexagonally on the focal plane and simultaneously covers a $2.6'$ diameter of field of view~(FOV) on the sky. Our observations have been completed in six single-pointings on the operational days 1299 and 1362. At each single-pointing position, we observe the sky in the intermediate spatial-sampling~(4 jiggle-pointings) and high spectral-resolution~($\Delta\sigma\,=\,0.04\,\mathrm{cm^{-1}}$ or $\Delta\nu\,=\,1.2\,\mathrm{GHz}$) modes. The information of our observation is listed in Table~\ref{table:obs}. For each jiggle-pointing, the sky has been observed in 8 scans, which add up to an effective exposure time of approximately $266\,\mathrm{s}$ across the FOV. The spectra presented in this work are calibrated against Uranus, {\it i.e.} point-source calibrated, using the SPIRE calibration tree, \texttt{spire\_cal\_14\_3}. Details of the SPIRE/FTS calibration can be found in \citet{Swinyard2014}. Beginning with the unapodized spectra, averaged from all scans, and their standard-deviation as the uncertainties, we describe our map-making strategy in the following subsection.

\begin{table*}
\centering
\caption{The \herschel~SPIRE/FTS observations presented in this work. }
\label{table:obs}
\begin{tabular}{cccccl}
\hline\hline
Pointing label   &   Obs\_ID  &  RA  &  Dec  &  Obs. Time  & Note  \\
        &      &   (J2000) & (J2000)    &       &    \\
\hline
1    &    1342262913    &    10:43:29.08    &    -59:34:29.96    &    2013-02-04 06:46:20    &    C\\
2    &    1342262914    &    10:43:34.47    &    -59:35:05.56    &    2013-02-04 07:24:13    &    C'\\
3    &    1342262915    &    10:43:39.85    &    -59:35:41.59    &    2013-02-04 08:02:06    &    SWS\\
4    &    1342262916    &    10:43:46.45    &    -59:36:23.36    &    2013-02-04 08:39:59    &    B'\\
5    &    1342256377    &    10:43:53.05    &    -59:37:04.77    &    2012-12-03 13:39:48    &    B\\
6    &    1342256376    &    10:43:59.03    &    -59:37:42.62    &    2012-12-03 13:01:54    &    A\\
\hline\hline
\end{tabular}
\tablefoot{The pointing labels are according to the positions indicated in Figure~\ref{fig:carina_zoom}. The Obs\_ID is the unique identifier for \herschel~observations. The labels used in \citet{onaka2008} are given in the Note column.}
\end{table*}

\subsection{Making maps of the \herschel~SPIRE/FTS lines}\label{subsec:mapmaking}

\begin{figure}[htbp!]
\centering
\includegraphics[width=0.5\textwidth]{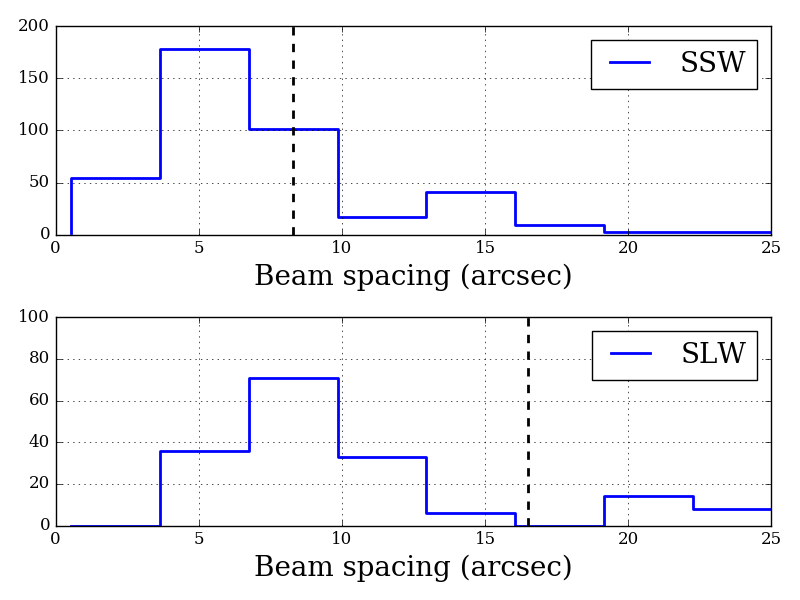}
\caption{Histograms of the distances for a given bolometer to its closest neighbor for the SSW~(top) and SLW~(bottom) arrays. The dashed lines mark half of the smallest beam FWHM listed in Table~\ref{table:linetable}, which correspond to $8.3''$ and $16.5''$ for the SSW and SLW arrays respectively.
\label{fig:beam_spacing}}
\end{figure}

In the intermediate-sampling mode, the bolometer spacings for the SSW and SLW arrays are respectively $16.3''$ and $25.3''$, which are about the sizes of individual beams at their minimum within the bandwidths. Although the individual single-pointing observation used in this work is not fully-Nyquist sampled, combining the footprints of bolometers from all six single-pointing positions~(see Figure~\ref{fig:carina_zoom}), the average bolometer spacings in our FOV achieve $8.6''$ and $12.2''$ for SSW and SLW arrays respectively. Histograms of distances between any given bolometer to its closest neighbor for SSW and SLW arrays are shown in Figure~\ref{fig:beam_spacing}, and they demonstrate that the observation within the effective FOV is approximately fully-Nyquist sampled.

Within the bandwidth~($194 < \lambda < 672 \, \mu\textrm{m}$) of the \herschel\ SPIRE/FTS, the relevant atomic and molecular lines used in this study are listed in Table~\ref{table:linetable}. As discussed in~\citet{Wu2013} and \citet{Swinyard2014}, the beam full-width-half-maxima~(FWHM) and sensitivities of bolometers vary across the instrument bandwidth. At each observed emission line, we list the correspondent instrument FWHM and mean $1\sigma$ uncertainty of the integrated line intensity~($\epsilon$) in Table~\ref{table:linetable}.

Around the frequency~($\nu_{\mathrm{line}}-15\,\ghz<\nu<\nu_{\mathrm{line}}+15\,\ghz$) of each line listed in Table~\ref{table:linetable}, the observed spectra from all unvignetted bolometers are assigned to a regular-grid map. The world coordinate system~(WCS) of the maps is centered at ($\alpha$,$\delta$) = (10:43:45.07, -59:36:19.63) with each pixel sized $15''\,\times\,15''$~($0.17\,\times\,0.17\,\mathrm{pc}$) and a total dimension of $7'\,\times\,7'$~($4.7\,\times\,4.7\,\mathrm{pc}$). We follow a similar map-making recipe detailed in \S\,2.1.2 of \citet{Wu2015} and briefly discuss the main steps here. The stand-alone map-making software, \texttt{HFTS\_mapping}, is written in Python and uses utilities publicly available in \texttt{astropy}\footnote{http://www.astropy.org/}~(version 1.1.1,~\citealt{Robitaille2013}), \texttt{scipy}~(version 0.17.0) and \texttt{numpy}~(version 1.12.1)\footnote{https://www.scipy.org/}. \texttt{HFTS\_mapping} is now available for public use\footnote{https://ism.obspm.fr/files/HFTS\_Mapping/HFTS\_mapping\_20171220.tgz}.

Our map-making procedure can be laid out in three steps:
\begin{itemize}
\item First, before coadding the level-2 bolometer spectra on the map, a continuum level is estimated between two frequency-intervals, ($\nu_{\mathrm{line}}-15\,\ghz<\nu<\nu_{\mathrm{line}}-5\,\ghz$) and ($\nu_{\mathrm{line}}+5\,\ghz<\nu<\nu_{\mathrm{line}}+15\,\ghz$), and removed as a parabola function. Weighted by the uncertainties, which are estimated as the standard deviation from all scans, the continuum-removed bolometer spectra are then assigned to the pixels, according to the bolometer pointings, to create spectral cubes over the FOV for each line listed in Table~\ref{table:linetable}.
\item Second, we measure the integrated line-intensity, assuming a sinc profile, from spectral cubes constructed in the previous step. For $\mathrm{CO\ J=7-6}$ and $\mathrm{[C\textsc{i}]^{3}\mathrm{P}_{1}-^{3}\mathrm{P}_{2}}$, which are only $2.7\,\mathrm{GHz}$ apart, their integrated line-intensities are measured simultaneously with two sinc profiles whose central frequencies are fixed to differ by $2.7\,\mathrm{GHz}$. At the end of this step, the maps of integrated line-intensities are produced~(see figures in Appendix\,\ref{appmaps}). 
\item The last step of map-making is to convolve the integrated line-intensity maps to the same spatial resolution, $42''$~($\sim\,0.5\,\mathrm{pc}$), which is the largest beam FWHM among all observed lines~(see Table~\ref{table:linetable}). We refer the curious readers to \citet{Wu2015} for more detailed description on the generation of kernels used in convolution.
\end{itemize}

Uncertainties for the final integrated line-intensities are estimated by a Monte-Carlo experiment which comprises 300 repetitions of the above three steps. For each repetition, the uncertainties used for coaddition of bolometer spectra in the first step are replaced by randomly generated noises which include random and systematic uncertainties from the observation. The random uncertainties are represented by a normal distribution whose standard-deviation is the root-mean-square~(rms) of residuals from a given bolometer spectrum after subtracting its continuum and the sinc line-profile(s).

For all bolometer spectra taken in the same single-pointing observation, one normal distribution that represents the systematic uncertainty has also been taken into account. The 1-$\sigma$ of the normal distribution for the systematic uncertainty is set to be $10\%$ of the bolometer spectrum~\citep{Swinyard2014}. The uncertainties used in the first step are then replaced by the rms of the generated random and systematic uncertainties. At the end of the Monte Carlo experiment, 300 convolved maps of integrated line-intensities for each line are produced. For a given line on each pixel, the standard-deviation between the 300 maps is taken as the final uncertainty.

Within our FOV, the $\mathrm{CO\ J=4-3}$ and $\mathrm{J=7-6}$ transitions have previously been observed with the $4\,\mathrm{m}$ NANTEN2 telescope. The FWHMs of the NANTEN2 observations are $38''$ and $26.5''$ for the $\mathrm{CO\ J=4-3}$ and $\mathrm{J=7-6}$ transitions, respectively, with $20\%$ of calibration uncertainty~\citep{Kramer2008}. After convolving the $\mathrm{CO\ J=4-3}$ and $\mathrm{J=7-6}$ images observed by NANTEN2 to the instrument FWHM of the \herschel~SPIRE/FTS, the data observed by the two telescopes show agreement within $30\%$, while the observed values by the~\herschel~SPIRE/FTS are systematically higher. Taking the uncertainty in the atmospheric transmission model~\citep{Pardo2001} into account, the observations compare reasonably well.

\begin{table*}[thbp!]
  \centering
  \caption{Emission lines used to constrain the physical conditions and observed by the \textsl{Herschel} SPIRE FTS. }
  \label{table:linetable}
  \begin{tabular}{lccccc}
      \hline\hline
      Transition & Frequency & E$_{up}$ & g$_{up}$  &FWHM  &  $\epsilon$\\
	               &	($\ghz$)&	    ($\kelvin$)    &&($''$)   &  ($10^{-9}\,\mathrm{W\,m^{-2}\,sr^{-1}}$)\\
      \hline
	$\mathrm{CO\ J=4-3    }$              &   $461.041$   &   $55.32$   &   $9$   &   $41.7$   &   $1.40$\\
	$\mathrm{[C\textsc{i}\,]^{3}\mathrm{P}_{0}-^{3}\mathrm{P}_{1}}$   &   $492.161$   &   $23.62$   &   $3$   &   $38.1$   &   $0.84$\\
	$\mathrm{CO\ J=5-4    }$              &   $576.268$   &   $82.97$   &   $11$   &   $33.5$   &   $1.27$\\
	$\mathrm{CO\ J=6-5    }$              &   $691.473$   &   $116.16$   &   $13$   &   $29.3$   &   $1.02$\\
	$\mathrm{CO\ J=7-6    }$              &   $806.652$   &   $154.87$   &   $15$   &   $33.0$   &   $1.02$\\
	$\mathrm{[C\textsc{i}\,]^{3}\mathrm{P}_{1}-^{3}\mathrm{P}_{2}}$   &   $809.342$   &   $62.46$   &   $5$   &   $33.0$   &   $1.01$\\
	$\mathrm{CO\ J=8-7    }$              &   $921.800$   &   $199.11$   &   $17$   &   $33.2$   &   $1.88$\\
	$\mathrm{CO\ J=9-8    }$              &   $1036.912$   &   $248.88$   &   $19$   &   $19.1$   &   $2.22$\\
	$\mathrm{CO\ J=10-9   }$              &   $1151.985$   &   $304.16$   &   $21$   &   $17.6$   &   $2.53$\\
	$\mathrm{CO\ J=11-10  }$              &   $1267.015$   &   $364.97$   &   $23$   &   $17.3$   &   $2.65$\\
	$\mathrm{CO\ J=12-11  }$              &   $1381.995$   &   $431.29$   &   $25$   &   $16.9$   &   $2.52$\\
	$\mathrm{CO\ J=13-12  }$              &   $1496.923$   &   $503.13$   &   $27$   &   $16.6$   &   $4.37$\\
      \hline\hline
  \end{tabular}
  \tablefoot{The two last columns correspond to the instrument FWHM and to the mean 1$\sigma$ uncertainty of the integrated line intensity.}
\end{table*}

\section{Data interpretation}\label{sec:interpretation}

We deduce the physical conditions at each pixel with version 1.5.2 of the Meudon PDR code \citep{LePetit2006}, which is available on the ISMServices webpage\footnote{http://ism.obspm.fr} of Paris Observatory. In this section, we first give an overview on the Meudon PDR code and discuss the strategies adopted in the data interpretation.

\subsection{The Meudon PDR code}\label{subsec:mod_overview}

The Meudon PDR code computes chemical densities and gas excitation in a 1-D stationary plane-parallel slab of gas and dust illuminated by a radiation field. At each point of the spatial grid, the code consistently solves the radiative transfer, the chemistry and the thermal balance, which computes individually the heating and cooling mechanisms. The chemical network used in this paper includes 224 species and 6311 chemical reactions.

H$_2$ formation on grains is a key process since it controls the transition between the neutral atomic and the molecular phases. The default prescription in the Meudon PDR code simulates the H$_{2}$ formation on dust grains through Eley-Rideal and Langmuir-Hinshelwood mechanisms, as described in \citet{LeBourlot2012}. Although a more sophisticated and computing time-demanding H$_2$ formation prescription that simulates the stochastic heating of grains by UV photons, its effect on the H$_2$ formation, and ortho/para conversion, on the grain surfaces is also available~\citep{Bron2014, Bron2016}, we opt for the \citet{LeBourlot2012} prescription to gain efficiency in computing time.
        
At each point of the spatial grid, the radiative transfer equation is solved considering absorption and diffusion by dust as well as absorption in continuum of some gas species as C. Self-shielding for critical species, such as H$_2$ and $\mathrm{CO}$, is computed with the \cite{Federman1979} formalism (see \citealt{LePetit2006} for details).

The most important heating processes taken into account are the photo-electric effect on grains~\citep{Bakes1994}, heating by collisional de-excitation of H$_2$, and cosmic rays as well as exothermic reactions. As for the cooling processes, the non-LTE level excitation of main coolants is computed at each position in the cloud, considering various micro-physical processes, such as collisions and radiative \mbox{(de-)excitation}~\citep{GonzalezGarcia2008}.

Another important aspect is the treatment of grains. Grains impact the photo-electric heating rate, the penetration of UV photons, the IR dust emission, and the formation of H$_2$. In the models used for this work, a MRN-like grain size distribution~\citep{Mathis1977} is adopted for carbonaceous grains and silicates, plus a log-normal distribution for polycyclic aromatic hydrocarbons~(PAHs). The photo-electric heating from PAHs is estimated with the prescriptions given in \citet{Bakes1994}. For every model, the code simulates and outputs detailed thermal (gas and grains) and chemical~(with level excitation for selected species) structures of the cloud, as a function of depths, as well as the integrated quantities, such as column densities and line intensities.

For this work, a non-uniform grid of 1367 PDR models has been produced. The explored physical space is the thermal pressure, the intensity of incident UV radiation fields and the depth of the cloud. The thermal pressure, $P$, ranges from 10$^5$ to 10$^9$ K cm$^{-3}$. The UV field scaling factor, $G_{\textrm{UV}}$, ranges from 1 to 10$^5$ in the Mathis units~\citep{Mathis1983}. The visual extinction, $A_\textrm{V}$ goes from 1 to 40 magnitudes. Other parameters, common to all models are presented in Tables~\ref{Tab:ParamCommon}. Elementary abundances used for the chemistry are the standard values in the ISM~(Table~\ref{Tab:ElemAbund}). These models are available in the Interstellar Medium Data Base~(ISMDB)\footnote{The ID of the grid of models in ISMDB is DM54.} at http://ismdb.obspm.fr.

\begin{table}[htbp]
\begin{center}
\caption{Common parameters used for the grid of PDR models.
\label{Tab:ParamCommon}}
\begin{tabular}{llll}
\hline
\hline
Quantity                                  & value                & unit                  & note \\  
\hline
$\zeta$                                   & 10$^{-16}$           & s$^{-1}$        & (1)  \\
N$_\textrm{H}$ / E(B-V)                   & 5.8$\times$10$^{21}$ & cm$^{-2}$ mag$^{-1}$  & (2)  \\
m$_{\textrm{grain}}$ / m$_{\textrm{gas}}$ & 0.01                 & no unit               & (3)  \\
$\alpha$                                  & $-$3.5               & no unit               & (4)  \\ 
a$_{\textrm{min}}$                        & 1$\times$10$^{-7}$   & cm                    & (5)  \\
a$_{\textrm{max}}$                        & 3$\times$10$^{-5}$   & cm                    & (6)  \\
m$_{\textrm{PAH}}$ / m$_{\textrm{grain}}$ & 4.6$\times$10$^{-2}$ & no unit               & (7)  \\
\hline
\end{tabular}
\end{center}
\tablefoot{(1): cosmic-ray ionization rate of H$_2$ per second. (2) Gas-to-grain ratio~\citep{Bohlin1978}. (3) grain to gas mass ratio. (4) MRN power law exponent. (5) minimum radius of grains. (6) maximum radius of grains. (7) PAH to grain mass ratio.}
\end{table}


\begin{table}[htbp]
\begin{center}
\caption{Elementary abundances used in the chemical network.
\label{Tab:ElemAbund}}
\begin{tabular}{lll}
\hline
\hline
Elements          & Value                 & Reference \\  
\hline
O/H               & 3.19$\times$10$^{-4}$ & (1)  \\
C/H               & 1.32$\times$10$^{-4}$ & (2)  \\
N/H               & 7.50$\times$10$^{-5}$ & (3)  \\
S/H               & 1.86$\times$10$^{-5}$ & (2)  \\
$^{12}$C/$^{13}$C &  70                   & (4)  \\
$^{16}$O/$^{18}$O & 560                   & (5)  \\
\hline
\end{tabular}
\tablebib{(1) \cite{Meyer1998}, (2) \cite{Savage1996}, (3) \cite{Meyer1997}, (4) \cite{Milam2005}, (5) \cite{Wilson1999}}
\end{center}
\end{table}

\subsection{Modeling strategy}\label{subsec:mod_strategy}

Alongside the transitions listed in Table~\ref{table:linetable}, observations of [C\textsc{ii}]\,$158\,\mum$, [O\textsc{i}]\,$63$ and $145\,\mum$ with the \herschel~Photodetector Array Camera and Spectrometer~(PACS,~\citealt{Poglitsch2010}) as well as the $^{13}\mathrm{CO}$ transitions covered by the \herschel~SPIRE/FTS bandwidth~($\mathrm{J=5-4}$ to $\mathrm{J=14-13}$) are also available. The Meudon PDR code can simulate line-intensities of all these transitions. That is, a priori, all of them could be used to find the best model that matches the observations. Nevertheless, we use only the CO and C lines listed in Table~\ref{table:linetable} to constrain the physical parameters based on the following three reasons.

First, it is not yet clear whether the atomic emissions associated with C$^+$ and O share the same origin with the observed CO and C transitions. A comparison of line profiles, resolved both spectrally and spatially, of [C\textsc{ii}]\,$158\,\mum$ and CO transitions~($\mathrm{J=2-1}$, $\mathrm{J=6-5}$, and $\mathrm{J=13-12}$) in M17\,SW shows that the emissions of C$^+$ and CO are hardly associated in their distributions~\citep{Perez-Beaupuits2012}. Spectrally-resolved observations of G$5.89-0.39$, a massive star-forming region in our Galaxy, have also shown a difference between the velocity distributions of [O\textsc{i}]\,$63\,\mum$~and the CO transitions~($\mathrm{J=7-6}$ and $\mathrm{J=6-5}$)~\citep{Leurini2015}.

Second, as discussed in \citet{Liseau2006}, to match well the model-predicted and observed intensity-ratio of [O\textsc{i}]\,$63\,\mum/$[O\textsc{i}]\,$145\,\mum$, one has to resort to either extremely high optical depths~($A_{\mathrm{V}}>100\,\mathrm{mag}$) or take into account the absorption from the source cloud itself or/and the clouds along the line of sight. With the spectrally-resolved [O\textsc{i}]\,$63\,\mum$ line enabled by SOFIA/GREAT~\citep{Heyminck2012}, it has indeed been observed in G$5.89-0.39$ that the emission of [O\textsc{i}]\,$63\,\mum$ is severely contaminated by self-absorption and absorption by clouds along the line of sight~\citep{Leurini2015}. Although observational evidence is still scarce at the moment of writing, we judge it best to exclude [C\textsc{ii}]\,$158\,\mum$ and [O\textsc{i}]\,$63\,\mum$ when searching for the best physical parameters. As to the [O\textsc{i}]\,$145\,\mum$ emission, since the observation of this transition covers only a small portion~($\sim\,20\%$) of the FOV, we opt to exclude this line as a direct constraint to the models for the consistency. However, a good comparison between the model-predicted and observed line intensities of [O\textsc{i}]\,$145\,\mum$, wherever it is possible, is ensured in our final results.

Third, in order to properly simulate the $^{13}$CO emissions, one has to consider the chemical effects due to mutual shielding by its isotopologue, $^{12}$CO. Theoretical computation has indicated that the $^{13}$CO lines can be more efficiently shielded by $^{12}$CO than by itself~\citep{Visser2009}. The Meudon PDR code is capable of simulating this process by solving the exact radiative transfer as described in \citet{Goicoechea2007}. However, this heavy computation would cost days in order to successfully simulate one model, {\it i.e.} one combination of $P$, $G_{\mathrm{UV}}$ and $A_\textrm{V}$. To efficiently produce the entire model-grid required for the analysis, we adopt the FGK approximation~\citep{Federman1979} that takes only self-shieldings into account. This simplification would result in the underestimate of $^{13}$CO line intensities. Therefore they are excluded from constraining the best physical parameters.

Prior to the search of best physical parameters on a given pixel based on the non-uniform grid of 1367 PDR models described above, we construct a numerical function that linearly interpolates the grid between its main free parameters, $P$, $G_{\textrm{UV}}$ and $A_\textrm{V}$, to convert the grid into a continuous 3-dimensional grid-function. We use radial-basis-functions~(\texttt{scipy.interpolate.Rbf}) to produce the grid-function, $M(P,\,G_{\textrm{UV}},\,A_\textrm{V})$. In addition to the three free parameters, $P$, $G_{\textrm{UV}}$, and $A_\textrm{V}$, an additional free parameter, named scaling factor~($\phi$) is introduced to describe several points. First, this parameter contains the information about the ratio of angular sizes of the observed clouds and beam, {\it i.e.} $\phi\,=\,\Omega_\mathrm{cl}/\Omega_\mathrm{beam}$, where $\Omega_\mathrm{cl}$~is the physical angular size of observed CO clouds and $\Omega_\mathrm{beam}$~is the beam area. Second, it also contains the information about the inclination angle of PDR components in the FOV with the line-of-sight. The Carina I-E region is known to host an edge-on PDR \citep{Brooks2003}, but all detected CO emission in our FOV may not come from edge-on PDRs. So we use face-on line intensities in our grid of PDR models and scale them with $\phi$ to consider limb-brightening effect. Third, this parameter also accounts for the possibility that several PDR components are superposed along the line of sight. At each pixel, we search for the combination of $\phi$, $P$, $G_{\textrm{UV}}$, and $A_\textrm{V}$, that best matches the observed $^{12}$CO line intensities by minimizing a normalized $\chi^2$:

\begin{equation}
\centering
\label{eq:chi2}
\chi^2 = \sum_{i=1}^{N_{\mathrm{obs}}}\,\frac{1}{(N_{\mathrm{obs}}-N_{\mathrm{p}}-1)}\,\frac{[I_{i}-\phi\,M_{i}(P,\,G_{\textrm{UV}},\,A_\textrm{V})]^{2}}{\sigma_{i}^{2}}.
\end{equation}

\begin{figure}[htbp!]
\centering
\includegraphics[width=0.45\textwidth]{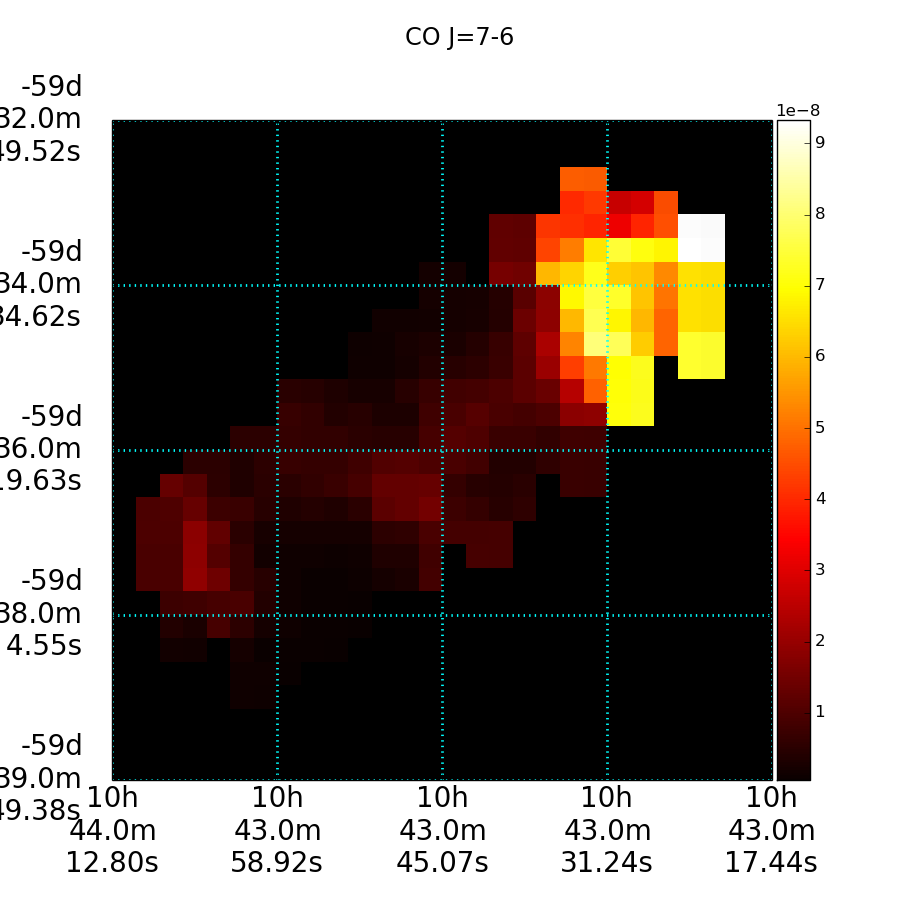}
\caption{Integrated intensities of CO $\mathrm{J=7-6}$ from our FOV. The displayed color-scale is in units of $\mathrm{W\,m^{-2}\,sr^{-1}}$.  \label{fig:co76}}
\end{figure}

\begin{figure*}[h!]
\includegraphics[width=0.5\textwidth]{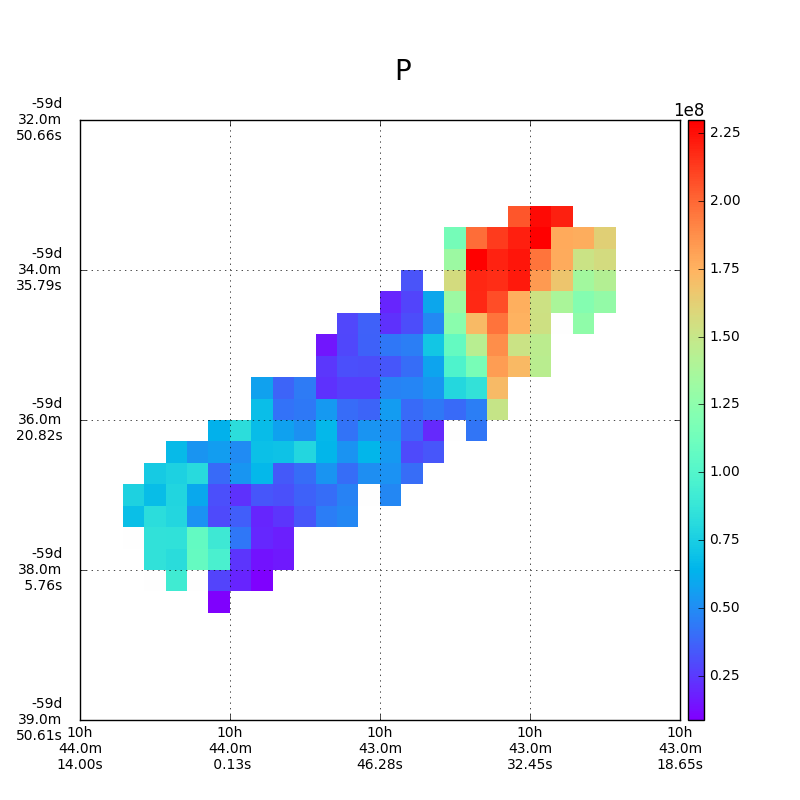}
\includegraphics[width=0.5\textwidth]{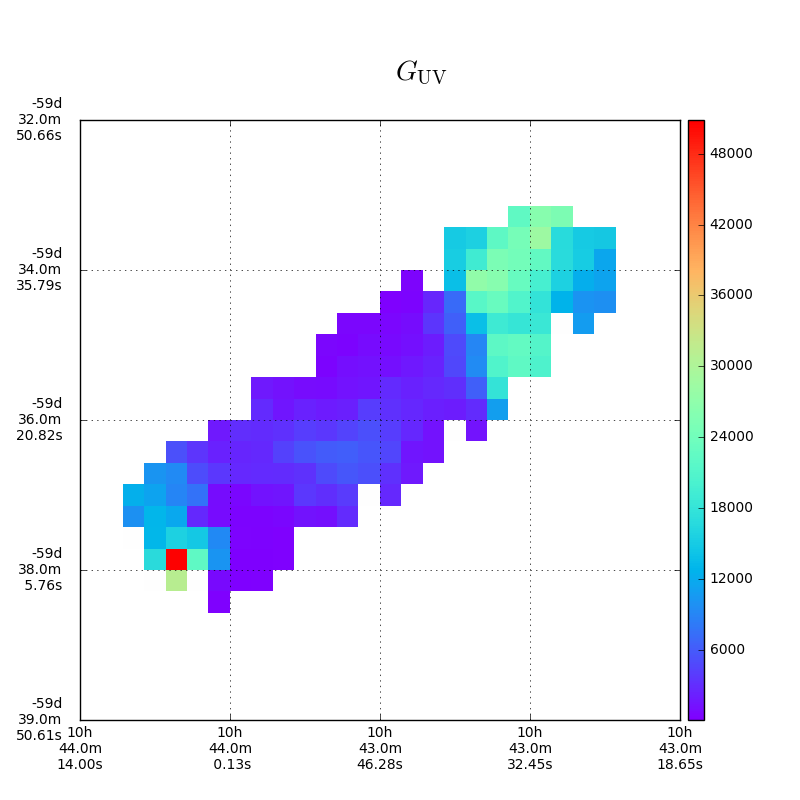}
\caption{Best physical parameters derived at each pixel from the comparison of observations with PDR models. On the left panel is shown the thermal pressure, $P$, in K cm$^{-3}$ and on the right panel the FUV scaling factor, $G_\mathrm{UV}$. \label{Fig:BestParamMap}}
\end{figure*}

\begin{figure}[htbp!]
\centering
\includegraphics[width=0.5\textwidth]{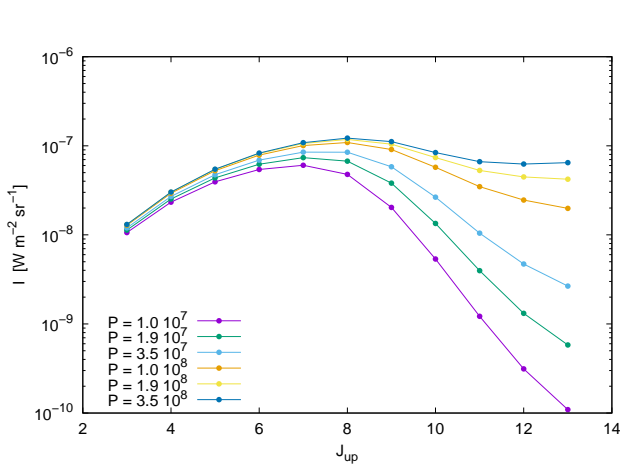}
\caption{$\mathrm{CO}$ SLEDs generated with the Meudon PDR code with $G_{\mathrm{UV}}$ and $A_\mathrm{V}$ fixed at $10^{4}$ and $10$ respectively and with different thermal pressures.}
\label{fig:high_pressure}
\end{figure}

In Equation~(\ref{eq:chi2}), $N_{\mathrm{obs}}$ and $N_{\mathrm{p}}$ represent the available numbers of the observed CO and C lines and the free parameters~($N_{\mathrm{p}}=4$), and $I_{i}$, $\sigma_{i}$, and $M_{i}(P,\,G_{\textrm{UV}},\,A_\textrm{V})$ denote the observed line intensity, its uncertainty, and the simulated line-intensity for the $i$-th transition.

As the $S/N$-ratio of the observed CO transitions generally decreases from $\mathrm{CO\ J=11-10}$ to $\mathrm{CO\ J=12-11}$ and to $\mathrm{CO\ J=13-12}$, we rule out the possibility that the intensities for high-J CO transitions beyond the bandwidth of \herschel\ SPIRE/FTS would be higher than the observed $\mathrm{CO\ J=13-12}$ transitions. An artificial penalty in the minimization procedure is forced to better constrain the best-fit solutions. When calculating $\chi^{2}$, a penalty is assigned to a given model that predicts $I(\mathrm{CO\ J=13-12})\,>\,I(\mathrm{CO\ J=12-11})$ and $I(\mathrm{CO\ J=12-11})\,>\,I(\mathrm{CO\ J=11-10})$.

We adopt the limited-memory BFGS~(L-BFGS) algorithm in our minimization procedure~(\texttt{scipy.optimize.minimize}) to search for the local minimum in the $\chi^{2}$-space. To ensure that the returned solution of ($\phi$, $P$, $G_{\textrm{UV}}$, $A_\textrm{V}$) indeed corresponds to the global minimum in the $\chi^{2}$-space, we randomly choose 100 sets of initial free parameters and output all the solutions found by \texttt{scipy.optimize.minimize}. For a given pixel, the solution that corresponds to minimum $\chi^{2}$ is chosen to be the final solution.
The uncertainties for the constrained solutions are then estimated with 300 Monte-Carlo experiments with an assumption of Gaussian distributions for the uncertainties of the integrated line-intensity for each transition.

\section{Results}\label{sec:results}

\subsection{CI and CO emission}\label{sec:obs_results}

The CO emission is observed from $\mathrm{J=4-3}$ to $\mathrm{J=13-12}$, along with $\cits$~and $\cison$~across the whole FOV~(see figures in the Appendix). In the Car I-E region at the northwest, the intensities of the CO and [C\textsc{i}] lines are generally higher by one order of magnitude than those observed in Car I-S and Car I/II. Beyond $\mathrm{J=10-9}$, the emission is concentrated in the Car I-E and locally in the Car I/II regions. To get a global view on the spatial distribution of the CO excitation over the FOV, we first approximate the excitation temperature~($T_{\mathrm{ex}}$) of CO transitions observed by the \herschel\ SPIRE/FTS under the LTE assumption. The reduced-$\chi^{2}$ for fitting each observed CO SLED with LTE is between 0.5 and 2.5, with an average value of $\sim\,1.5$ throughout the FOV. In the Car I-E region, the $T_{\mathrm{ex}}$ values are between $70$ and $100\,\mathrm{K}$. In the Car I-S region, the $T_{\mathrm{ex}}$ values are between $45$ and $55\,\mathrm{K}$. In the Car I/II region, the $T_{\mathrm{ex}}$ values are generally between $65$ and $70\,\mathrm{K}$, except a local~($\sim\,0.5\,\mathrm{pc}$ in diameter) observation of $T_{\mathrm{ex}}\,\sim\,80\,\mathrm{K}$. The observed ratio of $\cits/\cison$ is between 2.9 and 3.7 in the Car I-E region, corresponding to $T_{\mathrm{ex}}$ values between $33$ and $41\,\mathrm{K}$, which is consistent with the \textsl{NANTEN2} observations~\citep{Kramer2008}. In the Car I/II region, the observed ratio of $\cits/\cison$ is around 2.5, corresponding to $T_{\mathrm{ex}}\,\sim\,28\,\mathrm{K}$. In the Car I-S region, the $\cits$ and $\cison$ are not detected above $3\sigma$ threshold. The observed spatial trends of the [C\textsc{i}] and CO excitation are generally consistent with each other, showing molecular gas of higher-excitation states in the Car I-E region.

As the sensitivity of the instrument is at the highest around the $\mathrm{J=7-6}$ transition~(Table~\ref{table:linetable}), we present an overview on the morphology of CO emission based on the CO $\mathrm{J=7-6}$ line intensity. Figure \ref{fig:co76} shows the CO $\mathrm{J=7-6}$ line intensity distribution within our FOV. Comparing with the background optical image shown in Figure~\ref{fig:carina_zoom}, the observed CO gas fills up most of the optically-thick region. The most prominent CO and C-emission is found in the Car I-E region, with an average value of $5.3\times10^{-8}\,\mathrm{W\,m^{-2}\,sr^{-1}}$ for the CO $\mathrm{J=7-6}$ intensity, and peaks at the northwest corner of the FOV with a value of $9.3\times10^{-8}\,\mathrm{W\,m^{-2}\,sr^{-1}}$. All 77 pixels~($\sim2'$ or $\sim1.3\,\mathrm{pc}$ in diameter) covered in this region have observed CO $\mathrm{7-6}$ line intensities above $3\sigma$ threshold. This region sits at a projected distance of $\sim\,2.3\,\mathrm{pc}$ from Trumpler 14.

Moving to the Car I-S region, the area~($\sim4'$ or $\sim2.5\,\mathrm{pc}$ in diameter) is mostly dominated by the H\textsc{ii} region ionized by Trumpler 14. About $30\%$ of the Car I-S region has no observed CO $\mathrm{J=7-6}$ emission at above $3\sigma$ threshold, while the CO emission observed at above 3$\sigma$ is distributed along the dust-lane. One double-lined spectroscopic binary~(SB2) system, \textsl{ALS 15 204}~(O7.5Vz$+$O9: V), and a B0 star, \textsl{ALS 15 203}, which are identified as members of Trumpler 14 are found near the center of Car I-S~\citep{Apellaniz2016}. Along with other stellar members of Trumpler 14, these two stars can have important contribution to the excitation of CO gas in the region. The CO $\mathrm{J=7-6}$ emission detected in this region is the weakest within our FOV. Among the pixels with detected CO $\mathrm{J=7-6}$ emission~($>3\sigma$), the average intensity is $7.6\times10^{-9}\,\mathrm{W\,m^{-2}\,sr^{-1}}$, which is about an order of magnitude fainter than that in Car I-E.

The Car I/II region at the southeast corner of our FOV is sitting at the intersection of two interstellar clouds, Car I and Car II, as defined by the radio thermal continuum observed at $5$ and $0.843\,\mathrm{GHz}$~\citep{Gardner1968, Whiteoak1994}. About 70\% of the Car I/II area has CO $\mathrm{J=7-6}$ emission detected at above the $3\sigma$ threshold, and the detected CO $\mathrm{J=7-6}$ emission has an average intensity of $9.8\times10^{-9}\, \mathrm{W\,m^{-2}\,sr^{-1}}$. This region sits right in the middle between the two most massive star-clusters of the Carina Nebula, Trumpler 14 and Trumpler 16, both of which can provide UV-photons for the observed CO emission.

\subsection{Physical conditions}\label{sec:mod_results}
\begin{figure}[htbp!]
\centering
\includegraphics[width=0.45\textwidth]{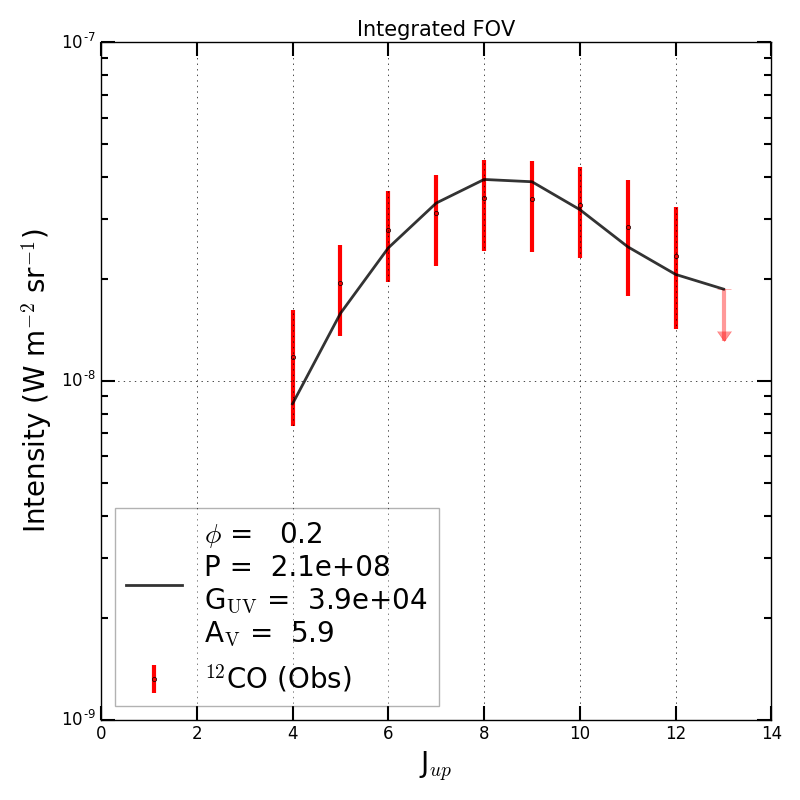}
\caption{CO SLED obtained by integration of the emission on the whole FOV (red bars). The black line represents the best PDR model in the grid of models that reproduces the observation.
\label{fig:sled_whole}
}
\end{figure}

\begin{figure}[htbp!]
\centering
\includegraphics[width=0.45\textwidth]{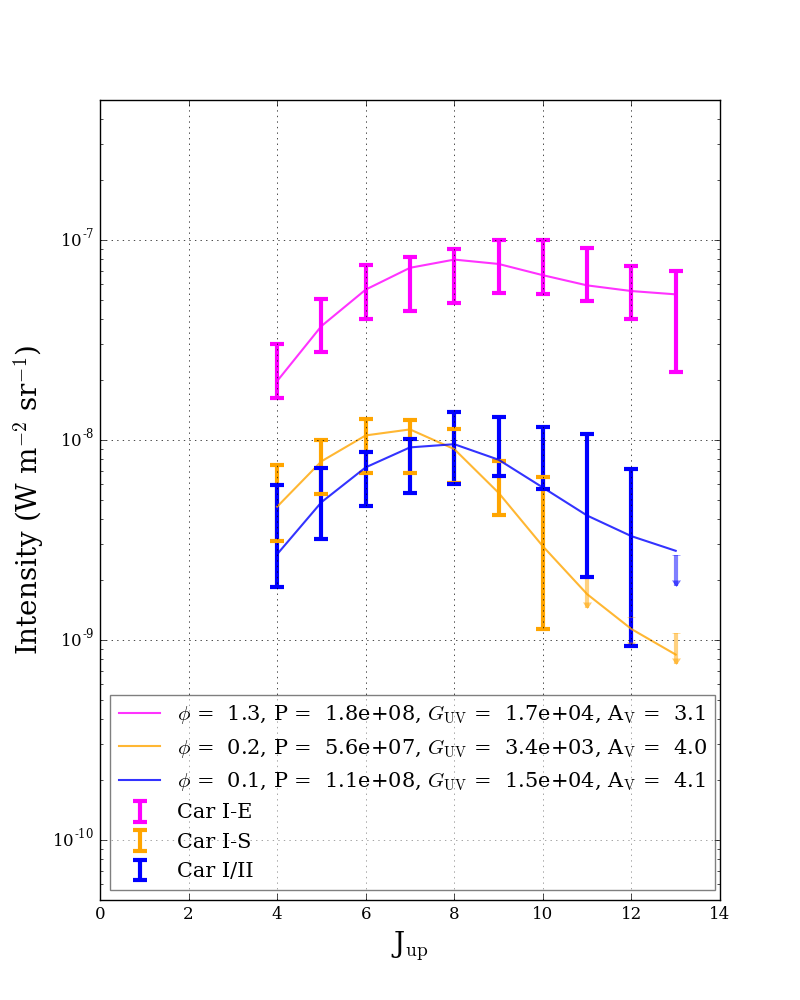}
\caption{
CO SLEDs observed from the three yellow-masked pixels chosen from the Car I-E, Car I-S, and Car I/II regions as indicated in Figure~\ref{fig:carina_zoom}.
The best-fit Meudon PDR models are overplotted the solid lines.
}
\label{fig:sled_3pix}
\end{figure}

\begin{figure*}[htbp!]
\centering
\includegraphics[width=0.31\textwidth]{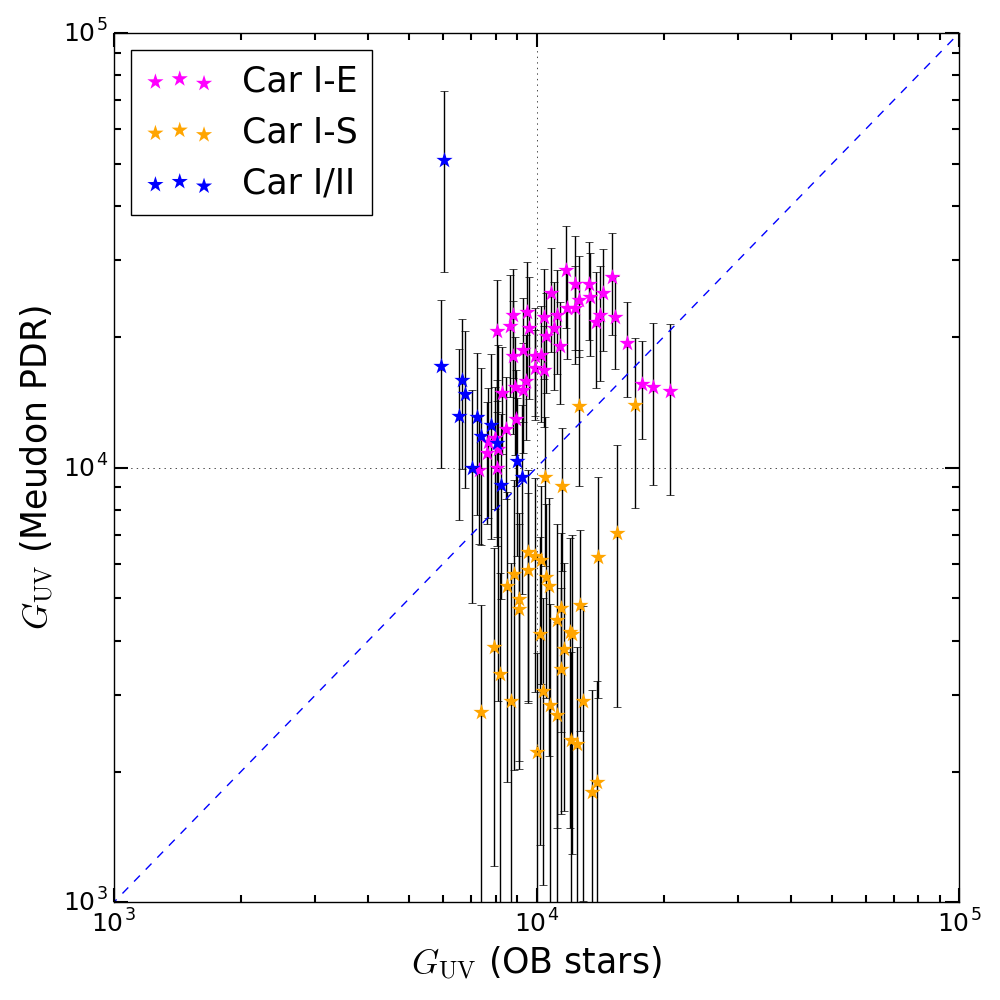}
\includegraphics[width=0.31\textwidth]{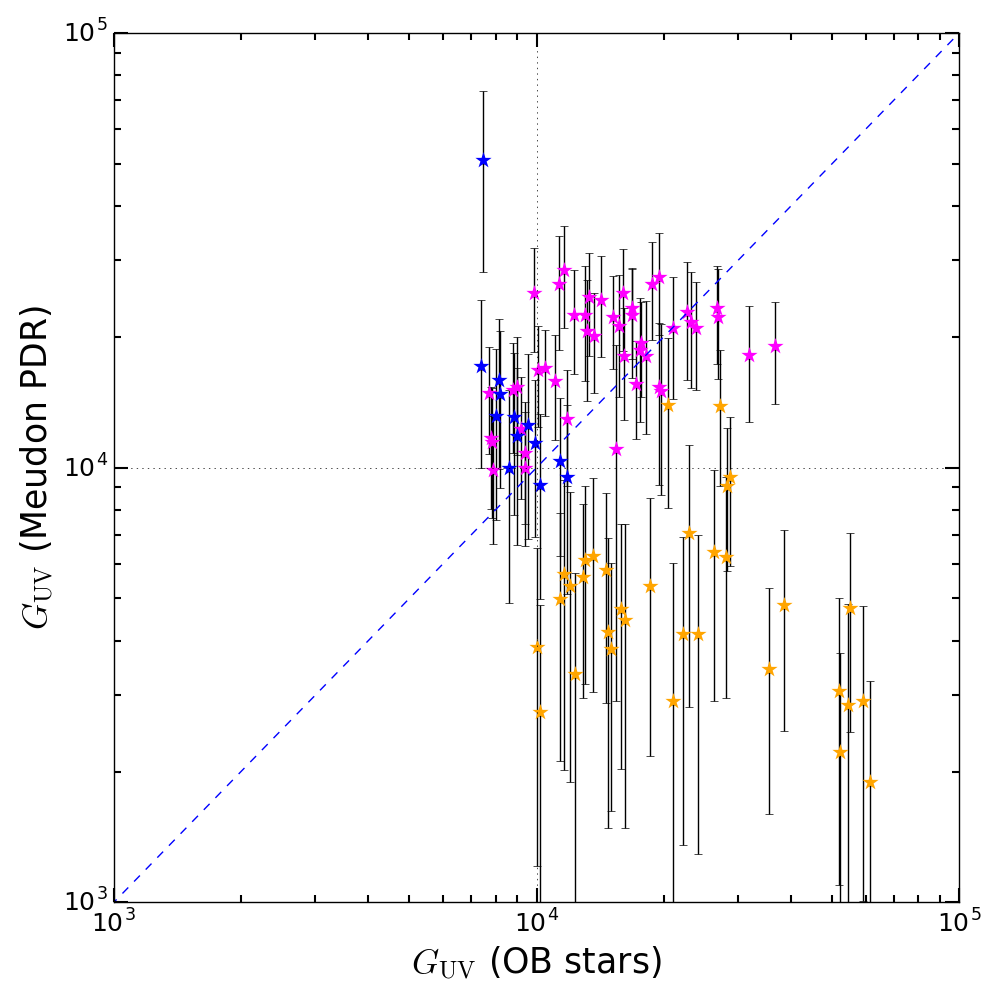}
\includegraphics[width=0.31\textwidth]{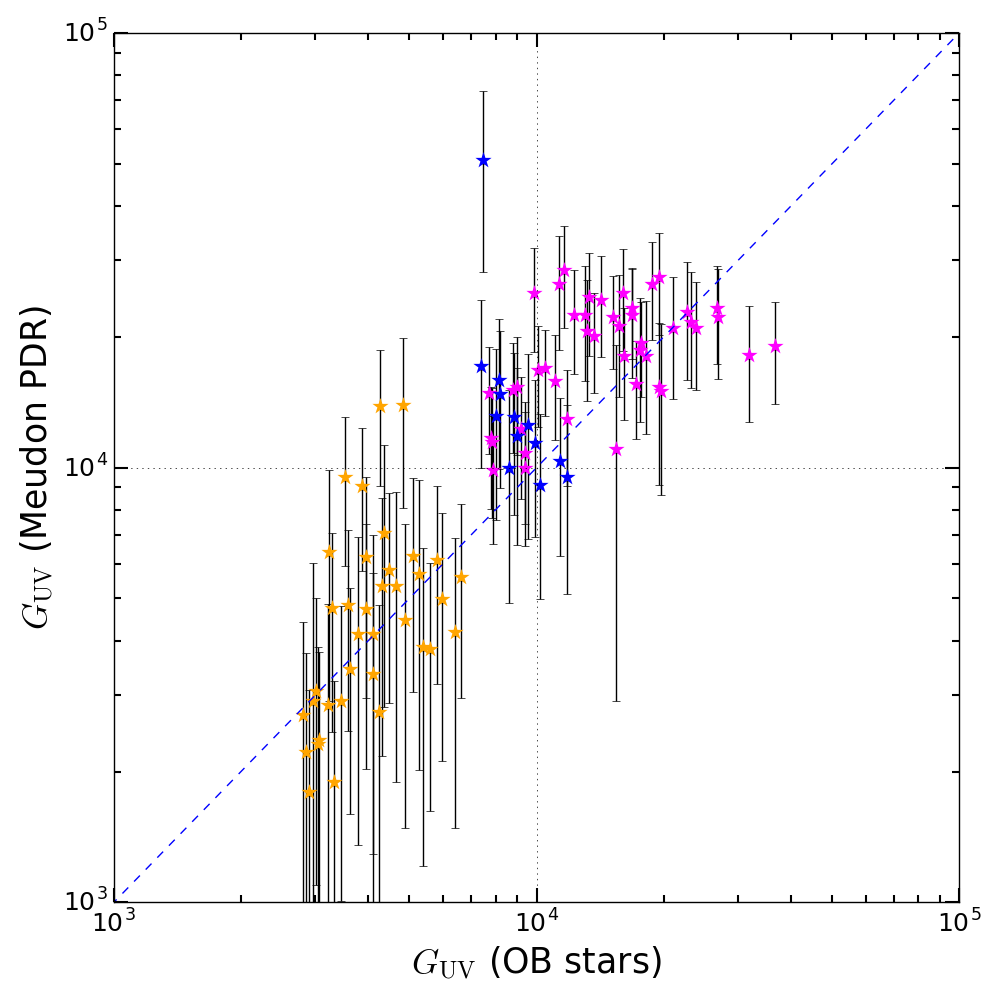}
\caption{Comparison of the $G_{\mathrm{UV}}$ values obtained with the Meudon PDR code and the values evaluated from the stellar composition within our FOV. Only pixels with good~($S/N\,>\,3$) detection of $\mathrm{CO\ J=7-6}$ are plotted. The dashed lines indicates a $1:1$ relationship between the quantities on the axes. The uncertainties of the $G_{\mathrm{UV}}$ values derived with the Meudon PDR models are estimated with the MC experiment as discussed in Section~\ref{subsec:mod_strategy}. {\it Left}: All OB-stars are assumed to be at the center of their host star-clusters, {\it i.e.}, Trumpler\,14 and 16. The observed CO gas is assumed to lie on the plane perpendicular to the line of sight at a $2.35\,\mathrm{kpc}$~distance. {\it Center}: The distribution of OB-stars is considered coplanar at their observed ($\alpha$, $\delta$) coordinates. The geometry of the CO gas is assumed to be coplanar as in the left panel. {\it Right}: The distribution of OB-stars is considered coplanar at their observed ($\alpha$, $\delta$) coordinates. The geometry of the CO gas is assumed as the shape indicated in Figure~\ref{fig:3dview}.
}
\label{fig:chiobstar_pdr}
\end{figure*}

\begin{figure*}[htbp!]
\includegraphics[width=\textwidth]{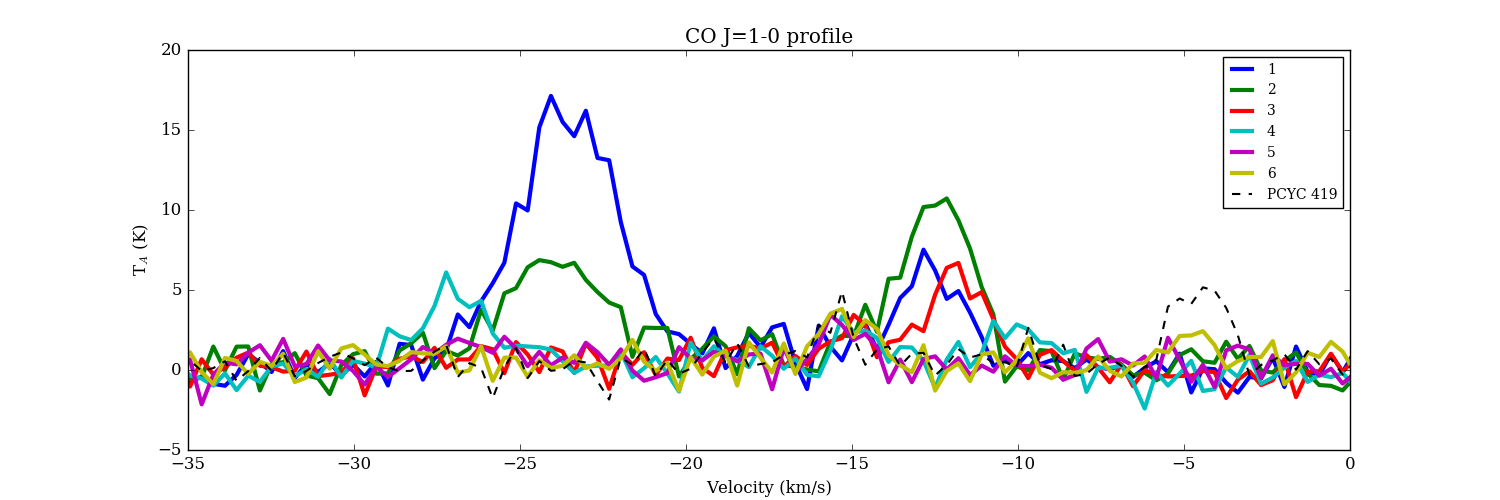}
\caption{Spectrally-resolved line profiles at six positions indicated in Figure~\ref{fig:carina_zoom}. The data is taken as a part of the Carina Parkes-ATCA Radio Continuum Survey~\citep{Rebolledo2016}. The dashed line indicates the line profile at a $\sim\,0.2\,\mathrm{pc}$ region from Car I/II (see Section~\ref{subsec:co_structure}).}
\label{fig:co10_profile}
\end{figure*}    

\begin{figure}[htbp!]
\centering
\includegraphics[width=0.5\textwidth]{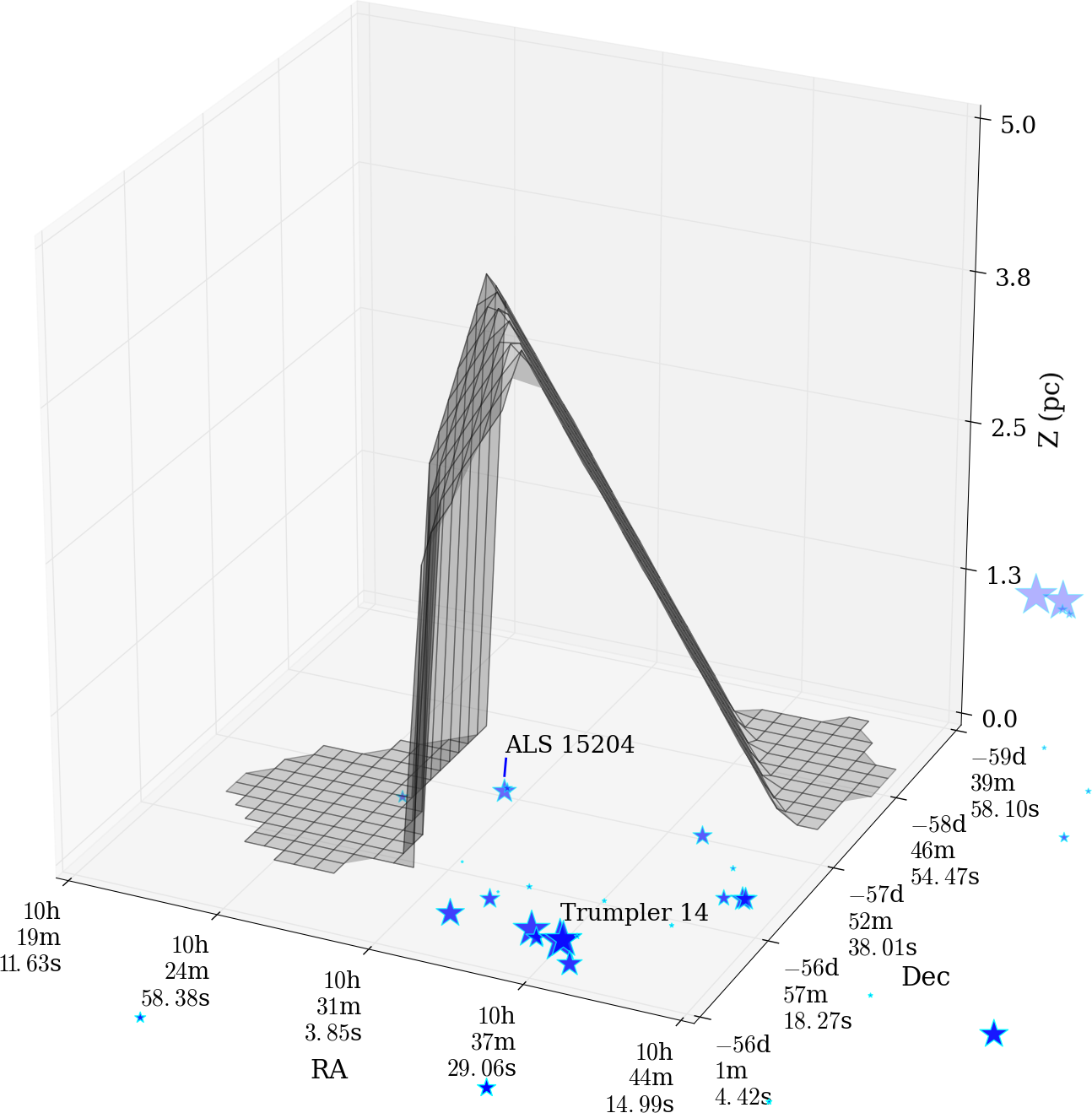}
\caption{In this figure, the gray mesh demonstrates the estimated 3D cloud structure in this work. The blue star symbols indicate the OB-stars in the FOV with their sizes in proportion to the FUV luminosities of stars. The Z-axis indicates the offset along the line of sight where $Z\,=\,0.0$ is at $2.3\,\mathrm{kpc}$ from us. Stars that are outside the FOV~(Figure~\ref{fig:carina_zoom}) toward the south are displayed in fainter colors.\label{fig:3dview}}
\end{figure}

We determine the physical conditions ($P$, $G_{\textrm{UV}}$, $A_\textrm{V}$ and $\phi$) at each pixel following the method described in section \ref{subsec:mod_strategy}. CO mid-J and high-J lines are good tracers of the gas pressure as it can be seen on Figure~\ref{fig:high_pressure} and we expect this parameter, as well as $G_\textrm{UV}$ that controls the energy input and photo-destruction rates be well constrained. As only $^{12}$CO lines with J$_{up}$ > 3 and C lines, which probe only the surface layer of molecular gas are used in constraining the solutions, the total $A_\textrm{V}$ parameter of our models is not reliably constrained. So, we do not discuss this parameter in the following.

A good agreement between PDR models and observed CO and [C\textsc{i}] line intensities is found at all pixels. To illustrate the general goodness of solutions, comparisons between the observed CO line intensities and those simulated by the Meudon PDR code are presented with the integration of CO lines on the whole FOV~(Figure~\ref{fig:sled_whole}), and with three typical pixels from the Car I-E, Car I-S, and Car I/II regions~(Figure~\ref{fig:sled_3pix}). The reduced-$\chi^{2}$ values vary between $0.2$ to $0.7$ across the FOV. CO and [C\textsc{i}] line intensities predicted by the Meudon PDR models are generally consistent with the \herschel~SPIRE/FTS observations. We also examine whether or not the observed intensities of $^{13}\mathrm{CO}$, [O\textsc{i}]\,$63$, [O\textsc{i}]\,$145\,\mum$, and [C\textsc{ii}]\,$158\,\mum$ computed by the PDR models but not used in the minimization procedure are consistent with the observations (see arguments in Section~\ref{subsec:mod_strategy}). As our constraints, the $^{12}$CO and [C\textsc{i}] emissions may not probe the full optical-depth of the PDRs, we expect the model-predicted $^{13}$CO intensities to be always lower than the observed values. In our observations, the $^{13}\mathrm{CO}$ emission is only detected up to the $\mathrm{J=8-7}$~(from $\mathrm{J=4-3}$) transition in the Car I-E region, and marginally in the Car I-S and Car I/II regions. In this region, comparing the model-predicted and observed intensities of $^{13}\mathrm{CO}$ transitions, the model-predicted values of $^{13}\mathrm{CO}$ transitions generally represent $\sim\,10\%$ of the observed intensities. Although the effects from mutual shielding of $^{12}$CO may account partly for this under-estimation, further investigations would be needed in order to understand the observed $^{13}\mathrm{CO}$ emission.

For the [C\textsc{ii}]\,$158\,\mum$ emission, the ratio of model-predicted and observed intensities varies between 0.02 and 0.3. Within the Car I-E region, which is dominated by the PDRs, the observed [C\textsc{ii}]\,$158\,\mum$ intensities are between 3 to 10 times larger than the model-predicted values. In Car I-S and Car I/II regions, where PDRs no longer dominate the FOV, the observed [C\textsc{ii}]\,$158\,\mum$ intensities are about 10 to 50 times larger than the model-predicted values. Generally, this spatial trend is consistent with the results found in \citet{Oberst2011}. As the [C\textsc{ii}]\,$158\,\mum$ emission can also originate from ionized gas and neutral gas at more diffuse states than the modeled molecular gas based on the CO and [C\textsc{i}] observations, we expect under-estimated [C\textsc{ii}]\,$158\,\mum$ intensities from the models. By approximating the \textsl{ISO}-observed [C\textsc{ii}]\,$158\,\mum$ emission as a linear combination of the [O\textsc{i}]\,$63\,\mum$ and [N\textsc{ii}]\,$122\,\mum$ emissions, \citet{Mizutani2004a} estimate that the contribution to the [C\textsc{ii}]\,$158\,\mum$ emission by the ionized gas is about $20\%$ in the Carina Nebula. Based on the \textsl{ISO}-observed ratio of [N\textsc{ii}]\,$122\,\mum$ and [N\textsc{ii}]\,$205\,\mum$ emissions, \citet{Oberst2011} estimate this contribution to be $37\%$. However, the direct association of the observed [O\textsc{i}]\,$63\,\mum$ emission to the PDRs is uncertain, and a large fraction of the [C\textsc{ii}]\,$158\,\mum$-emitting neutral gas may not be spatially associated to the denser CO-emitting cloud, as recently revealed in the Horsehead Nebula by the SOFIA/GREAT observations~\citep{Pabst2017}. The adopted isobaric PDR models, which well describe the observed CO and [C\textsc{i}] emissions, may only account for a small fraction of the [C\textsc{ii}]\,$158\,\mum$ emission originated from the PDRs. Without spectrally-resolved observations, as urged by the SOFIA/GREAT observations from N\,159 of the Large Magellanic Cloud~(LMC)~\citep{Okada2015}, it is unattainable to pin down the [C\textsc{ii}]\,$158\,\mum$ emission directly associated with the CO-emitting clouds.

For the [O\textsc{i}]\,$63\,\mum$ transition, as discussed in Section~\ref{subsec:mod_strategy}, the observed interstellar [O\textsc{i}] emission at $63\,\mum$ can suffer largely from foreground absorption~\citep{Liseau2006}, which leads to the generally over-estimated [O\textsc{i}]\,$63\,\mum$ emission with PDR models. Within our FOV, the ratio of model-predicted and observed intensities is always larger than $10$. The model-predicted [O\textsc{i}]\,$145\,\mum$ transitions, on the other hand, are generally reasonable, and the ratio of model-predicted and observed line intensities are between $1$ and $3$ over the FOV.

Figure~\ref{Fig:BestParamMap} shows the maps of derived $P$ and $G_{\textrm{UV}}$. As each pixel is modeled independently in the process, the smooth appearance on the $P$ and $G_{\textrm{UV}}$ maps reassures that our solutions are not strongly degenerated. Based on the solutions, Car I-E is dominated by high pressure gas ($P\,\sim 1.5-2.2\times10^8\,\mathrm{K\,cm^{-3}}$) heated by strong UV radiation fields ($G_{UV}\,\sim 3\times10^4$). In the Car I-S region, lower $P$ ($\le 7\times10^7\,\mathrm{K\,cm^{-3}}$) and $G_{UV}$~($3000\,<\,G_{UV}\,<\,1.2\times10^4$) values are found. The pressure and the intensity of the radiation field then slightly rise into the Car I/II region~($P\,\sim\,10^8\,$ K cm$^{-3}$ and $G_{\mathrm{UV}}\,\sim\,10^4$). The $\phi$ parameter found by our procedure varies between 0.2 and 1.4 in the FOV. The higher values are found in the Car I-E region. Since in our procedure we use face-on values for simulated line intensities, this is consistent with a PDR seen edge-on that fills the beam as mentioned by previous authors \citep{Brooks2003, Kramer2008}. The lower values are found in the Car I-S area, consistent with a region where molecular gas fills up the beam partially.

\begin{table*}
\centering
\caption{A list of massive stars found within a projected distance of $7\,\mathrm{pc}$ from the center of FOV~($10:43:45.07,\ -59:36:19.63$).}
\label{table:stars}
\begin{tabular}{llllccl}
\hline\hline
Name & RA & Dec & Stellar type  & T &$L_{\mathrm{FUV}}$ & Binary\\
            & (J2000) & (J2000)    &       & ($10^{3}\,\mathrm{K}$) & ($10^{5}\,\mathrm{L_{\odot}}$) & \\
\hline
HDE 303312    &    $10:43:30.842$    &    $-59:29:23.8$    &    O9.7 IV    &    29.5\tablefootmark{a}    &    0.31\tablefootmark{a}    &    VB:0-1 SBE\\
ALS 15204    &    $10:43:41.237$    &    $-59:35:48.18$    &    O7.5 V    &    35.88    &    1.50    &    VB:1-3 SB2?\\
CPD -58 2611    &    $10:43:46.695$    &    $-59:32:54.82$    &    O6 V    &    38.87    &    2.13    &    VB:1-2 SB1?\\
ALS 15207    &    $10:43:48.707$    &    $-59:33:24.1$    &    O9 V    &    32.88    &    1.03    &    VB:0-1\\
HD 93128    &    $10:43:54.372$    &    $-59:32:57.37$    &    O3.5 V    &    43.86    &    3.63    &    VB:3-22\\
Trumpler 14-9    &    $10:43:55.354$    &    $-59:32:48.61$    &    O8.5 V    &    33.88    &    1.17    &    VB:many\\
HD 93129 AaAb    &    $10:43:57.462$    &    $-59:32:51.27$    &    O2 I    &    43.90    &    4.85    &    VB:many\\
HD 93129 B    &    $10:43:57.638$    &    $-59:32:53.5$    &    O3.5 V    &    43.86    &    3.63    &    VB:many\\
CPD -58 2620    &    $10:43:59.917$    &    $-59:32:25.36$    &    O7 V    &    36.87    &    1.70    &    VB:1-6\\
ALS 15206    &    $10:44:00.927$    &    $-59:35:45.74$    &    O9.2 V    &    32.48    &    0.98    &    VB:0-2\\
CPD -58 2627    &    $10:44:02.445$    &    $-59:29:36.77$    &    O9.5 V    &    31.89    &    0.90    &    VB:1-2\\
HD 93160    &    $10:44:07.267$    &    $-59:34:30.61$    &    O7 III    &    36.08    &    0.70    &    VB:1-2\\
HD 93161 A    &    $10:44:08.84$    &    $-59:34:34.49$    &    O7.5 V    &    35.88    &    1.50    &    VB:1-2 SB2\\
HD 93161 B    &    $10:44:09.08$    &    $-59:34:35.3$    &    O6.5 IV    &    37.93\tablefootmark{b}    &    1.40\tablefootmark{b}    &    VB:1-3 SB1?\\
HD 93162    &    $10:44:10.389$    &    $-59:43:11.09$    &    O2.5 If*/WN6    &    42.96\tablefootmark{c}    &    4.67\tablefootmark{c}    &    VB:1 SB2\\
ALS 15210    &    $10:44:13.199$    &    $-59:43:10.33$    &    O3.5 I    &    41.08    &    4.30    &    VB:2\\
HDE 303311    &    $10:44:37.463$    &    $-59:32:55.44$    &    O6 V    &    38.87    &    2.13    &    VB:1\\
\hline
Tr 14-30    &    $10:43:33.35$    &    $-59:35:11.1$    &    B1 Ia    &    26.00\tablefootmark{a}    &    0.40\tablefootmark{a}    &    --\\
ALS 15203    &    $10:43:41.21$    &    $-59:35:53.3$    &    B0 V    &    31.50    &    0.26    &    --\\
Tr 14-28    &    $10:43:43.56$    &    $-59:34:03.5$    &    B2 V    &    20.60    &    0.01    &    --\\
Tr 14-22    &    $10:43:48.81$    &    $-59:33:35.2$    &    B2 V    &    20.60    &    0.01    &    --\\
Tr 14-24    &    $10:43:50.9$    &    $-59:33:50.6$    &    B1 V    &    26.00    &    0.08    &    --\\
Tr 14-18    &    $10:43:57.96$    &    $-59:33:53.7$    &    B1.5 V    &    24.50    &    0.05    &    --\\
Tr 14-19    &    $10:43:58.46$    &    $-59:33:01.6$    &    B1 V    &    26.00    &    0.08    &    --\\
Tr 14-29    &    $10:44:05.11$    &    $-59:33:41.47$    &    B1.5 V    &    24.50    &    0.05    &    ERO 21\\
Tr 16-124    &    $10:44:05.83$    &    $-59:35:11.7$    &    B1 V    &    26.00    &    0.08    &    --\\
Tr 16-245    &    $10:44:13.8$    &    $-59:42:57.1$    &    B0 V    &    31.50    &    0.26    &    SB2\\
Tr 16-246    &    $10:44:14.75$    &    $-59:42:51.8$    &    B0.5 V    &    29.00    &    0.18    &    --\\
Tr 16-11    &    $10:44:22.52$    &    $-59:39:25.8$    &    B1.5 V    &    24.50    &    0.05    &    --\\
LS 1840    &    $10:44:24.62$    &    $-59:30:35.9$    &    B1 V?    &    26.00    &    0.08    &    SB2\\
Tr 16-122    &    $10:44:25.49$    &    $-59:33:09.3$    &    B1.5 V    &    24.50    &    0.05    &    --\\
Tr 16-94    &    $10:44:26.47$    &    $-59:41:02.9$    &    B1.5 V    &    24.50    &    0.05    &    --\\
Tr 16-18    &    $10:44:28.97$    &    $-59:42:34.3$    &    B2 V    &    20.60    &    0.01    &    --\\
Tr 16-12    &    $10:44:29.42$    &    $-59:38:38.1$    &    B1 V    &    26.00    &    0.08    &    --\\
Tr 16-10    &    $10:44:30.37$    &    $-59:37:26.7$    &    B0 V    &    31.50    &    0.26    &    SB2\\
Tr 16-17    &    $10:44:30.49$    &    $-59:41:40.6$    &    B1 V    &    26.00    &    0.08    &    --\\
Tr 16-13    &    $10:44:32.9$    &    $-59:40:26.1$    &    B1 V    &    26.00    &    0.08    &    --\\
Tr 16-14    &    $10:44:37.19$    &    $-59:40:01.5$    &    B0.5 V    &    29.00    &    0.18    &    --\\
Tr 16-16    &    $10:44:40.31$    &    $-59:41:49$    &    B1 V    &    26.00    &    0.08    &    --\\
\hline\hline
\end{tabular}
\tablefoot{
For O stars, the luminosity and effective temperature~($T_{\mathrm{eff}}$) are calculated based on the calibration of solar-metallicity O stars of type \textsc{i}, \textsc{iii}, and \textsc{v}~\citep{Martins2005}.
The calibration for B stars are based on \citet{Pecaut2013}.}
\raggedright
\tablefoottext{a}{from \citet{Povich2011}}
\tablefoottext{b}{from \citet{Gagne2011}}
\tablefoottext{c}{The listed luminosity for \textsl{HD\,93\,162} is calculated as if it is an O\,2.5\,I star. However, the luminosity for this star is uncertain and can possibly be five times brighter~\citep{Hamann2006}.}
\end{table*}

The $G_{\mathrm{UV}}$ values obtained by our comparison between PDR models and observations are larger than the ones cited in previous works. In particular \cite{Brooks2003} estimated the FUV flux in several ways. First, based on the \cite{Kaufman1999} PDR models, they find a FUV scaling factor between $600$ and $10^4$ in the Habing unit. They also considered the most massive stars in the Tr 14 cluster and estimated the resulting flux at the [C\textsc{ii}] peak location (10:43:22\,-59:34:45), finding $1.4\times10^4$ in the Habing unit, {\it i.e.} $10^4$ in the Mathis units.

We re-investigate the FUV luminosity that illuminates the molecular gas observed in the FOV. The spatial variation of our derived $G_{\mathrm{UV}}$ across the FOV suggests that the CO clouds may be distributed at different distances from the illuminating stars. As the massive stellar members of Trumpler 14 and 16 are well identified~\citep{Vazquez1996, Sana2010, Hur2012, Sota2014, Apellaniz2016, Alexander2016}, we are able to compare the derived $G_{\mathrm{UV}}$ with the estimated UV-photons input by the nearby OB-stars. 

Table~\ref{table:stars} lists all observed OB-stars within a projected distance of $7\,\mathrm{pc}$~($\sim7.5'$) from the center of our FOV with  updated FUV luminosity. If one is unaware of the spatial distributions of stars and naively assumes that all the listed stars are located at the center of their host star-clusters, and that the observed CO gas lies at the same plane perpendicular to our line of sight at a $2.3\,\mathrm{kpc}$ distance, the estimated stellar $G_{\mathrm{UV}}$ values and the $G_{UV}$ values derived from the Meudon PDR models agree poorly with each other (left panel in Figure~\ref{fig:chiobstar_pdr}). However, when the sky coordinates of the nearby OB-stars are taken into account (still coplanar with the CO gas), the comparison between the estimated stellar $G_{\mathrm{UV}}$ values and the $G_{UV}$ values derived from the Meudon PDR models slightly improves in the Car I-E and Car I/II regions~(center panel in Figure~\ref{fig:chiobstar_pdr}).

Two stellar members of Trumpler 14, an O-stars binary system, \textsl{ALS 15 204}, and a B0 star, \textsl{ALS 15 203}, are observed in the Car I-S region (near the pointing \#3 in Figure~\ref{fig:carina_zoom}). Observations with the 4m Anglo-Australian Telescope (AAT) have suggested that a separate subcluster may be formed around these two stars~\citep{Alexander2016}. Judging from the spectrally-resolved CO $\mathrm{J=1-0}$ observations\footnote{Taken from the Carina Parkes-ATCA Radio Continuum Survey publicly available at http://mopra.org/data/}~\citep{Rebolledo2016} across the six central pointings~(see Figure~\ref{fig:carina_zoom}) from our FOV in Figure~\ref{fig:co10_profile}, the CO $1-0$ emission at Car I-S region is dominated by the velocity component at around $-13\,\mathrm{km\,s^{-1}}$~(pointings \#\,3, \#\,4, and \#\,5). Assuming that the subcluster carves out a cavity of CO gas in the Car I-S region, the observed CO emission from the region may be  associated to the cloud sitting further away from the observer along the line of sight. With a simple cloud geometry demonstrated in Figure~\ref{fig:3dview}, which is created by moving the cloud directly over \textsl{ALS 15 204} and \textsl{ALS 15 203} in Car I-S further by $\sim\,3\,\mathrm{pc}$\footnote{This distance is motivated by the assumption that the cavity created by the subcluster has a similar dimension along the line of sight as the approximate size of the Car I-S region.}, ensuring the spatial continuity of the cloud along constant slopes into Car I-E and Car I/II, and taking the coordinates of nearby OB-stars into account, the $G_{\textrm{UV}}$ values derived with the Meudon PDR code come to an excellent agreement with the expected $G_{UV}$ contributed by the OB-stars in the field without scaling~(right panel in Figure~\ref{fig:chiobstar_pdr}). Even if we assume that the distance to Trumpler\,14 is $\sim2.9\,\mathrm{kpc}$~\citep{Hur2012}, in which case the expected $G_{UV}$ contributed by the OB-stars would decrease by a factor of $\sim\,1.5$, the agreement still holds within the estimated uncertainties.

In our simple and crude geometry assumption, all the stars in the vicinity are located on the same plane, which is unlikely the reality. Unfortunately, given the uncertainties in our observations and models, we do not have sufficient information to accurately constrain the positions of stars and clouds along our line of sight. However, the excellent agreement displayed in Figure~\ref{fig:chiobstar_pdr} demonstrates that our simple assumption may be a good approximation to the reality.

\section{Discussion}
\label{sec:discussion}
%
\subsection{UV radiation field in Carina}
\label{subsec:co_structure}

The CO emitting gas observed in this work only cover a small area~($\sim\,2\,\times\,5\,\mathrm{pc^{2}}$) of the Carina Nebula~(Figure~\ref{fig:carina_fov}). In this subsection, we discuss our findings under the frame of the whole Carina Nebula. The Carina Nebula is famous for its bipolar structure with its major axis almost perpendicular to the Galactic plane. In its northern part, expansion along the Galactic plane caused by the formation of young star clusters is inhibited by the dense molecular gas~\citep{Smith2000a}. One of the biggest mysteries in the Carina Nebula is its supernova history.

Based on the observed line-splitting of ionized gas and the total gas mass estimated from the cool dust mass, \citet{Smith2007a} estimated the total kinetic energy of the nebula to be $8\,\times\,10^{51}\,\mathrm{erg}$, which represents $30\%$ of the total mechanical energy input by the stellar wind during the past $3\,\mathrm{Myr}$~\citep{Smith2006a}. The lack of radio or X-ray synchrotron emission indicates that a recent supernova activity is unlikely. 

However, confirmation of the diffuse X-ray emission which offsets from the young massive star clusters and the discovery of an isolated neutron star, $2\mathrm{XMM\,J}104608.7-594306$, have led to a possible supernova history in the Carina Nebula~\citep{Townsley2011, Pires2015}. In the vicinity of our FOV, an observed B-type star~(Tr\,14-29 in Table~\ref{table:stars}, also known as Trumpler\,14-MJ\,218) is moving with a high velocity~($\sim$ 100 km s$^{-1}$) toward the center of Trumpler\,14, directly away from $2\mathrm{XMM\,J}104608.7-594306$. If Tr\,14-29 is the runaway star associated with $2\mathrm{XMM\,J}104608.7-594306$ during a past supernova explosion, its estimated flight time is $\sim\,(1.1-3)\,\times\,10^{4}\,\mathrm{yr}$~\citep{Pires2015}. In an alternative scenario that Tr\,14-29 originates from the open star cluster, Trumpler\,16, which sits right between Tr\,14-29  and $2\mathrm{XMM\,J}104608.7-594306$, its flight time would be $10^{5}\,\mathrm{yr}$~\citep{Ngoumou2013}. Both estimated flight times are roughly consistent with the absence of a supernova remnant in the Carina Nebula. All these characteristics suggest that the region in our FOV may be associated with a Galactic superbubble. The observed physical conditions of PDRs can thus provide a suitable analogue for the observations of more faraway starburst regions hosted in similar interstellar environments, such as the famous 30 Doradus.

Comparing the CO SLEDs from the Car I-E, Car I-S, and Car I/II regions~(see Figure~\ref{fig:sled_3pix}), it is clear that the Car I-E region is dominated by CO gas of high thermal pressure and strong radiation fields~($P\,=\,1.8\,\times\,10^{8}\,\mathrm{K\,cm^{-3}}$ and $G_{\mathrm{UV}}\,=\,1.7\,\times\,10^{4}$). The thermal pressure and the strength of radiation decrease~($P\,=\,5.6\,\times\,10^{7}\,\mathrm{K\,cm^{-3}}$ and $G_{\mathrm{UV}}\,=\,3.4\,\times\,10^{3}$) in the Car I-S region and then increase again in the Car I/II region~($P\,=\,1.1\,\times\,10^{8}\,\mathrm{K\,cm^{-3}}$ and $G_{\mathrm{UV}}\,=\,1.5\,\times\,10^{4}$). Within our FOV, the PDRs in the Car I-E region have been previously studied by several groups. Based on the approximate stellar composition and the far-infrared~(FIR) continuum observed in Car I, \citet{Brooks2003} and \citet{Mizutani2004a} have estimated the $G_{\mathrm{UV}}$ to be $\sim\,10^{4}$ in the Habing unit. On the other hand, estimations with PDR models have been thus far giving somehow lower $G_{\mathrm{UV}}$ values. Constrained by the [C\textsc{ii}]\,$158\,\mum$, [O\textsc{i}]\,$63$ and [O\textsc{i}]\,$145\,\mum$ emissions observed with \textsl{ISO}, the interpretations with the PDR Toolbox~\citep{Pound2008, Kaufman2006} estimate the $G_{\mathrm{UV}}$ values to be around $1390$, in the Habing unit~($\sim\,1000$ in the Mathis unit), in Car I~\citep{Oberst2011}. With a clumpy PDR model, using the observed CO $\mathrm{J=2-1, J=4-3, J=7-6}$, and the two atomic carbon lines, [C\textsc{i}] $370$ and $609\,\mum$, the strength of the radiation field in Car I-E is estimated to be $\sim\,3200$ in the Habing unit~($\sim\,2500$ in the Mathis unit)~\citep{Kramer2008}. In a big picture, these results indicate that the UV-photons and stellar winds provided by nearby massive stars are more than sufficient in supporting the global heating of CO and [C\textsc{i}] emissions observed in this region.

Using the Meudon PDR models, the $G_{\mathrm{UV}}$ values constrained from the \herschel~SPIRE/FTS observations in the Car I-E region range between $10^{4}$ and $3\,\times\,10^{4}$, which are generally higher than the values found in the literature. However, the excellent agreement of the model-derived $G_{\mathrm{UV}}$ values with the estimated radiation fields from the nearby stellar composition appears throughout our FOV, as demonstrated in Figure~\ref{fig:chiobstar_pdr}. Our analysis has confirmed that the chief energy source for the observed CO emissions is the UV-photons input by the massive stars in the vicinity. We would like to note from our results that in the Car I-E region, the model-derived $G_{\mathrm{UV}}$ values tend to be marginally higher than the $G_{\mathrm{UV}}$ contributed solely by the OB-stars. The emission of CO observed along the ionization front may also be supported by local heating sources. In Car I-E, narrow-band imaging-observations from the optical to the mid-infrared have revealed that many massive~(B4 to A0) young stellar objects~(YSOs) are distributed on or just behind the ionization front~\citep{Tapia2006}. Along the ionization front, the model-derived thermal pressure values are the highest in our FOV~($\sim\,2\,\times\,10^{8}\,\mathrm{km\,cm^{-3}}$, see Figure~\ref{Fig:BestParamMap}). Interestingly, along the high-pressure ionization front, at the very edge of our FOV in the northwest, the highest values for the CO and [C\textsc{i}] emissions are observed~(see figures in the Appendix). At this location~(10:43:23.25, -59:33:56.9), a massive class I YSO, V723 Car, has been observed to go into outburst just before 2003~\citep{Tapia2015}. All $^{12}$CO transitions covered by the \herschel~SPIRE/FTS and $^{13}$CO, up to $\mathrm{J=8-7}$, is unambiguously detected here. This result suggests a potential contribution of local heating sources. Unfortunately, we cannot derive the physical conditions from this location due to the variation of instrument resolution, but it will be an interesting target for future high-resolution observation for understanding the interplay between the formation of massive YSOs and their host environment.

The location of Car I/II region has almost the same distances to the centers of Trumpler 14 and 16 and lies at the intersection of Car I and Car II. Based on the projected distances from the O- and B-stars in the vicinity~(see Table~\ref{table:stars}), the $G_{\mathrm{UV}}$-values at the Car I/II region should be comparable or weaker than the Car I-E region. This is the case for most of the pixels in the region. Globally in Car I/II, the dust mass and temperature, derived from the photometry data in our FOV, are quite low, however CO is unambiguously observed~(Wu et al. in prep.). This result is in agreement with the comparison of dust-mass, estimated with the continuum at $1.2\,\mathrm{mm}$~($9\,\mathrm{M_{\odot}}$, \citealt{Brooks2005}), and gas-mass, estimated with $\mathrm{CO\ J=2-1}$~($91\,\mathrm{M_{\odot}}$~(LTE) or $411\,\mathrm{M_{\odot}}$~(virial), \citealt{Rathborne2002}). The large discrepancy between the gas-mass estimated under LTE and virial assumption implies that the observed cloud is not gravitationally bound but supported by external radiation fields~\citep{Rathborne2002}. 

One thing that has been puzzling from our results is the local high $G_{\mathrm{UV}}$ values~($\sim\,5\,\times\,10^{4}$) derived in the Car I/II region of our FOV~(see Figure~\ref{Fig:BestParamMap}). At this location~(10:44:03.89, -59:37:57.63), CO emission is well detected up to $\mathrm{J=12-11}$. Under the LTE assumption, the $T_{\mathrm{ex}}$ value derived locally is $\sim\,80\,\mathrm{K}$, while $T_{\mathrm{ex}}$ is generally below $70\,\mathrm{K}$ in the Car I/II region. This point also appears as the only outlier in Figure~\ref{fig:chiobstar_pdr}, hinting that the UV-photons from nearby OB-stars may not be able to support the CO excitation. Observations in $\mathrm{CO\ J=1-0}$ with \textsl{Mopra} show a velocity component centered at $-5\,\mathrm{km s^{-1}}$, which appears only at this point in our FOV~(see dashed-line in Figure~\ref{fig:co10_profile}), implying that the structure of CO gas may differ locally. The CO cloud observed here is likely heated partially by stellar members in Trumpler 14 and partially by the two nearby early-type stars, \textsl{HD\,93\,162} and \textsl{ALS\,15\,210}, of Trumpler 16. \textsl{ALS\,15\,210} has been classified as an O\,3.5\,I star. The Wolf-Rayet star, \textsl{HD\,93\,162}, also known as \textsl{WR\,25}, is possibly the most luminous known star in our Galaxy~(up to $L\,=\,6.3\,\times\,10^{6}\,L_{\odot}$, \citealt{Hamann2006}), but it is still not sufficient to account for the local high $G_\mathrm{UV}$ value. The UV-photons input by the nearby massive stars simply cannot explain the model-derived $G_{\mathrm{UV}}$ values locally. Based on the modeling results of the $1\,-\,24\,\mathrm{\mum}$ spectral energy distribution in the Carina Nebula, a type 0/I YSO of intermediate-mass with hard X-ray signature has been detected at (10:44:03.086, -59:37:48.41)~\citep{Povich2011} and listed in the Pan-Carina Young Stellar Object Catalog~(PCYC) as \textsl{PCYC\,419}. However, at the physical scale of $\sim\,0.2\,\mathrm{pc}$, we do not expect the outflow from an intermediate-mass YSO to dominate the heating of CO gas. To further understand the local heating mechanisms for the observed CO gas, high-resolution observations from the region is necessary.

\subsection{CO at PDR interfaces}
\label{subsec:co_pdr}

For more than 30 years, PDRs have been studied to understand the transition between atomic and molecular gas as well as the impact of the radiative feedback of massive stars on their parent cloud~\citep{Tielens1985, Sternberg1989}. Before \herschel~observations, the most accessible lines probing the very edge of PDRs and the physics and chemistry that take place in these interfaces were the atomic lines of O, C$^+$ and C as well as molecular lines of H$_2$ and a few CO lines. Comparisons of these observations to PDR models lead to the conclusion that the edge of PDRs can be seen as a clumpy medium in which dense clumps are embedded in a more diffuse medium \citep{Stutzki1990, Parmar1991, Meixner1993, Tauber1994, Hogerheijde1995, Andree-Labsch2017}. In this scenario, UV photons can penetrate deeply into the cloud through the interclump medium ($n_\textrm{H} \sim 10^4-10^5$ cm$^{-3}$) and heat the edge of the dense clumps ($n_\textrm{H} \sim 10^6-10^7$ cm$^{-3}$) where H$_2$ vibrational lines and CO mid and high-J lines are produced. More recently, \herschel~observations of CO rotational emission in protostars, PDRs and extragalactic regions also showed high-J excitation. Shapes of CO SLEDs in different environments have been compared by \cite{Indriolo2017}. In protostellar systems as Orion KL, the very high-J excitation (J$_{up} \sim 40$), indicates energy inputs through outflows and shocks \citep{Goicoechea2015a}. In extragalactic regions, the sources of energy responsible for CO excitation are less clear because of the mixture of several emitting sources in the beam. \cite{Rosenberg2015} suggest that CO excitation in some Class I galaxies may be dominated by UV heating whereas Class III galaxies would require mechaninal heating (shocks and turbulence) or X-rays heating.

Here, we find that highly illuminated ($G_{UV} \sim 10^4$) PDR models with high thermal pressures ($P \sim 10^8$ K cm$^{-3}$) can successfully explain CO excitation in the Carina Nebula without introducing a clumpy cloud-structure. UV photons, via the photo-electric effect on grains, are the main heating mechanism. Isobaric models have been proposed as an alternative to the classical clump approach for the Orion Bar~\citep{Marconi1998, Allers2005}. An isobaric model is found to successfully explain the CH$^+$ emission in the Orion Bar~\citep{Nagy2013}. More recently, it is suggested that the PDRs can support several tens of atomic and molecular lines including H$_2$ lines and CO lines from $\mathrm{J=4-3}$ to $\mathrm{J=23-22}$ in two prototypical PDRs, the Orion Bar and NGC~7023 NW~\citep{Joblin2018}.

It is suggested that hot chemistry takes place at the edge of PDRs and is responsible for the mid-J and high-J CO lines that are observed~\citep{Joblin2018}. In our models, we simulate the formation of H$_2$ with the formalism described in \citet{LeBourlot2012} that considers the Langmuir-Hinshelwood and Eley-Rideal mechanisms. This detailed treatment increases the formation rate of H$_2$ compared to the classical value determined in diffuse interstellar gas by Copernicus and FUSE, $3\times10^{17}$ cm$^3$ s$^{-1}$ \citep{Jura1974, Gry2002} that is often used in astrochemical models. Typically, in the Car I-E area, we find a 3 to 4 times higher formation rate at the edge of the PDR. This enhanced formation rate allows H$_2$ to form close to the edge of the PDR, in warm gas (a few hundred Kelvins depending on physical conditions). In this warm gas, some key chemical reactions with H$_2$ usually inhibited because of large activation energies can be efficient because of, first, the relative velocity between the reactants and, second, the internal energy stocked in H$_2$ ro-vibrational levels. That is the case of the C$^+$ + H$_2$ reaction that forms CH$^+$ with an activation energy of 4537~K. We compute the rate of this reaction with the formalism of \citet{Agundez2010}.

\begin{figure}[htbp!]
\centering
\includegraphics[width=0.45\textwidth]{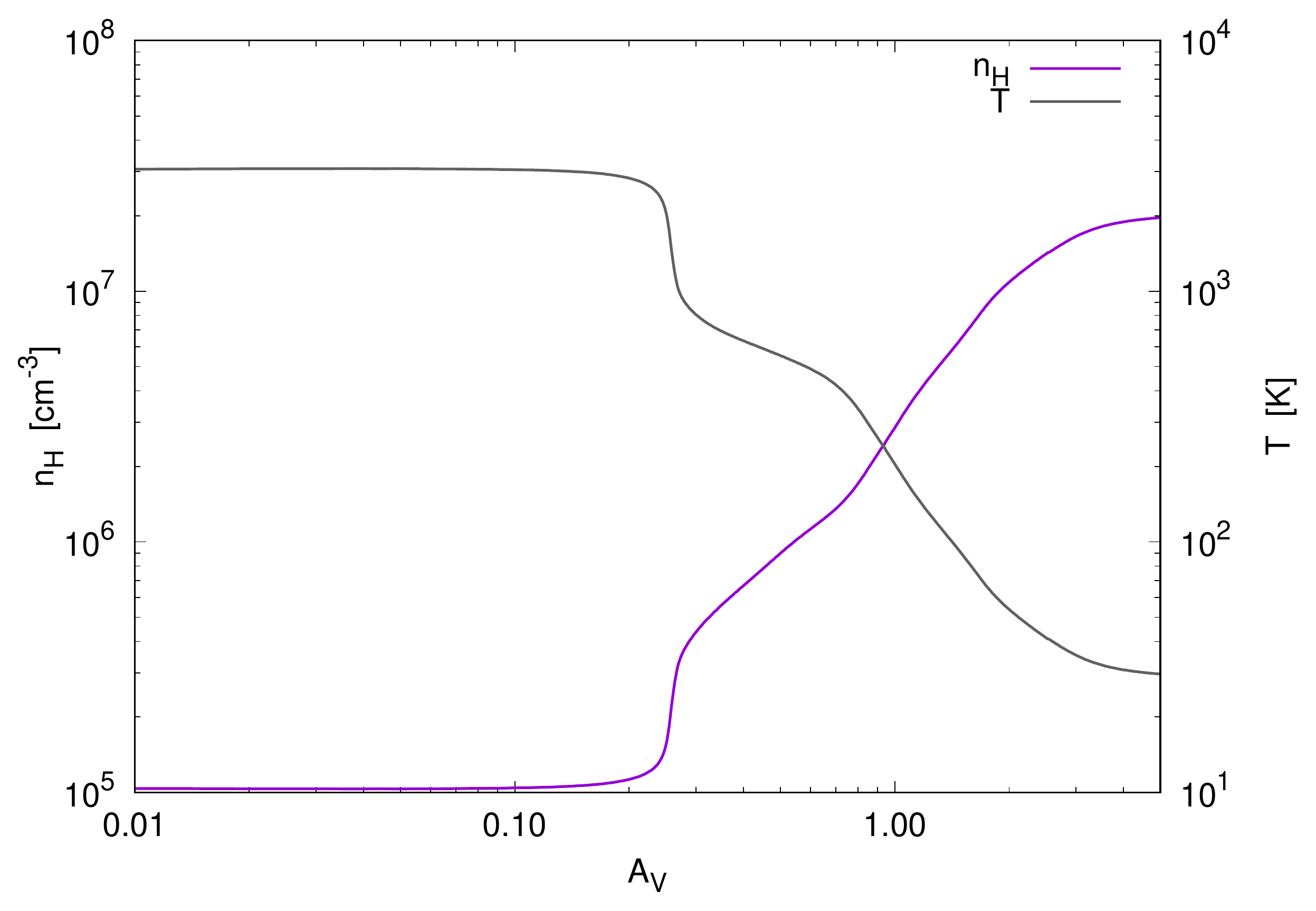}
\includegraphics[width=0.47\textwidth]{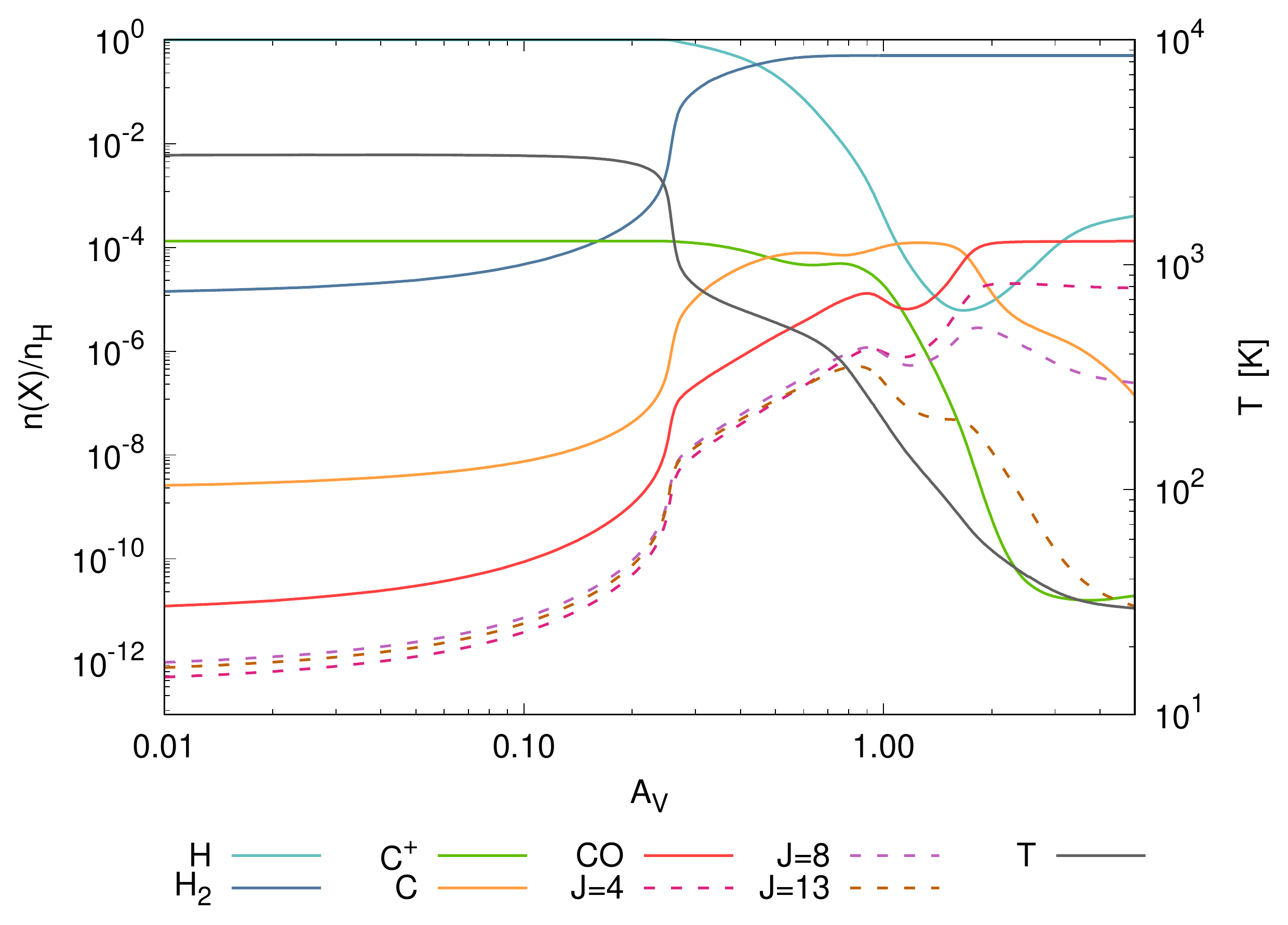}
\caption{Example of density and temperature profiles computed by the Meudon PDR code for a model with $P = 3.5\times10^8$ K cm$^{-3}$, $G_{UV} = 2\times10^4$ and with a total depth $A_\textrm{V}$ = 10. Top panel presents the proton density and the gas temperature profiles. Bottom panel presents the relative densities of H, H$_2$, C$^+$, C and CO as well as the relative densities of CO $\mathrm{J=4, 8}$ and 13.\label{Fig:PDRModel_Profiles}}
\label{fig:PDRProfile}
\end{figure}

Then, several chemical routes that start with CH$^+$ lead to the formation of CO. As a consequence, in bright PDRs, the formation of CO starts as soon as H$_2$ self-shielding becomes in effect. Figure \ref{fig:PDRProfile} presents the density profiles of H, H$_2$, C$^+$, C and CO (total and in its levels $\mathrm{J=4, 8}$ and 13) as a function of $A_\textrm{V}$ for a model with $P = 3.5\times10^8$ K cm$^{-3}$ and $G_\textrm{UV} = 2\times10^4$. CO level $\mathrm{J=13}$ is populated before the C/CO transition. So, a fraction of the emission of CO in mid-J and high-J levels comes from a slab of the PDR in front of the C/CO transition. Recent \textsl{ALMA} observations of the Orion Bar presented in \citet{Goicoechea2016, Goicoechea2017} show the same trend of chemical stratification. In particular, SH$^+$ emission whose formation path is similar to the one of CH$^+$ but with an activation energy threshold of 9860 K, shows a clear front close to the edge of Bar.

\subsection{P-G$_{UV}$ relation}
\label{subsec:PG0}

\begin{figure}[htbp!]
\centering
\includegraphics[width=0.4\textwidth]{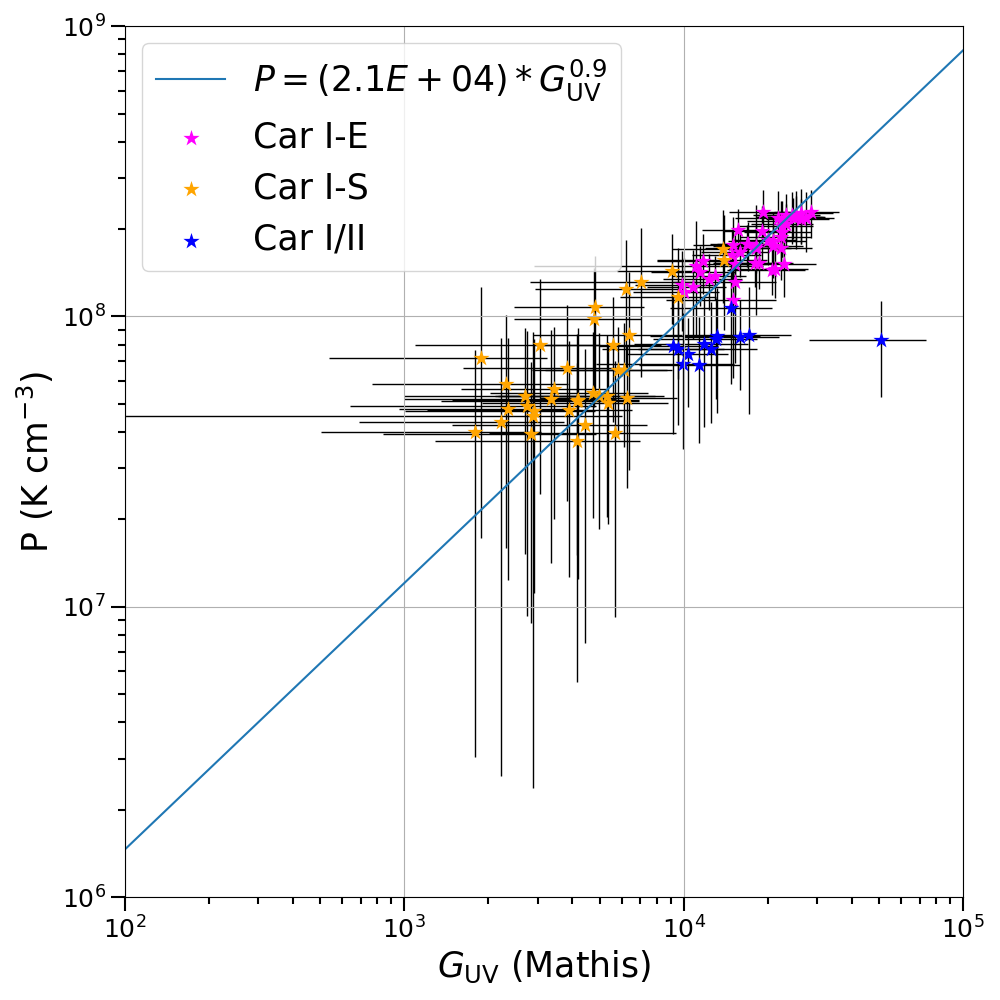}
\caption{Relationship between the derived $P$ and $G_{\mathrm{UV}}$ for our FOV. The solid blue line shows the empirical power-law relationship between $P$ and $G_{UV}$ values.}
\label{fig:P_chi_pdr}
\end{figure}

Figure \ref{fig:P_chi_pdr} presents the thermal pressure as a function of $G_{UV}$ for all the pixels of our Carina FOV.  As the pressure, $P$, is derived from CO and C lines in this work, it is generally associated with neutral gas. Except in the Car I-S region, which is not very bright in the CO emission and dominated by the H\textsc{ii} region, we fit a power-law relationship between $P$ and $G_{UV}$ values in Figure \ref{fig:P_chi_pdr}. In the Car I/II region, some pixels seem to have the same thermal pressure for different $G_{UV}$ while for other ones $P$ increases with $G_{UV}$. In the Car I-E region, which is dominated by the bright PDR, the relationship is the most prominent. From our observations, the empirical relationship between $P$ and $G_{UV}$ is:
\begin{equation}
P =  2.1\times10^4 \,\, G_\textrm{UV}^{0.9}.
\end{equation}

In the extragalactic observation, a similar relationship has been suggested in \citet{Wu2015}. In the Galactic PDRs, a similar relationship between $P$ and $G_\textrm{UV}$ values spanning over more than 3 orders of magnitude has also been observed~\citep{Joblin2018}. The observed relationship suggests that the UV radiation field of massive stars may create a high pressure layer at the edge of PDRs and the physical mechanism at play could be the UV induced photo-evaporation as described in \citet{Bertoldi1989,Bertoldi1996}. As the Meudon PDR model describes a stationary system, a hydrodynamical PDR code is required to understand the exact mechanism leading to the compression of the gas and to quantify its effect~\citep{Bron2018}.

\section{Conclusion}
The \textsl{Herschel} SPIRE FTS uses a pioneering design to simultaneously observe a broad range of CO transitions in spectral imaging mode which enables the derivation of physical properties of molecular gas in a macroscopic field of view. Combined with the state-of-the-art Meudon PDR models, we obtain the following results:
\begin{enumerate}
\item The CO emission is observed in the Carina Nebula, near the young star cluster, Trumpler\,14, from $\mathrm{J=4-3}$ to $\mathrm{J=13-12}$ across the Car I-E, Car I-S and Car I/II regions~(Figure~\ref{fig:carina_zoom} and figures in the Appendix). In the Car I-E region at the northwest of our FOV, the intensities of the observed CO are generally higher by one order of magnitude than those observed in Car I-S and Car I/II. Beyond $\mathrm{J=10-9}$, the emission is concentrated in the Car I-E and locally in the Car I/II regions, hinting higher excitation temperatures of CO gas in these regions.
\item We successfully account for the observed CO and [C\textsc{i}] emission with high pressure and high FUV illumination PDR models. Comparing with previous PDR modeling results found in the same region, the $G_{\mathrm{UV}}$ values derived in this work are generally higher~\citep{Brooks2003, Mizutani2004a, Kramer2008, Oberst2011}. However, the excellent relationship observed between the model-derived $G_{\mathrm{UV}}$ values with the estimated radiation fields~(Figure~\ref{fig:chiobstar_pdr}) from nearby stellar compositions suggest the reliability of the derived $G_{\mathrm{UV}}$ values presented in this work. In correspondence with the derived $G_{\mathrm{UV}}$ values, the estimated thermal pressure is of the order of magnitude~$\sim\,10^{8}\,\mathrm{K\,cm^{-3}}$. With the modeling results, we confirm that the chief energy sources for the observed CO emissions are the UV-photons and stellar feedback provided by the massive stars in the vicinity.
\item Although the observed $^{12}\mathrm{CO}$, [C\textsc{i}], and [O\textsc{i}]\,$145\,\mum$ transitions can be well accounted for with the Meudon PDR models, direct comparisons of the model-predicted intensities to the observed $^{13}\mathrm{CO}$, [O\textsc{i}]\,$63\,\mum$, and [C\textsc{ii}]\,$158\,\mum$ intensities are not straightforward. Caution is urged when constraining the physical conditions of PDRs with combinations of $^{12}\mathrm{CO}$, $^{13}\mathrm{CO}$, [C\textsc{i}], [O\textsc{i}]\,$63\,\mum$, and [C\textsc{ii}]\,$158\,\mum$ observations.
\item The thermal pressure derived in this work is of the order of magnitude~$10^{8}\,\mathrm{K\,cm^{-3}}$ across our FOV, with the highest values found at the dissociation front in the Car I-E region. Although the spatial resolution of our observations prevents us from obtaining a detailed cloud structure at the dissociation front, our results are in general agreement with the recent \textsl{ALMA} observations of the Orion Bar that the observed cloud edge is compressed by a high-pressure~($2\,\times\,10^{8}\,\mathrm{K\,cm^{-3}}$) wave propagating into the molecular cloud~\citep{Goicoechea2016, Goicoechea2017}.
\item We find a relation between the thermal pressure at the edge of the PDR and the intensity of the FUV radiation field, $P\,\sim\,2\times10^4\,G_{UV}$. This relation is similar to the one found by \citet{Wu2015} and \citet{Joblin2018}. It suggests that massive stars may create a high pressure slab at the edge of PDRs.
\item Two local~($\sim\,0.2\,\times\,0.2\,\mathrm{pc}$) regions in our FOV deserve further investigation with higher-resolution observations: (1) At the very edge of our FOV~(10:43:23.25, -59:33:56.9), a massive class I YSO, V723 Car has been observed to go into outburst just before 2003~\citep{Tapia2015}. All $^{12}$CO transitions covered by the \herschel~SPIRE/FTS and $^{13}$CO, up to $\mathrm{J=8-7}$, is unambiguously detected here. Unfortunately, we cannot derive the physical conditions from this location due to the variation of instrument resolution, but it will be an interesting target for understanding the interplay between the formation of massive YSOs and their host environment. (2) In the Car I/II region, we find a high $G_{\mathrm{UV}}$ values~($\sim\,5\,\times\,10^{4}$) locally at~(10:44:03.89, -59:37:57.63). At this position, CO emission is well detected up to $\mathrm{J=12-11}$, and the LTE-approximated $T_{\mathrm{ex}}$ value also implies locally excited gas. We cannot identify the possible heating source(s) based on the available observations. To further understand the local heating mechanisms at these locations, high-resolution observations from the region is necessary.
\end{enumerate}

\section*{Acknowledgements}
\begin{itemize}
\item We offer our sincere gratitude to Dr. I. Sakon, Dr. D. Ishihara, Dr. T. Shimonishi, Dr. R. Ohsawa, and, Dr. K. Arimatsu for their success in obtaining the \herschel~observations presented in this work and to Dr. T. I. Mori for investing her time in the \textsl{ISO} data analysis. The authors sincerely thank Dr. D. Rebolledo for sharing the \textsl{Mopra} data, Dr. C. Kramer for sharing the \textsl{NANTEN2} data, Dr. L. K. Townsley and Dr. P. Broos for sharing the \textsl{Chandra} data, and Dr. T. Preibisch for sharing the \herschel~photometry data. We thank Dr. A. Heays and Dr. S. Hony for useful discussions and the anonymous referee for useful suggestions.
\item This research was supported in part by the Grant-in-Aid for Scientific Research for the Japan Society of Promotion of Science (140500000638), the PRC 1311 between France (CNRS) and Japan (JSPS), the Agence Nationale de la Recherche (ANR) through the programme SYMPATICO (Program Blanc Projet (NR-11-BS56-0023), the Programme National Physique et Chimie du Milieu Interstellaire (PCMI) of CNRS/INSU with INC/INP co-funded by CEA and CNES, and the EU FP7 project DustPedia (Grant No.\ 606847).
\item This research has made use of the SIMBAD database~\citep{Wenger2000} and the VizieR catalogue access tool~\citep{Ochsenbein2000}, operated at CDS, Strasbourg, France.
\end{itemize}

\bibliographystyle{aa}
\bibliography{carina_pdr} 
\newpage
\appendix
\section{Line maps observed by the \textsl{Herschel} SPIRE FTS in M83}
\label{appmaps}
\par{We show the observed maps of the lines listed in Table~\ref{table:linetable} by the SPIRE FTS. All the maps are displayed in the observed spatial resolution and in units of $\mathrm{W\,m^{-2}\,sr^{-1} GHz^{-1}}$~(line profile maps) and $\mathrm{W\,m^{-2}\,sr^{-1}}$~(integrated intensity maps).
}
		\begin{figure*}[htbp!]
			\centering
			\subfloat[][]{
				\includegraphics[width=0.45\textwidth]{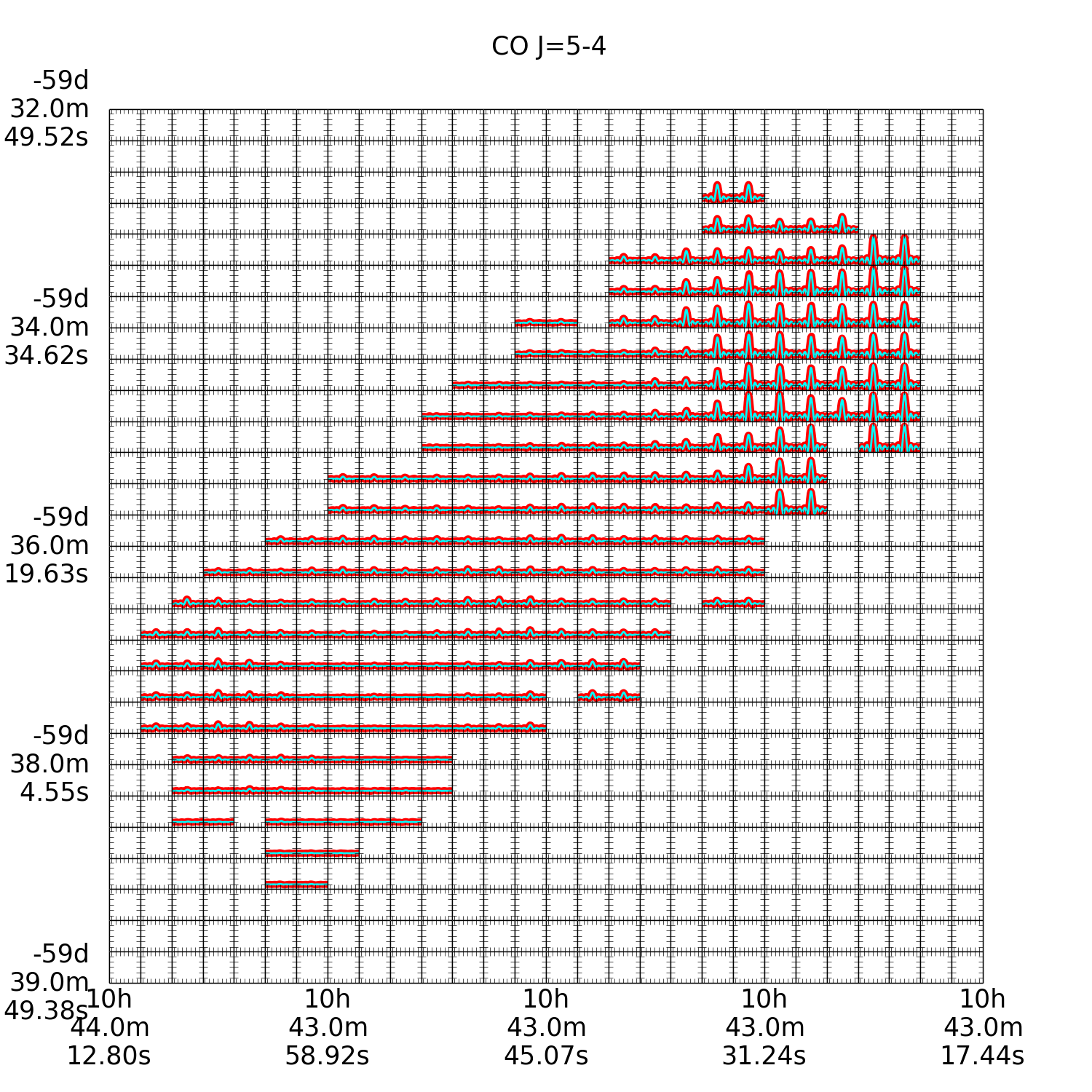}
				\label{co54_linestack}
			}
			\quad
			\subfloat[][]{
				\includegraphics[width=0.45\textwidth]{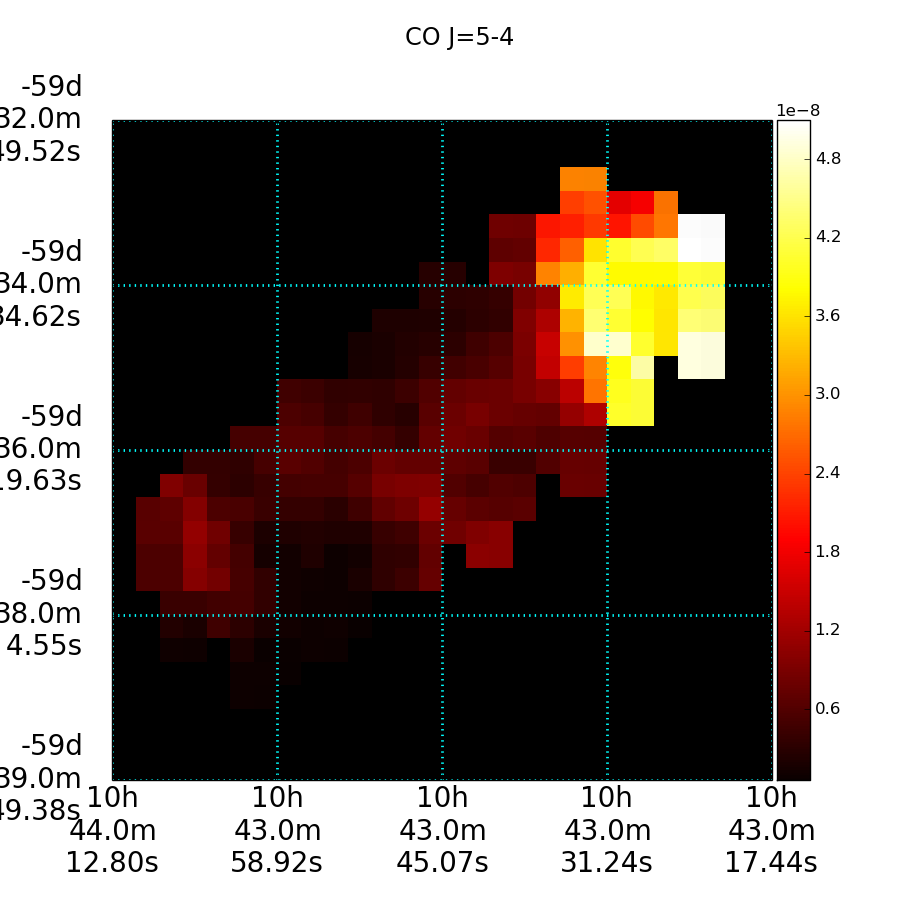}
				\label{co54_intmap}
			}
			\caption{Illustration of the spatial distribution of the observed CO $\mathrm{J}=5-4$ line. The left panel shows the continuum-removed coadded spectrum on every pixel within a range of $568<\nu<582\ \mathrm{GHz}$. The vertical axis in each pixel ranges between $-1.0\times\,10^{-17}$ and $5.2\,\times\,10^{-17}\,\mathrm{W\ m^{-2}\ sr^{-1}\ Hz^{-1}}$. The color map on the right is in units of $\mathrm{W\,m^{-2}\,sr^{-1}}$.}			
		\end{figure*}
		\begin{figure*}[htbp!]
			\centering
			\subfloat[][]{
				\includegraphics[width=0.45\textwidth]{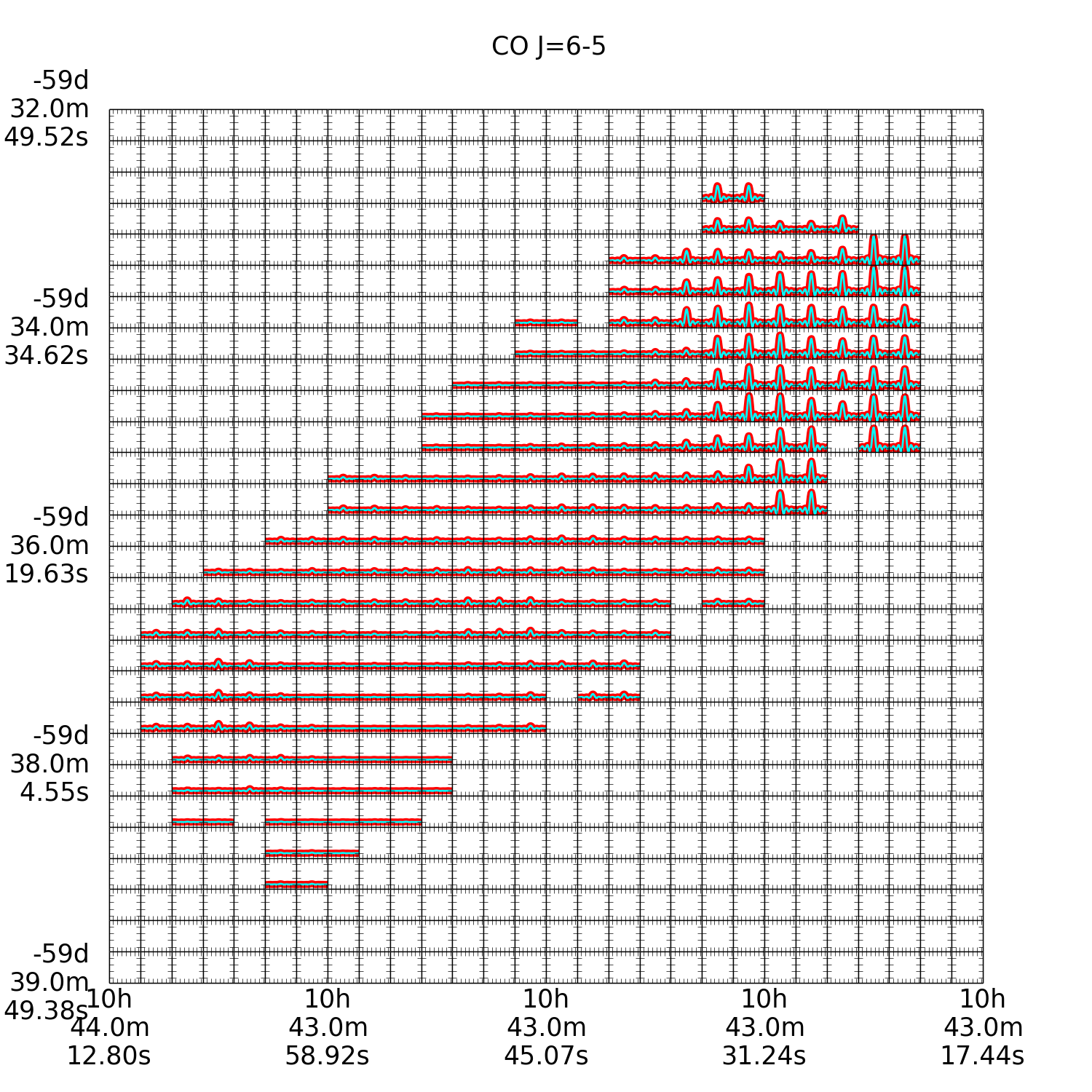}
				\label{co65_linestack}
			}
			\quad
			\subfloat[][]{
				\includegraphics[width=0.45\textwidth]{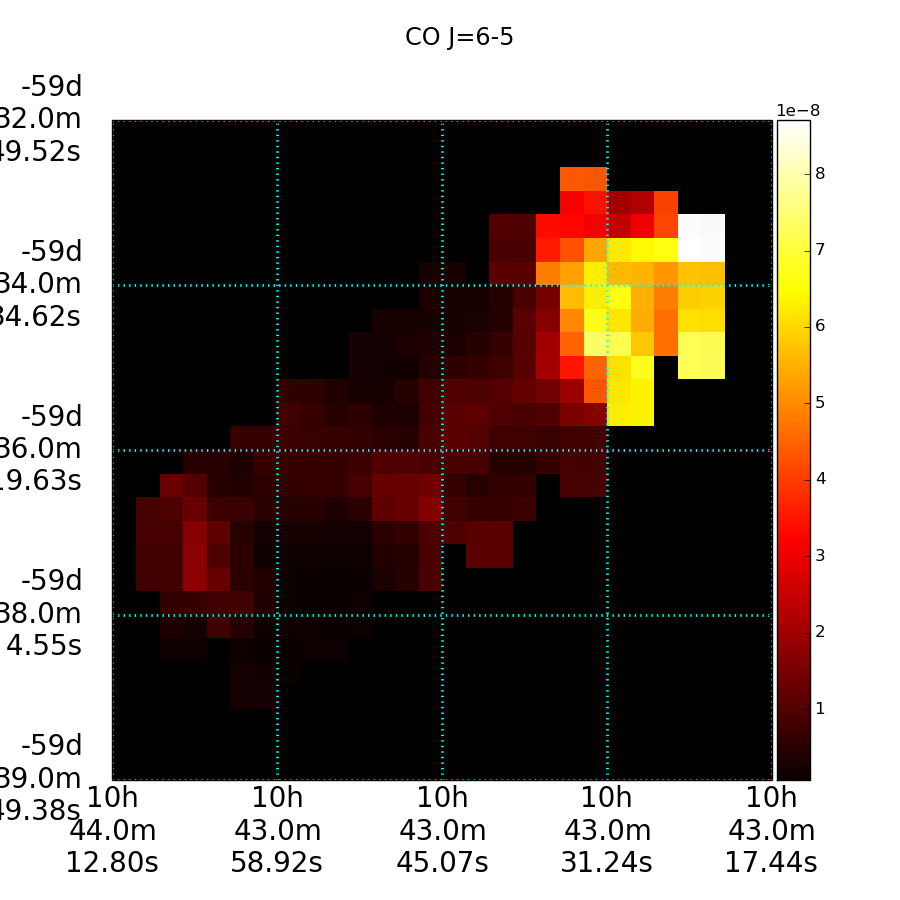}
				\label{co65_intmap}
			}
			\caption{Illustration of the spatial distribution of the observed CO $\mathrm{J}=6-5$ line. The left panel shows the continuum-removed coadded spectrum on every pixel within a range of $683<\nu<698\ \mathrm{GHz}$. The vertical axis in each pixel ranges between $-8.0\,\times\,10^{-17}$ and $8.8\,\times\,10^{-16}\,\mathrm{W\ m^{-2}\ sr^{-1}\ Hz^{-1}}$. The color map on the right is in units of $\mathrm{W\,m^{-2}\,sr^{-1}}$.}			
		\end{figure*}
		\begin{figure*}[htbp!]
			\centering
			\subfloat[][]{
				\includegraphics[width=0.45\textwidth]{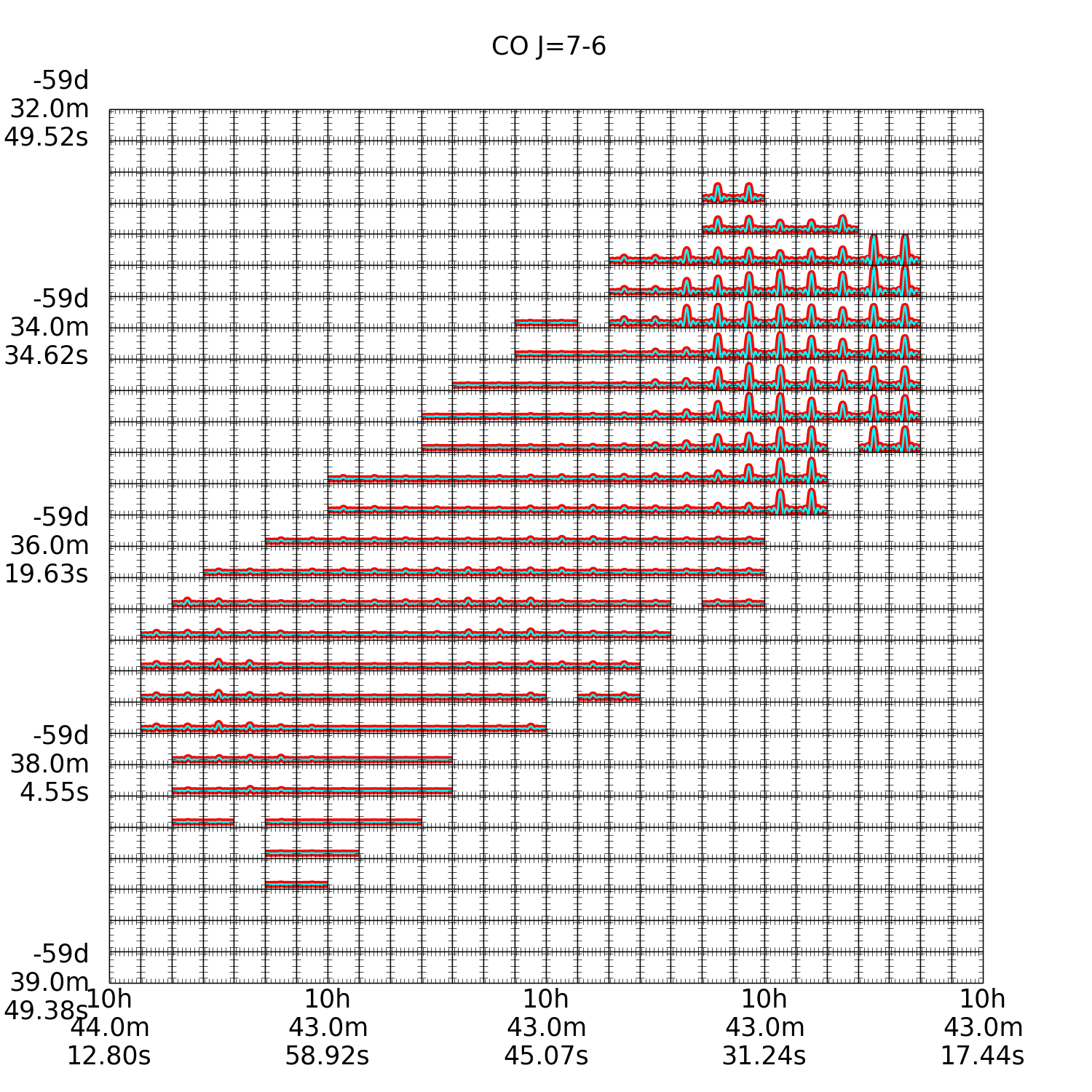}
				\label{co76_linestack}
			}
			\quad
			\subfloat[][]{
				\includegraphics[width=0.45\textwidth]{./CO_maps/COJ7-6_intmap_v14_point_mu.png}
				\label{co76_intmap}
			}
			\caption{Illustration of the spatial distribution of the observed CO $\mathrm{J}=7-6$ line. The left panel shows the continuum-removed coadded spectrum on every pixel within a range of $798<\nu<813\ \mathrm{GHz}$. The vertical axis in each pixel ranges between $-2.0\,\times\,10^{-17}$ and $1.0\,\times\,10^{-16}\,\mathrm{W\ m^{-2}\ sr^{-1}\ Hz^{-1}}$. The color map on the right is in units of $\mathrm{W\,m^{-2}\,sr^{-1}}$.}			
		\end{figure*}
		\begin{figure*}[htbp!]
			\centering
			\subfloat[][]{
				\includegraphics[width=0.45\textwidth]{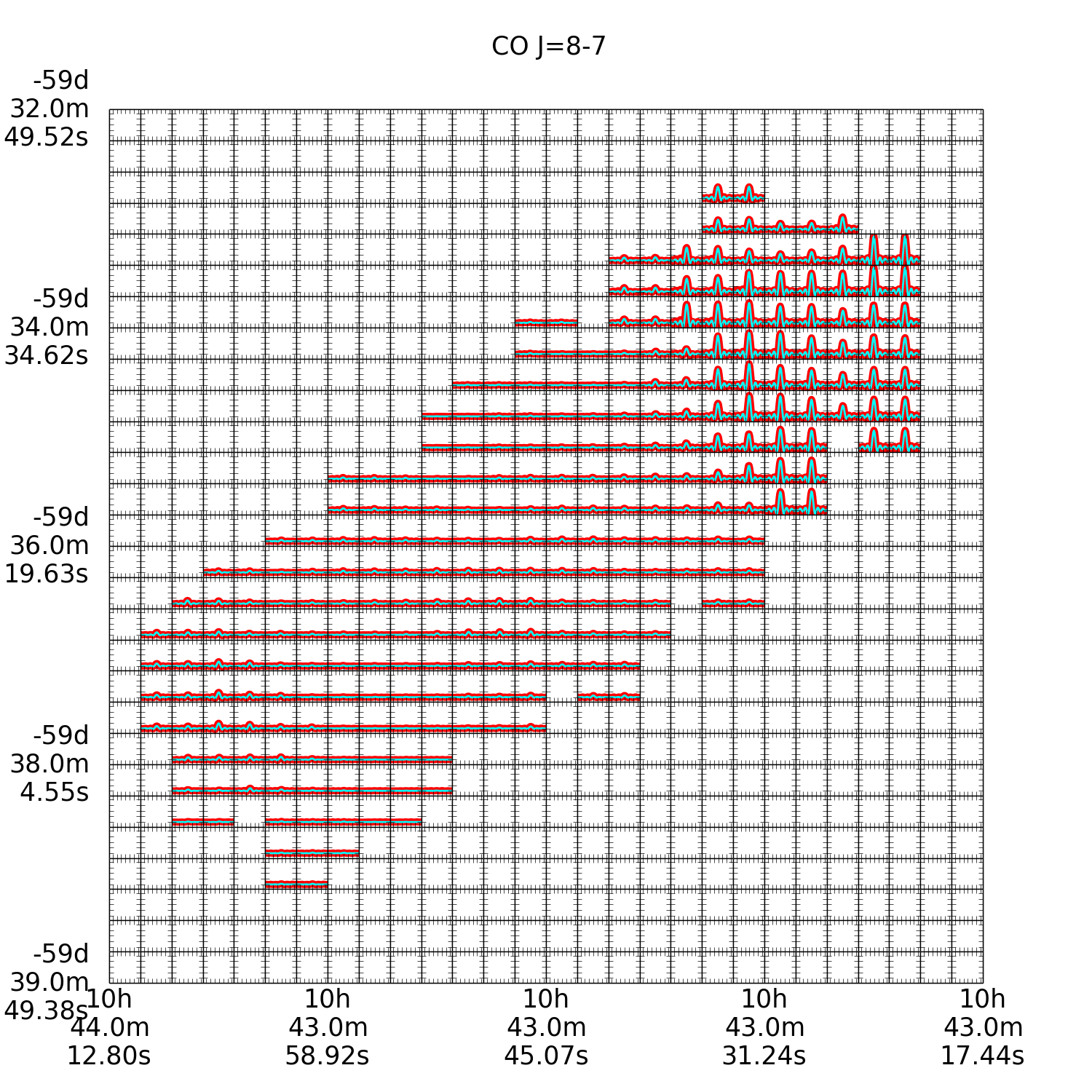}
				\label{co87_linestack}
			}
			\quad
			\subfloat[][]{
				\includegraphics[width=0.45\textwidth]{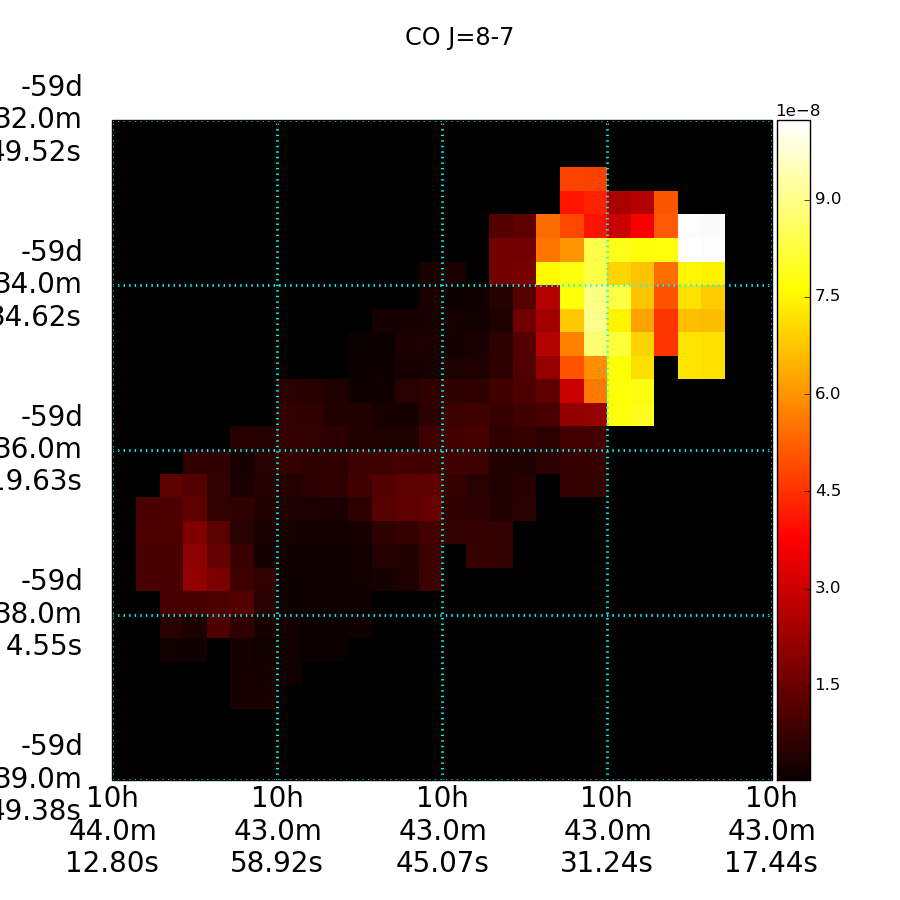}
				\label{co87_intmap}
			}
			\caption{Illustration of the spatial distribution of the observed CO $\mathrm{J}=8-7$ line. The left panel shows the continuum-removed coadded spectrum on every pixel within a range of $913<\nu<928\ \mathrm{GHz}$. The vertical axis in each pixel ranges between $-1.0\,\times\,10^{-17}$ and $1.0\,\times\,10^{-16}\,\mathrm{W\ m^{-2}\ sr^{-1}\ Hz^{-1}}$. The color map on the right is in units of $\mathrm{W\,m^{-2}\,sr^{-1}}$.}			
		\end{figure*}
		\begin{figure*}[htbp!]
			\centering
			\subfloat[][]{
				\includegraphics[width=0.45\textwidth]{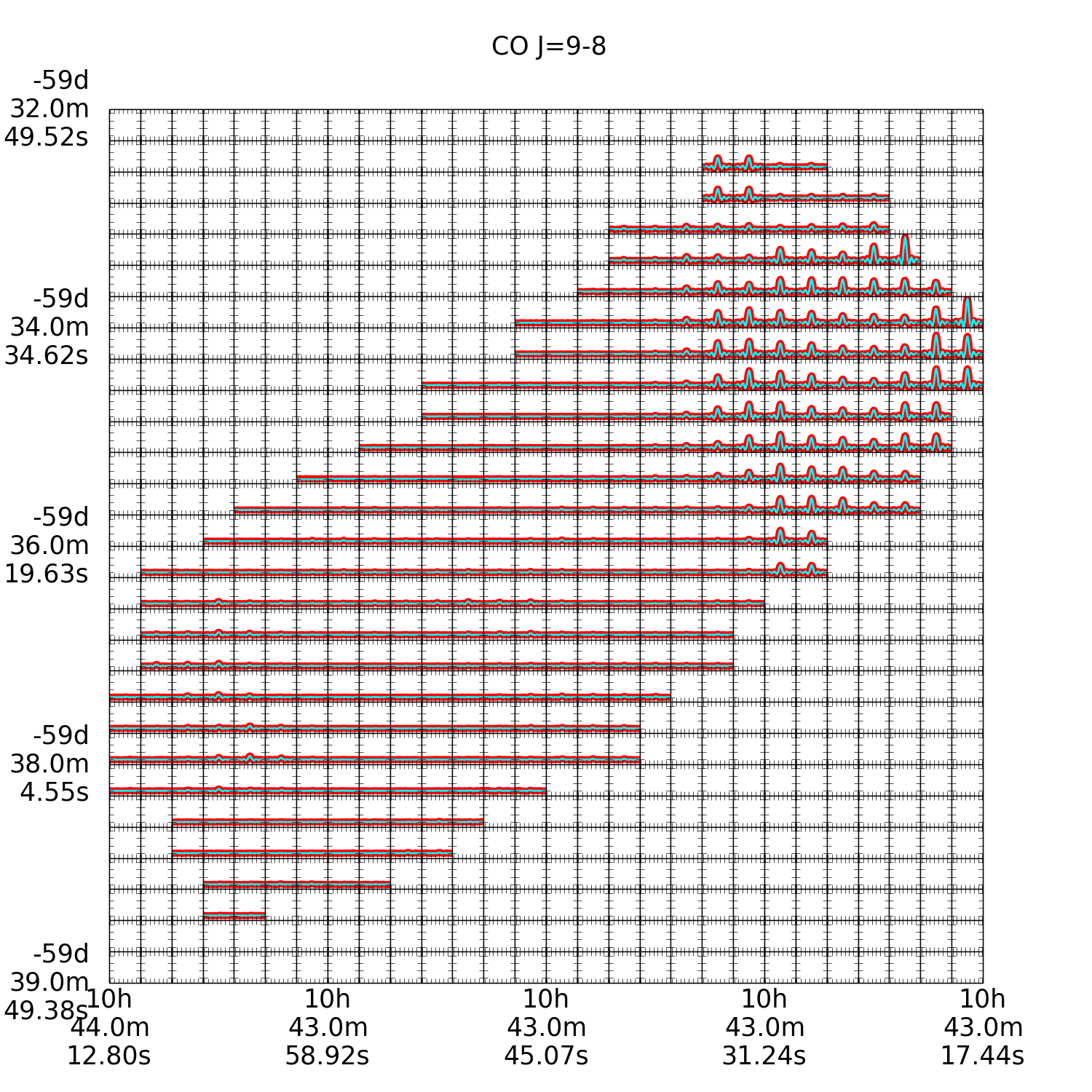}
				\label{co98_linestack}
			}
			\quad
			\subfloat[][]{
				\includegraphics[width=0.45\textwidth]{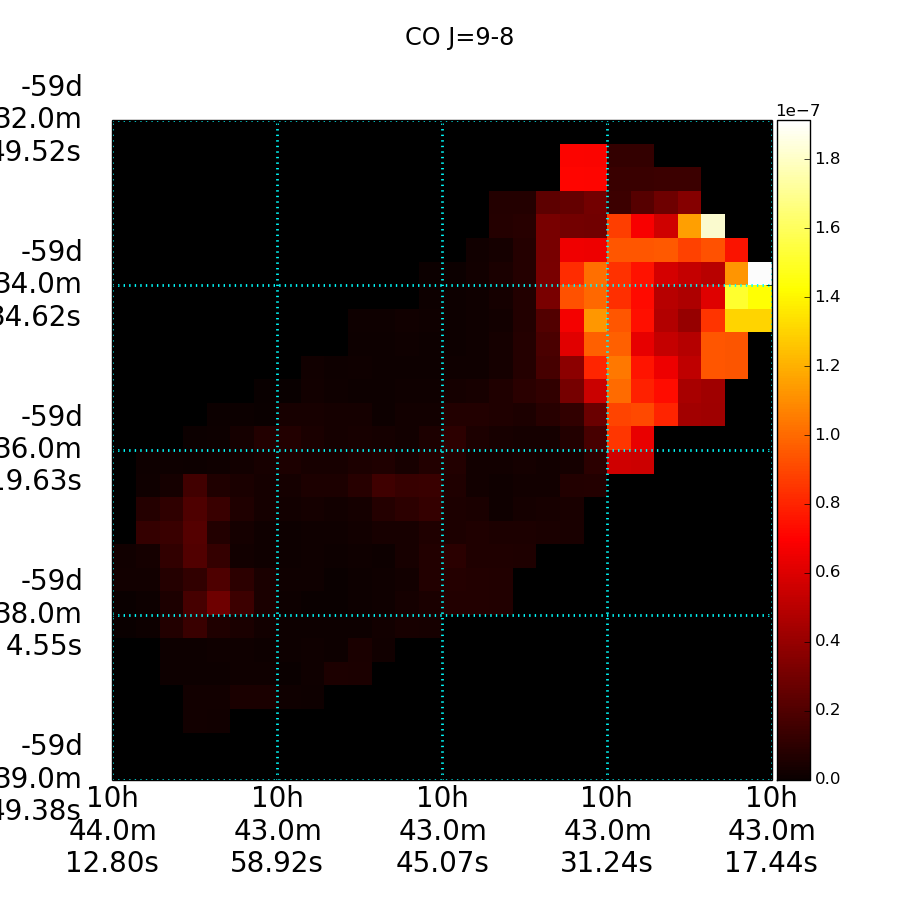}
				\label{co98_intmap}
			}
			\caption{Illustration of the spatial distribution of the observed CO $\mathrm{J}=9-8$ line. The left panel shows the continuum-removed coadded spectrum on every pixel within a range of $1028<\nu<1042\ \mathrm{GHz}$. The vertical axis in each pixel ranges between $-4.0\,\times\,10^{-17}$ and $1.9\,\times\,10^{-16}\,\mathrm{W\ m^{-2}\ sr^{-1}\ Hz^{-1}}$. The color map on the right is in units of $\mathrm{W\,m^{-2}\,sr^{-1}}$.}			
		\end{figure*}
		\begin{figure*}[htbp!]
			\centering
			\subfloat[][]{
				\includegraphics[width=0.45\textwidth]{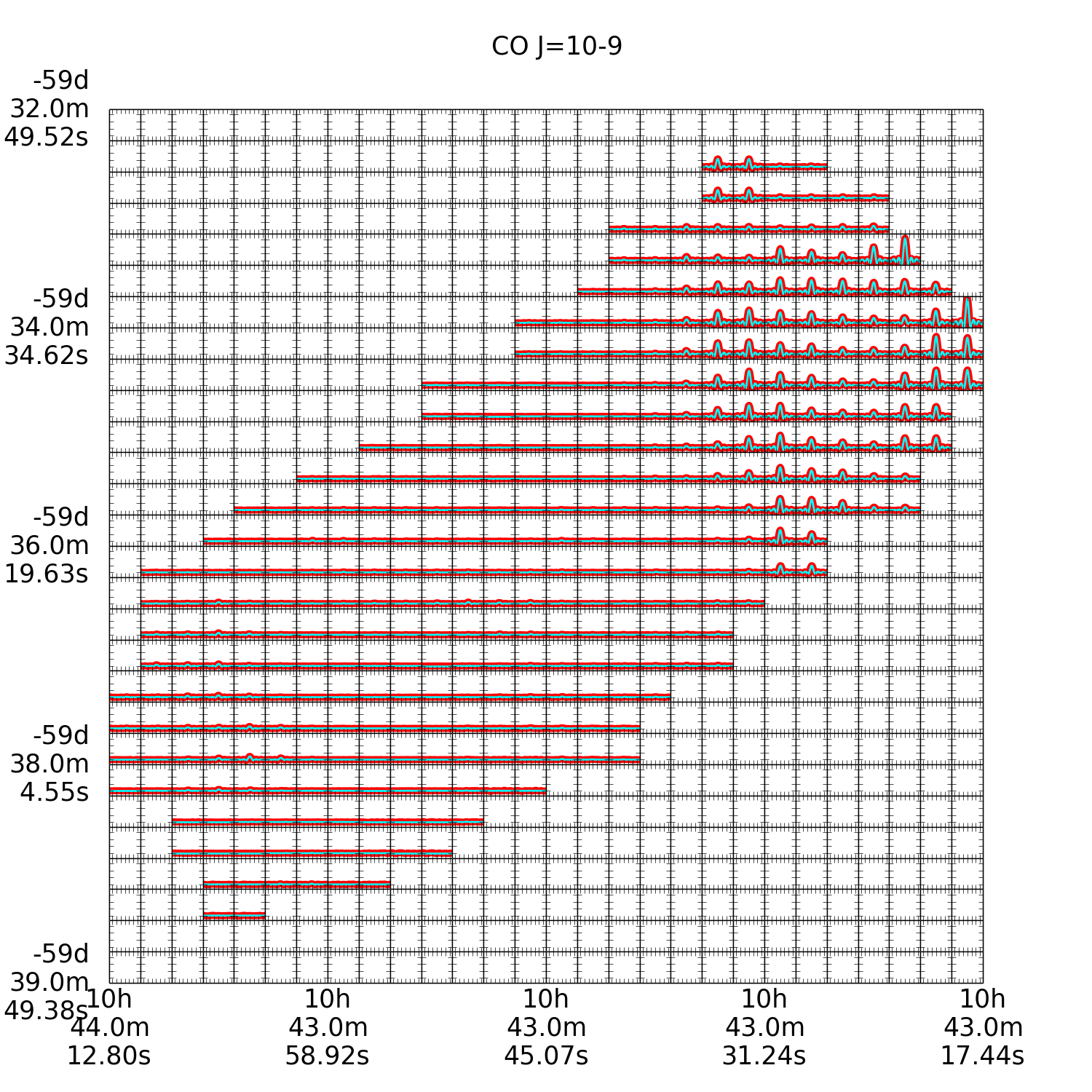}
				\label{co109_linestack}
			}
			\quad
			\subfloat[][]{
				\includegraphics[width=0.45\textwidth]{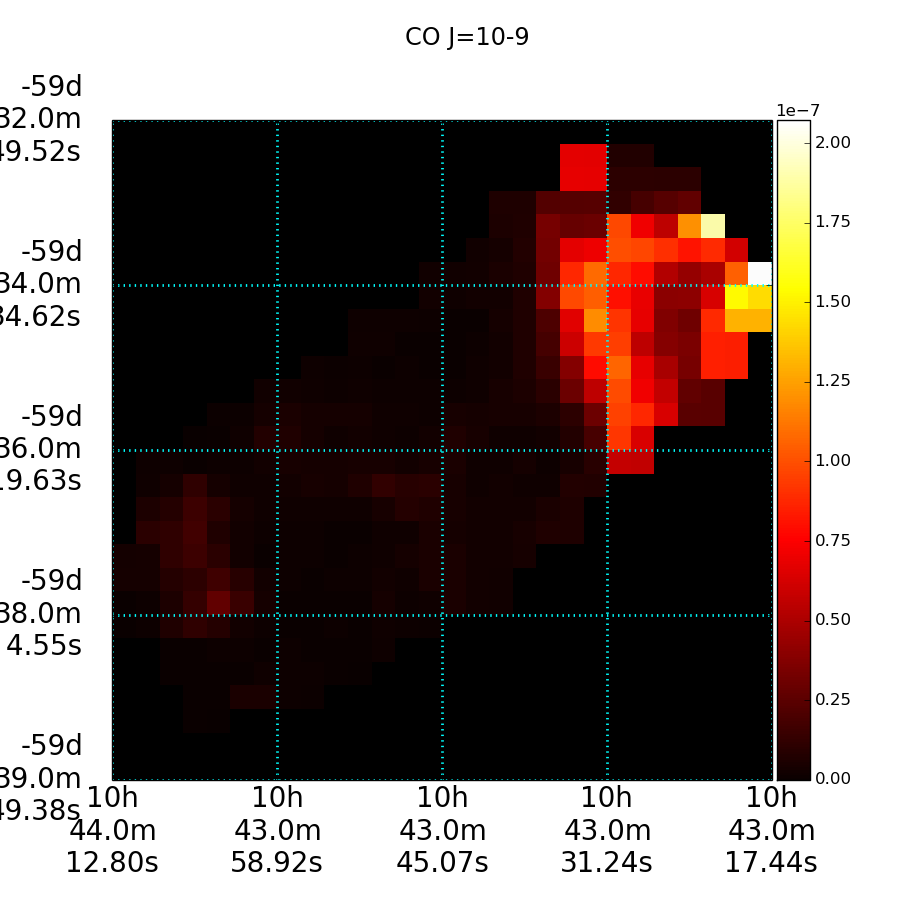}
				\label{co109_intmap}
			}
			\caption{Illustration of the spatial distribution of the observed CO $\mathrm{J}=10-9$ line. The left panel shows the continuum-removed coadded spectrum on every pixel within a range of $1143<\nu<1157\ \mathrm{GHz}$. The vertical axis in each pixel ranges between $-3.0\,\times\,10^{-17}$ and $2.1\,\times\,10^{-16}\,\mathrm{W\ m^{-2}\ sr^{-1}\ Hz^{-1}}$. The color map on the right is in units of $\mathrm{W\,m^{-2}\,sr^{-1}}$.}			
		\end{figure*}
		\begin{figure*}[htbp!]
			\centering
			\subfloat[][]{
				\includegraphics[width=0.45\textwidth]{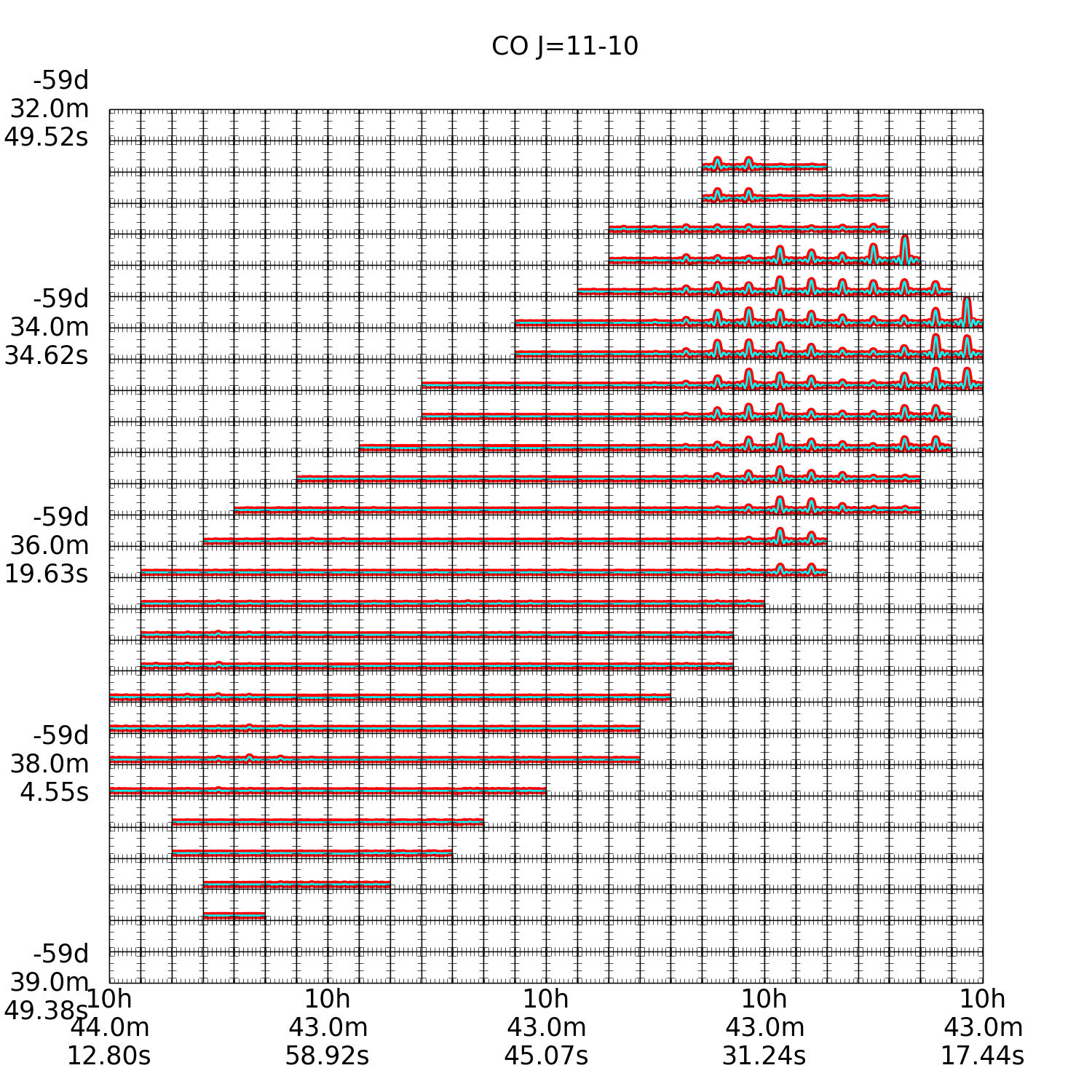}
				\label{co1110_linestack}
			}
			\quad
			\subfloat[][]{
				\includegraphics[width=0.45\textwidth]{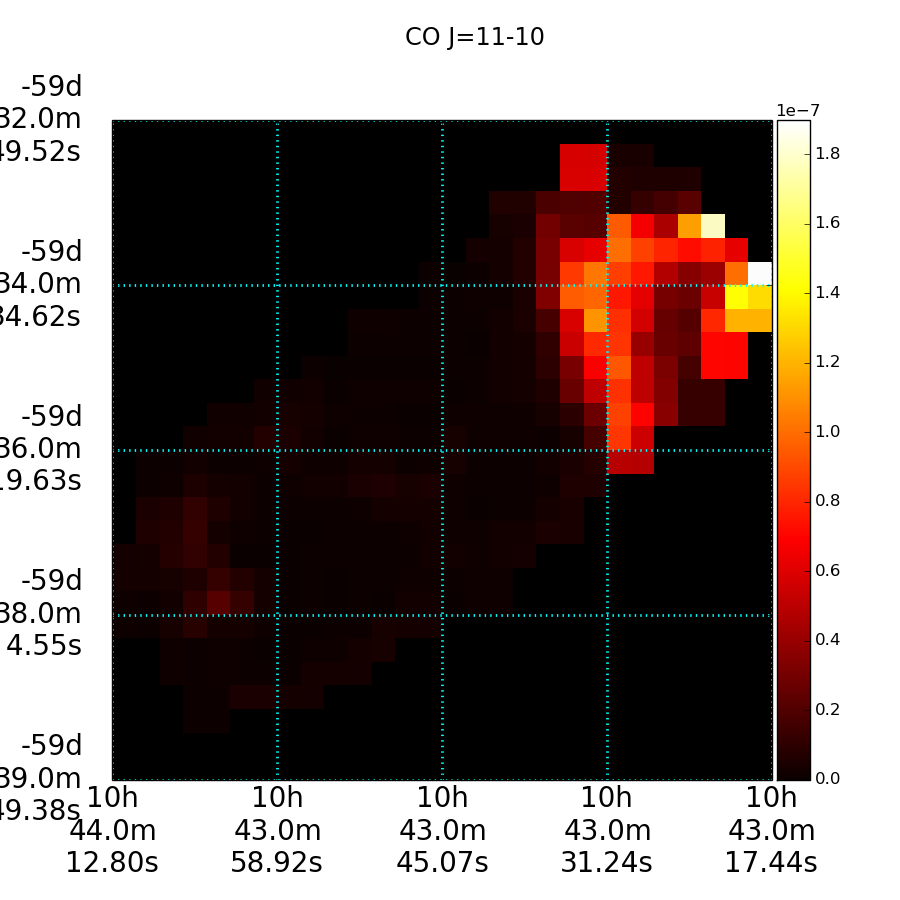}
				\label{co1110_intmap}
			}
			\caption{Illustration of the spatial distribution of the observed CO $\mathrm{J}=11-10$ line. The left panel shows the continuum-removed coadded spectrum on every pixel within a range of $1258<\nu<1272\ \mathrm{GHz}$. The vertical axis in each pixel ranges between $-3.0\,\times\,10^{-17}$ and $2.0\,\times\,10^{-16}\,\mathrm{W\ m^{-2}\ sr^{-1}\ Hz^{-1}}$. The color map on the right is in units of $\mathrm{W\,m^{-2}\,sr^{-1}}$.}			
		\end{figure*}
		\begin{figure*}[htbp!]
			\centering
			\subfloat[][]{
				\includegraphics[width=0.45\textwidth]{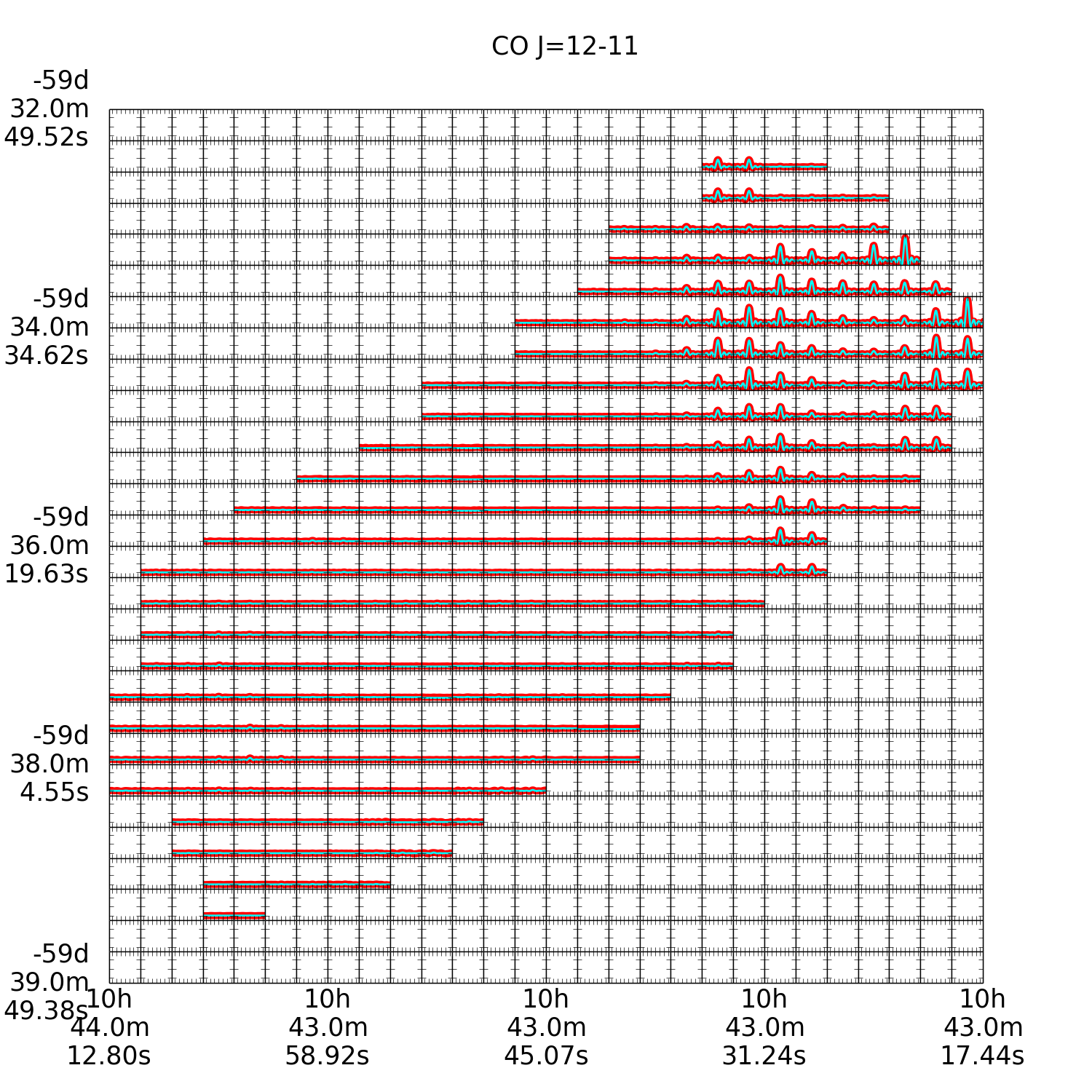}
				\label{co1211_linestack}
			}
			\quad
			\subfloat[][]{
				\includegraphics[width=0.45\textwidth]{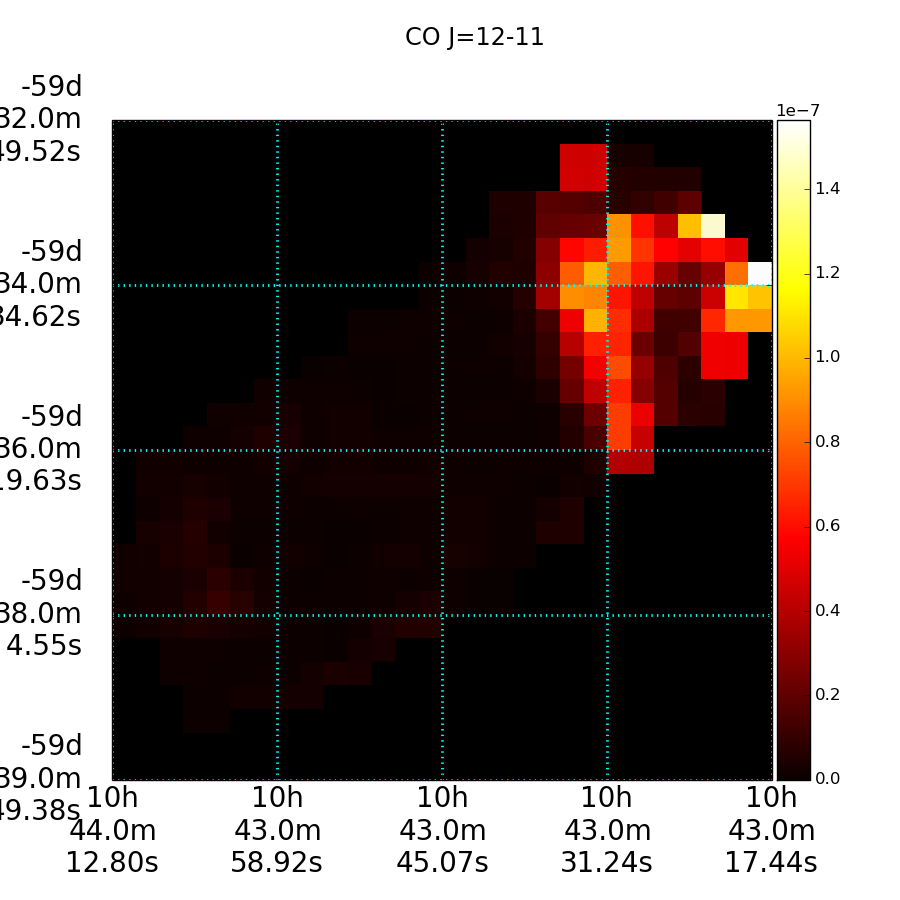}
				\label{co1211_intmap}
			}
			\caption{Illustration of the spatial distribution of the observed CO $\mathrm{J}=12-11$ line. The left panel shows the continuum-removed coadded spectrum on every pixel within a range of $1372<\nu<1387\ \mathrm{GHz}$. The vertical axis in each pixel ranges between $-3.0\,\times\,10^{-17}$ and $1.6\,\times\,10^{-16}\,\mathrm{W\ m^{-2}\ sr^{-1}\ Hz^{-1}}$. The color map on the right is in units of $\mathrm{W\,m^{-2}\,sr^{-1}}$.}			
		\end{figure*}
		\begin{figure*}[htbp!]
			\centering
			\subfloat[][]{
				\includegraphics[width=0.45\textwidth]{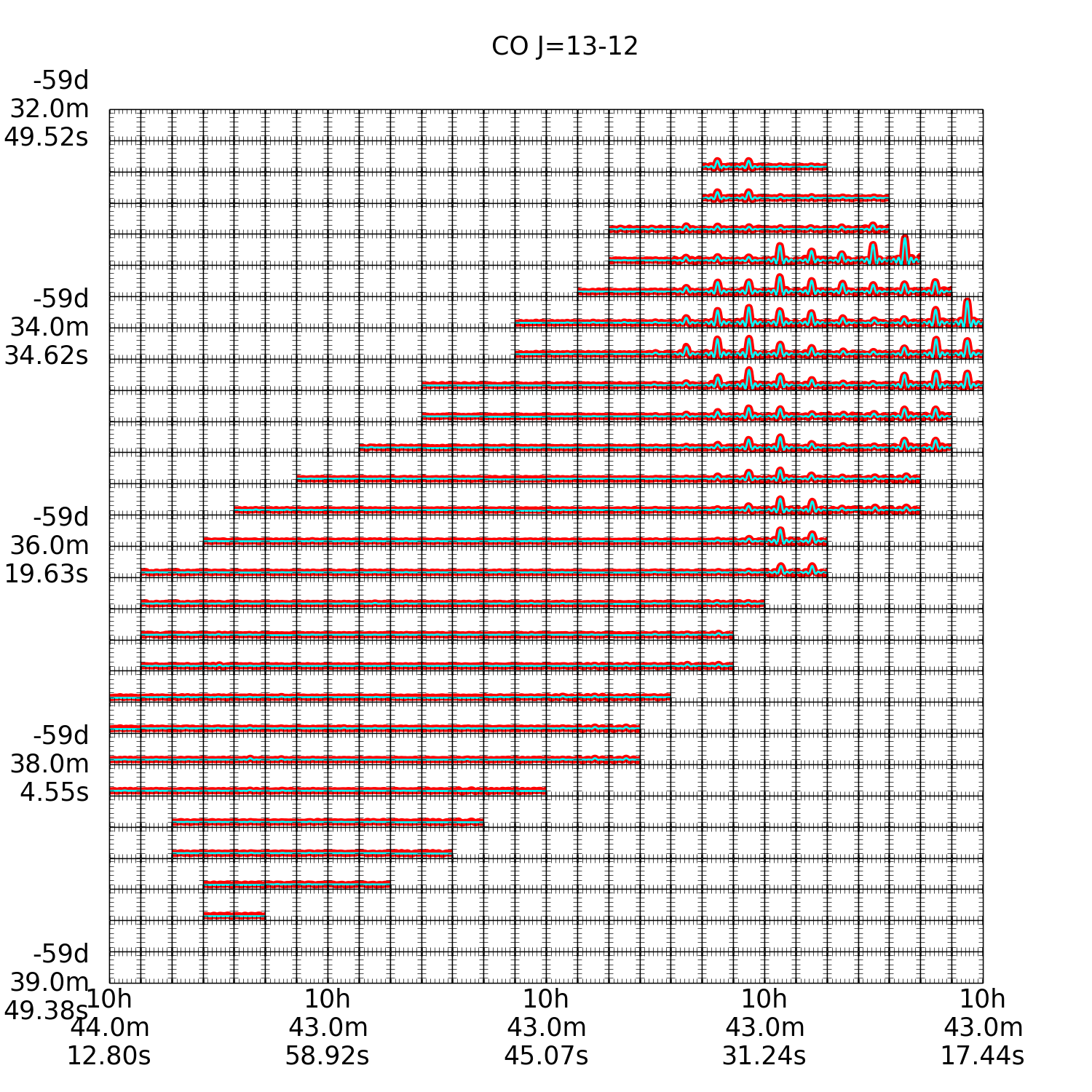}
				\label{co1312_linestack}
			}
			\quad
			\subfloat[][]{
				\includegraphics[width=0.45\textwidth]{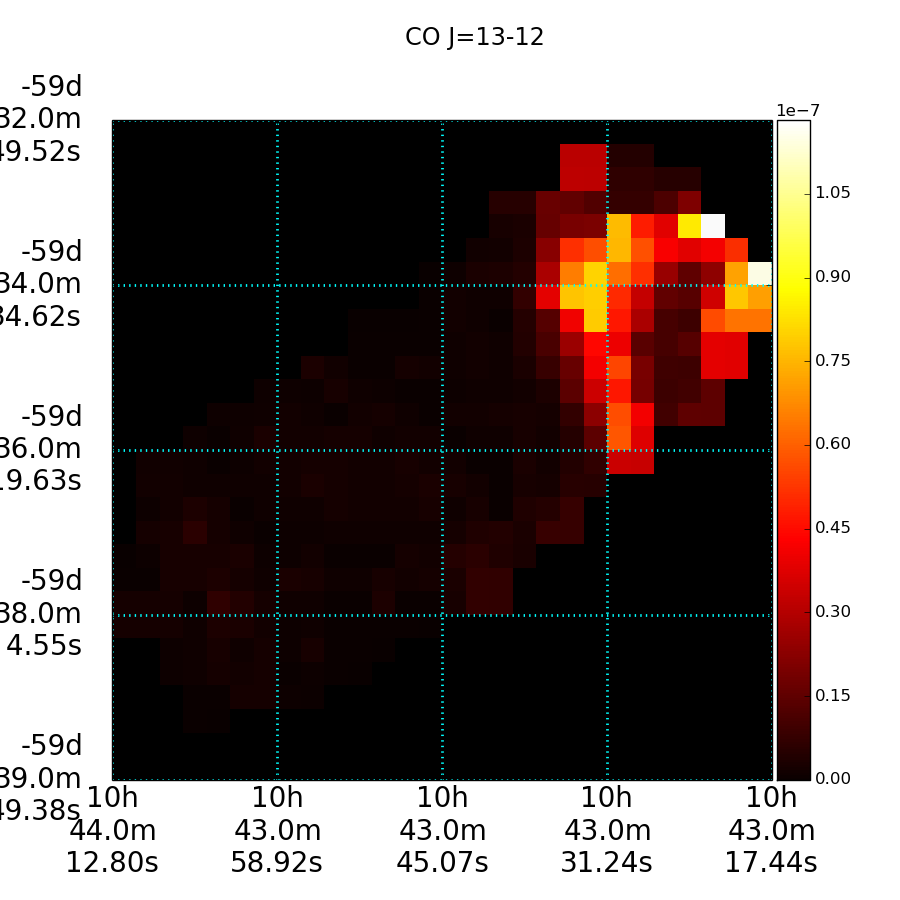}
				\label{co1312_intmap}
			}
			\caption{Illustration of the spatial distribution of the observed CO $\mathrm{J}=13-12$ line. The left panel shows the continuum-removed coadded spectrum on every pixel within a range of $1487<\nu<1502\ \mathrm{GHz}$. The vertical axis in each pixel ranges between $-2.0\,\times\,10^{-17}$ and $1.2\,\times\,10^{-16}\,\mathrm{W\ m^{-2}\ sr^{-1}\ Hz^{-1}}$. The color map on the right is in units of $\mathrm{W\,m^{-2}\,sr^{-1}}$.}			
		\end{figure*}
		\begin{figure*}[htbp!]
			\centering
			\subfloat[][]{
				\includegraphics[width=0.45\textwidth]{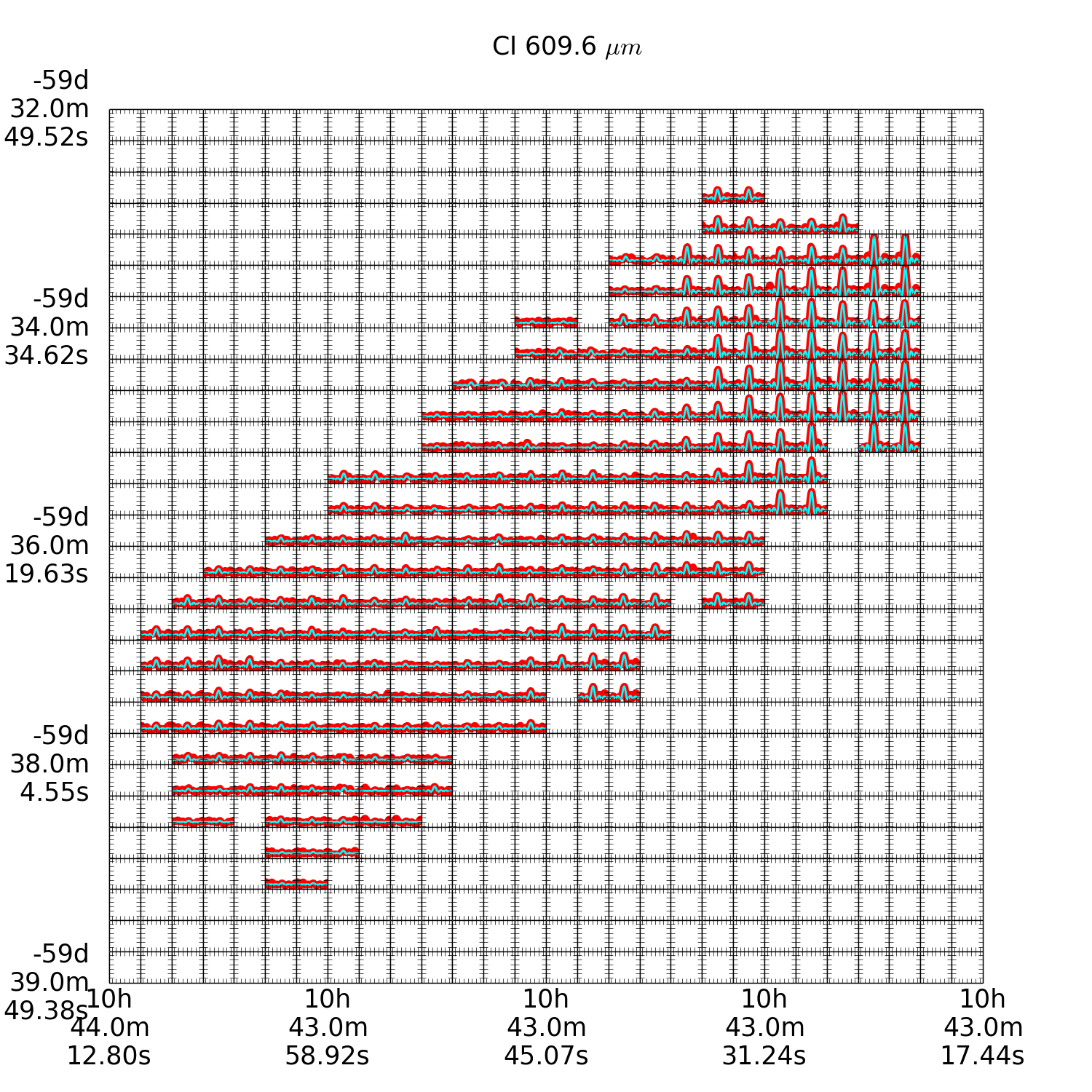}
				\label{ci609_linestack}
			}
			\quad
			\subfloat[][]{
				\includegraphics[width=0.45\textwidth]{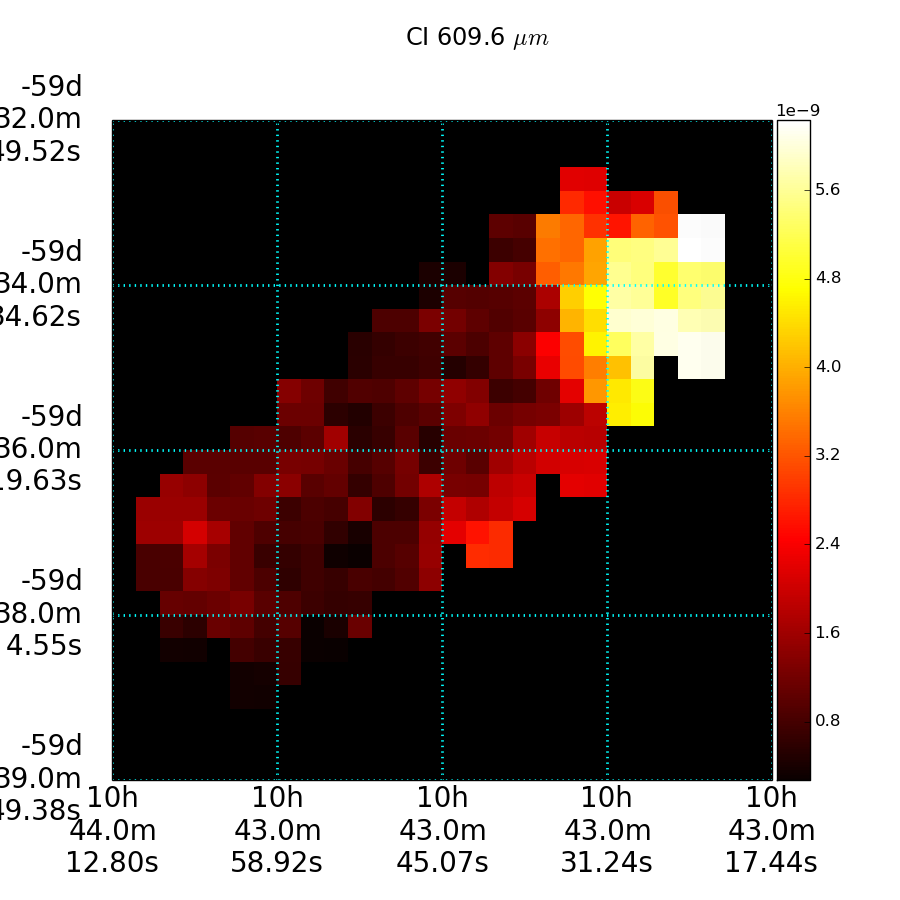}
				\label{ci609_intmap}
			}
			\caption{Illustration of the spatial distribution of the observed $\cison$ line. The left panel shows the continuum-removed coadded spectrum on every pixel within a range of $484<\nu<499\ \mathrm{GHz}$. The vertical axis in each pixel ranges between $-1.0\,\times\,10^{-18}$ and $6.2\,\times\,10^{-18}\,\mathrm{W\ m^{-2}\ sr^{-1}\ Hz^{-1}}$. The color map on the right is in units of $\mathrm{W\,m^{-2}\,sr^{-1}}$.}			
		\end{figure*}
		\begin{figure*}[htbp!]
			\centering
			\subfloat[][]{
				\includegraphics[width=0.45\textwidth]{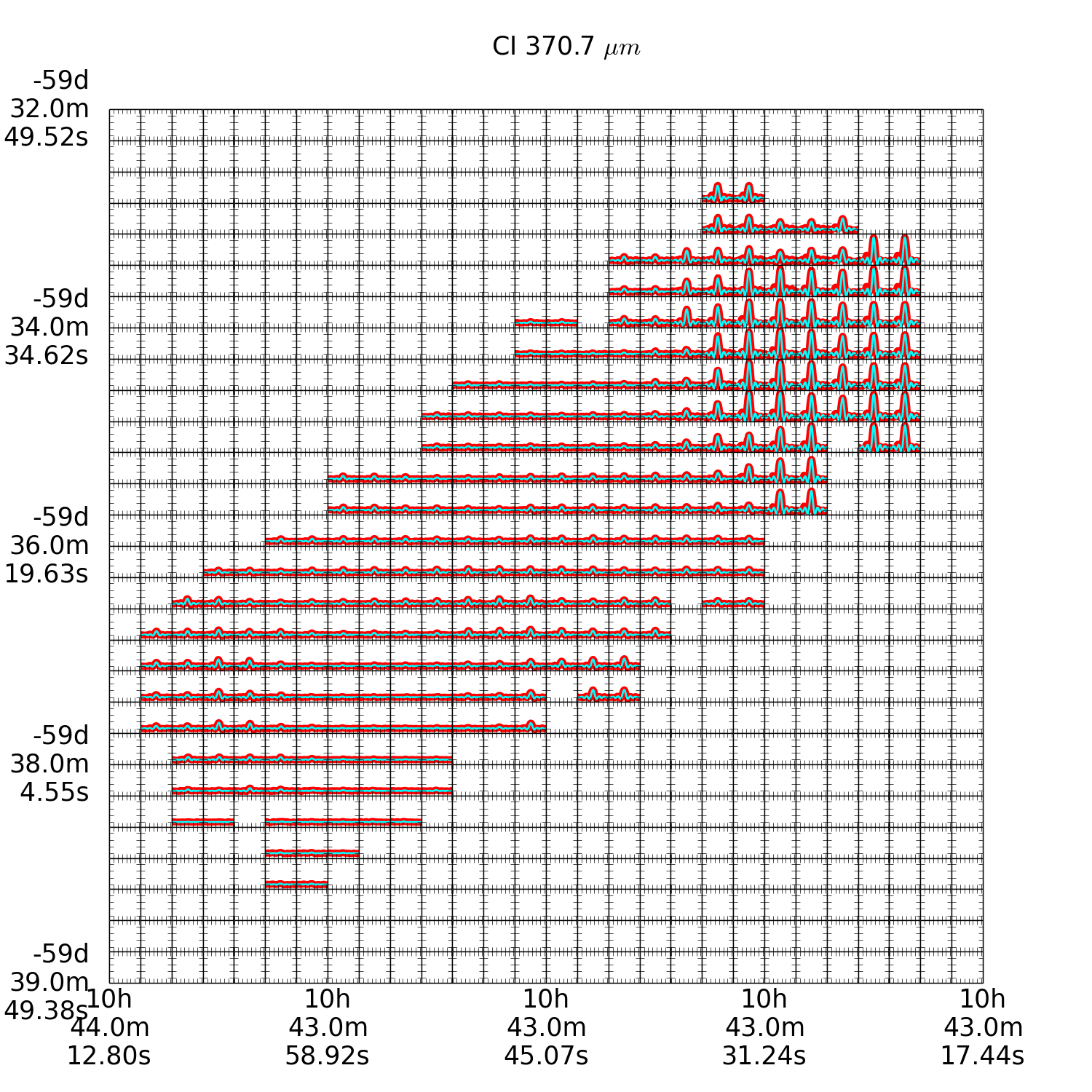}
				\label{ci370_linestack}
			}
			\quad
			\subfloat[][]{
				\includegraphics[width=0.45\textwidth]{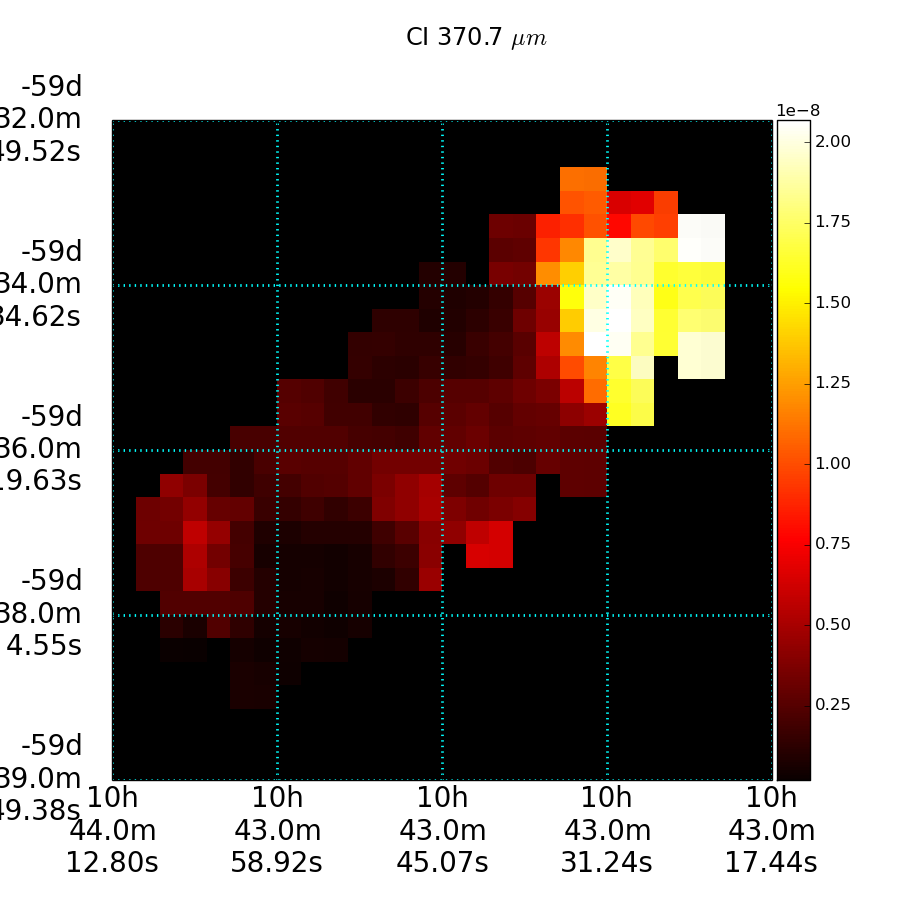}
				\label{ci370_intmap}
			}
			\caption{Illustration of the spatial distribution of the observed $\cits$ line. The left panel shows the continuum-removed coadded spectrum on every pixel within a range of $801<\nu<815\ \mathrm{GHz}$. The vertical axis in each pixel ranges between $-5.0\,\times\,10^{-18}$ and $2.2\,\times\,10^{-17}\,\mathrm{W\ m^{-2}\ sr^{-1}\ Hz^{-1}}$. The color map on the right is in units of $\mathrm{W\,m^{-2}\,sr^{-1}}$.}			
		\end{figure*}

%
%
%


\end{document}